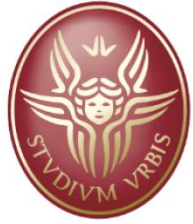

FACOLTÀ DI SCIENZE MATEMATICHE, FISICHE E NATURALI

Thesis work for the Master in Astronomy and Astrophysics
presented in July 2011 and updated in January 2013

# THE MEASUREMENT OF SOLAR DIAMETER AND LIMB DARKENING FUNCTION WITH THE ECLIPSE OBSERVATIONS

Author : Dr. Andrea Raponi
Supervisor : Prof. Paolo De Bernardis
Assistant Supervisor : Dr. Costantino Sigismondi

# Contents

# Introduction

Is the Sun's diameter variable over time? This question first requires the definition of a solar edge. This is why a discussion of the solar diameter and its variations must be linked to the discussion of the so-called Limb Darkening Function (LDF) i.e. the luminosity profile at the solar limb.
Despite the observation of the Sun begins with Galileo Galilei and despite the advent of space technology, the problem of variation of the solar diameter seems still far from being resolved. The reason is that we are interested in changes that (if they exist) are extremely small compared to the solar radius, and many factors prevent the achievement of such precise astrometry.
So why are we interested in such measures? Any additional information on the behavior of the Sun can help us to better understand its internal structure that still has many uncertain aspects. Certainly the investigation of the variability of the diameter can greatly help to discern between different solar models available up today.
We have some clues about the solar variability thanks to the recent discovery of the variation of so-called Total Solar Irradiance over a solar cycle of 11 years. This subject is thus extremely important also to better understand the influence of the Sun on Earth's climate.
The goal of this study is the introduction of a new method to perform astrometry of high resolution on the solar diameter from the ground, through the observation of eclipses. Since the definition of the exact position of the solar edge passes through the detection of the limb profile (LDF), we have also obtained a good opportunity to get information on the solar atmosphere. This could be another use of the method that we introduce here.

# Summary

**Chapter 1.** The subject of solar variability is introduced.
In section 1.1 the magnetic phenomena and Total Solar Irradiance are analyzed. Their variations, linked in some way, provide us with the main lines of investigation to explore the variability of the Sun's parameters.
Section 1.2 shows how the topic is of interest to study the Earth's climate. In particular, it addresses the issue of little ice ages.
Section 1.3 addresses possible causes of TSI variability: surface magnetism (1.3.1), photospheric temperature (1.3.2) and radius variation (1.3.3).
Section 1.4 shows how luminosity, temperature and radius are linked each other regarding the Sun's variability (1.4.1). Moreover an analytical model (1.4.2) and a numerical simulation (1.4.3) are shown as examples to show how theoretical efforts can be compared with the measures of radius variation.
Section 1.5 introduces the interesting topic of the past measures of solar diameter. Measurements of 17th century are analyzed, and the surprising high values are highlighted.

**Chapter 2.** The observation of the LDF needs to take into account many aspects.
Section 2.1 shows how the true limb profile could be different from the observed limb profile especially because of the Earth's atmosphere. Here the solar limb is defined.
In section 2.2 the effect of the instruments of observation on the LDF is discussed.
The detection of a LDF requires the specification of the band pass of observation. The dependence of the LDF on the wavelength is discussed in section 2.3.
Solar features like asphericity (2.4.1) and magnetic structures (2.4.2) have to be taken into account in the comparison of different LDF profiles. This question is discussed in section 2.4.

**Chapter 3.** The methods for the measurement of the solar diameter are shown.
In section 3.1 the principles of the heliometers (3.1.1) and its space versions (3.1.2) are explained.
Section 3.2 shows the drift-scan methods.
Section 3.3 shows how to measure the solar diameter with the planetary transits. The problem of the so called black-drop is discussed as well.

**Chapter 4.** Solar diameter can be inferred by eclipse observations. Here the ancient, the modern and the new methods for this goal are shown.
In section 4.1 we show how a lower limit of the Sun's radius can be inferred just from a right eyewitness, in particular from the eclipse in 1567 observed by Clavius (4.1.1) and from the eclipse in 1715 observed by Halley (4.1.2).
In section 4.2 we summarize the path of the modern observation of eclipse taking advantage of the Baily's beads (4.2.1) and the new lunar map by Kaguya space mission (4.2.2). Problems about recent observations are discussed (4.2.3).
In section 4.3 we propose the new method. The LDF achieved is discretized in solar layers. The goal of this method is the detection of the Inflection Point Position, but the LDF obtained can be useful also for studies about solar atmosphere.

**Chapter 5.** Here it is shown how to put into practice the method explained in section 4.3.
Section 5.1 shows the application of the method for the annular eclipse in January 15, 2010 observed by R. Nugent in Uganda and A. Tegtmeier in India. The results are discussed.
Section 5.2 shows how the method proposed can be greatly simplified. A list of steps is suggested to make this method accessible .
In section 5.3 we do some recommendations for future observations.

**Chapter 6.** Here the assumption of negligible errors on ephemeris is discussed. We consider the sources of Occult 4 software in section 6.1.
In section 6.2 the role of the atmosphere parameters in the determination of the topocentric ephemeris is studied.
In section 6.3 it is discussed how to treat the measures without the assumption of negligible errors on ephemeris.

**Chapter 7.** Section 7.1 shows how to get information on solar atmosphere thanks to the LDF measure. The temperature profile is also discussed.
Section 7.2 mentions the construction of a solar atmospheric model (7.2.1). A work on LDF models is shown as example to show how theoretical efforts can be compared with the measures of LDF to choose the better solar atmospheric model (7.2.2).
Finally the solar network pattern and its influence on the LDF is discussed in section 7.3.

# CHAPTER 1
# The variability of the solar parameters

## 1.1. Magnetic phenomena and Total Solar Irradiance

The Sun has lost its image as an immutable star since Galileo pointed his telescope and identified sunspots, refuting the idea that they were objects transiting in front of the Sun, but rather part of the surface of the Sun itself.

To date, we conceive the variability of the Sun at all time scales: from the evolutionary time scale, to the variability of hours due to magnetic phenomena on the surface. But despite being more accessible to human time scales, the variability on shorter time scales still represent a challenge in terms of general understanding of their events. This is due to the complication that arises from the fact that small-scale variability, such as the 11-year cycle, are more due to complicated magnetic phenomena.

Despite the observation of the changing nature of the solar surface dates back to the 17th century, the idea of the immutability of the solar luminosity has been unhinge only recently. "Solar constant" was in fact the name that A. Pouillet gave in 1837 to the total electromagnetic energy received by a unit area per unit of time, at the mean Sun-Earth distance (1AU). Today we refer to this parameter more correctly as Total Solar Irradiance (TSI). The difficulties in measuring the TSI through the ever-changing Earth atmosphere precluded a resolution of the question about its immutability prior to the space age. Today there is irrefutable evidence for variations of the TSI. Measurements made during more than two solar cycles show a variability on different time-scales, ranging from minutes up to decades, and likely to last even longer. Despite of the fact that collected data come from different instruments aboard different spacecrafts, it has been possible to construct a homogeneous composite TSI time series, filling the different gaps and adjusted to an initial reference scale. The most prominent discovery of these space-based TSI measurements is a 0.1% variability over the solar cycle, values being higher during phases of maximum activity (Pap 2003).

The link between solar activity and TSI has certainly paved the way for a deeper understanding of the solar physics, and for a debate on the role of the Sun in Earth's climate. The study of the Sun has always provided by nature in this dual role.

## 1.2. Little Ice Ages

The relation between solar activity and TSI give a direct link between solar activity variation and terrestrial climate variation. The most intriguing example is the overlap between the long-term sunspot absences and the Little Ice Ages.

The Maunder Minimum was a period of almost a century during the 17th century when European astronomers observed a very low solar activity. The same time is remembered as Little Ice Age because it recorded average temperatures about 1.3 K lower than the current.

The lack of solar activity observations prior to 1610 is bypassed by other indirect measures, such as 14C in tree rings or 10B in ice cores. High concentration of these isotopes indicate increase in the flux of cosmic rays and then a decrease in solar activity. These measures still show an agreement between solar activity and temperature of the period and in addition, extending the time horizon, they show a quasi-periodic ~ 200 years cycle of activity.

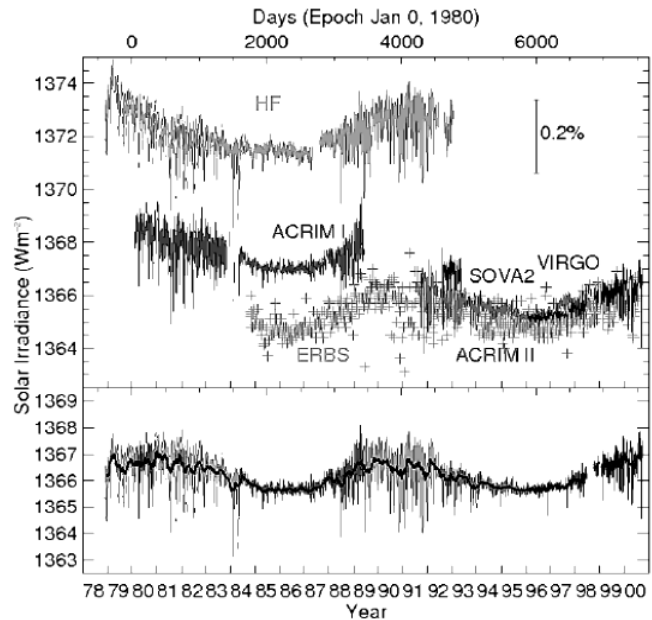

Figure 1.1.1. The various total irradiance time series are presented on the upper panel, the composite total solar irradiance is shown on the lower panel. (J. Pap 2003)

It 's interesting then make predictions on the next minimum of solar activity, of which it seems there are already early signs.

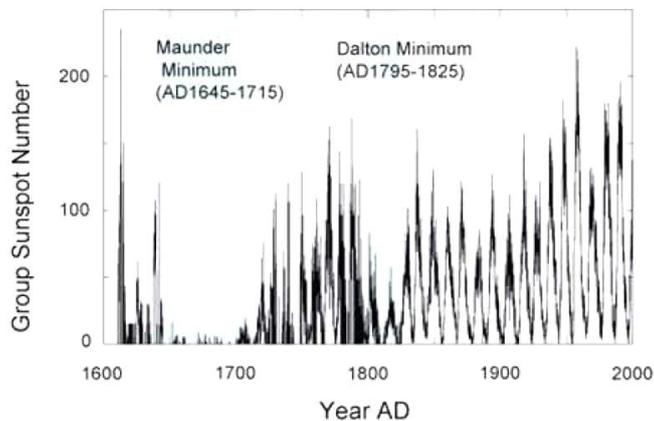

Figure 1.2.1. Observational record of sunspot numbers since AD 1610 (Hoyt & Schatten 1998). Large-scale weakening of solar activity, recognized as a substantial reduction in sunspot numbers, is seen in AD 1645–1715 and in AD 1795–1825.

### 1.3. Causes of variability of the TSI

**1.3.1. Surface magnetism.** The different time-scales found in the TSI data are linked with different physical mechanisms. While p-modes variability may justify variability at the scale of the minute (Woodward and Hudson 1983), granulation and supergranulation may explain variations at the scale of the hour (Frohlich et al 1997).

Presently, the most successful models assume that surface magnetism is responsible for TSI changes on time-scales of days to years (Foukal 1992, Solanki et al. 2005). For example, modeled TSI versus TSI measurements made by the VIRGO experiment aboard SOHO, between January

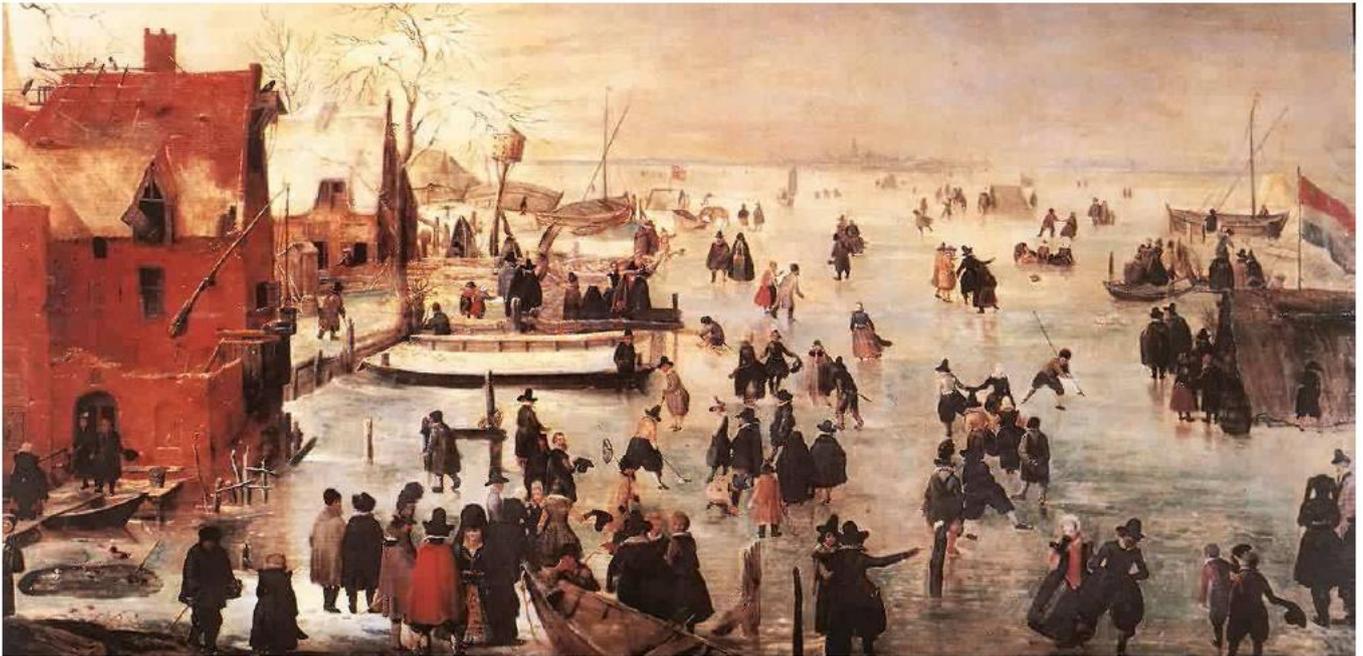
Figure 1.2.2. Hendrick Avercamp (1585-1643) painted many scenes of life on the frozen canals of Amsterdam during the Maunder Minimum. Those canal do not freeze anymore or not to the same extent.

1996 and September 2001, show a correlation coefficient of 0.96 (Krivova et al. 2003). However, significant variations in TSI remain unexplained after removing the effect of sunspots, faculae and the magnetic network (Pap et al. 2003). Moreover, there is a phase shift between TSI and magnetic variations at the beginning of solar cycles 22 and 23. In addition, solar cycle 23 is much weaker than the two previous cycles, in terms of magnetic strength, while TSI remains at about the same level (Pap 2003). In the same way, de Toma et al. (2001) reported that TSI variations in solar cycle 23 implied additional mechanisms, other than surface magnetic features alone. Finally, Kuhn (2004) concluded that "sunspots and active region faculae do not explain the observed irradiance variations over the solar cycle" (Rozelot et al. 2004).

**1.3.2. Observed photosferic temperature variation.** The variability of the photosferic temperature with the solar cycle is still a matter of study. Detection of solar temperature variation with the required precision is hard to set up. Gray & Livingston (1997), using the ratios of spectral line depths as indicators of stellar effective temperature, showed that the observed variation of photospheric temperature is in phase with the solar cycle. The amplitude $dT = 1.5 \pm 0.2$ K found by these authors can account for nearly the entire variations of total solar irradiance during the solar cycle. However, the measurements of Gray & Livingston while free from the effects surface magnetic feature, depend on a calibration coefficient that relates the variation of the photospheric temperature to the variation of the depth of the observed spectral lines. They obtained this correlation coefficient empirically from observations of six stars with colors identical to the Sun. However, Caccin & Penza (2002) noted that the gravitational acceleration g for all these stars was not the same, and through theoretical calculations they found a g-dependence of the correlation coefficient.
Recently, monitoring the spectrum of the quiet atmosphere at the center of the solar disk during thirty years at Kitt Peak, Livingston and Wallace (2003) and Livingston et al. (2005) have shown a "nearly immutable basal photosphere temperature" during the solar cycle within the observational accuracy, i.e. dT = 0 with error bars of 0.3 K (Fazel et al. 2005).

**1.3.3. Observed radius variation.** The solar radius is the global property with the most uncertain determination of the solar cycle–related changes. The results are very contradictory, and the problem is far from being settled: e. g., Gilliland (1981) suggested (a) a 76-year period with a half-amplitude of about 0.2 arcsec and maxima in 1911 and 1987, (b) an 11-year period with a half-amplitude of about 0.1 arcsec and minima at times with maximum sunspot number, and (c) a secular decrease of about 0.1 arcsec per century over the past 265 years. Sofia, Demarque, and Endal (1985) proposed a 90-year cycle with a half-amplitude of nearly 0.5 arcsec and maxima in 1887 and 1977. Sofia et al. (1983) found that at the solar eclipse on 24 January 1925 the solar radius was 0.5 arcsec larger than it was at the eclipse on 26 February, 1979. More recently, Wittmann, Alge, and Bianda (1993) reported an increase of about 0.4 arcsec in the 10-year interval between 1981 and 1990 -1992 (Li et al. 2003).

The lack of agreement between near simultaneous groundbased measurements at different locations suggests that atmospheric contamination is so severe that it prevents any meaningful measurement of solar diameter variations. Moreover the lack of coherence of the set of the values obtained from different observers can also be explained by the lack of strategy: different instrumental characteristics, different choice of wavelength and different processing methods. The measurements from space are very few and they also have many sources of doubt. The Michelson Doppler Imager (MDI) on board the Solar and Heliospheric Observatory (SOHO) indicates a change of no more than 23 ±9 mas (milliarcseconds) in phase with solar cycle (Bush, Emilio, and Kuhn, 2010), but because the design of the instrument was not optimized for astrometry, significant corrections had to be applied to the measurements, which introduce some uncertainty in the results.

The Solar Disk Sextant (SDS) was specifically designed for this purpose (Sofia, Heaps, & Twigg 1994), and five flights have been carried out. The measurements between 1992 and 1996 show a variation of 0.2 arcseconds in anti phase with respect to the solar cycle (Egidi et al. 2006). Although milliarcsecond sensitivity was achieved this result seems to be too large according to the studies on helioseismology, in particular to the f-mode frequencies (although it should be noted that the radius determined from f-mode oscillations represents changes at depths from 4 to 10 Mm below the solar surface, so comparisons must be made with care).

In chapter 3 it will be exposed in detail the methods of measurement from the ground and space mission currently underway.

## 1.4. Theoretical models

**1.4.1. Global parameters.** Because the surface magnetism can't explain all the variation of the TSI (or equivalently the Luminosity L) over a solar cycle, it is presented the need of a model that take in account the variation of the temperature and the radius of the Sun. We have the simple relation:

$$L_\odot = 4\pi R_\odot^2 \sigma T_{eff}^4$$

Being Δ the Stefan–Boltzmann constant. The relative variation is:

[1]$\Delta L/L = 2\Delta R/R + 4\Delta T_{eff}/T_{eff}$

---

[1] Hereafter we would imply ☉

For a full understanding of changes in solar parameters the goal then is to find another relation that binds the temperature to the radius, through the so called "luminosity-asphericity parameter":

$$\omega = \Delta \ln R / \Delta \ln L$$

The first contribution in this direction is due to S. Sofia and A.S. Endal that in 1979 identified in the convective envelope the possible cause of a solar variability that involve Radius and Temperature: *"Because of the unstable nature of turbolence, stochastic changes in the efficiency of convection may occur. An increase of the efficiency of convection leads to a shrinkage of the convective region, thus releasing gravitational potential energy and adding to the radiative solar flux. Conversely, a decrease of the convective efficiency leads to an expansion of the convective envelope, thus providing a temporary sink of radiative energy."*

**1.4.2. Analytical model.** Among the works that followed this idea, the analytical effort of Callebaut, Makarov and Tlatov (2002) is presented. Their self-consistent calculation starts with the assumption that the expansion interests the layer above αR, with 0 < α < 1. The increase in height in the interval (αR,R) is given by:

$$1)\ h(r) = \frac{(r-\alpha R)^n \Delta R}{R^n (1-\alpha)^n}$$

With *r* the current radial coordinate and *n* = 1,2,3,.. respectively for a linear, quadratic, cubic expansion.
The relative increase in thickness for an infinitesimal thin layer situated at *r* = *R* is:

$$2)\ \left(\frac{dh}{dr}\right)_R = \frac{n\Delta R}{(1-\alpha)R}$$

According to the ideal gas law $p = \rho kT/m$ and the polytropic law $p = k\rho^\Gamma$ this leads to the relative change in temperature at the solar surface:

$$3)\ \left(\frac{\Delta T}{T}\right)_R = (\Gamma - 1)\left(\frac{d\rho}{\rho}\right)_R = -(\Gamma - 1)\left(\frac{dh}{dr}\right)_R$$

Using (2) it can be expressed in terms of the relative change in solar radius:

$$4)\ \left(\frac{\Delta T}{T}\right)_R = -\frac{(\gamma-1)n\Delta R}{(1-\alpha)R}$$

where Γ is replaced by γ, the ratio of the specific heats (supposing an adiabatic process). Using the relation for ΔL/L seen in section 1.4.1 we have:

$$5)\ \frac{\Delta L}{L} = 4\left(\frac{\Delta T}{T}\right)_R + 2\frac{\Delta R}{R} = -2\left(\frac{2n(\gamma-1)}{1-\alpha} - 1\right)\frac{\Delta R}{R}$$

The gravitational energy required for the expansion is given by:

$$6)\ \Delta E_{exp} = 4\pi \int_{\alpha R}^{R} \rho g(r) h(r) r^2 dr$$

where the density in the upper region of the Sun, according to Priest (1985) may be approximated by:

$$7)\ \rho = 8.91 \cdot 10^3 (1-r)^{2.28}\ kg/m^3$$

Integrating (6), using (5) and assuming $n = 1$ one obtain:

$$8)\ \Delta E_{exp} = \frac{2.60 \cdot 10^{41}(1-\alpha)^{4.28}}{(1/3+\alpha)} \frac{\Delta L}{L}$$

The variation of emitted energy during half a solar cycle is:

$$9)\ \Delta E_{rad} = \Delta L \cdot 5.5y = 7 \cdot 10^{34} \frac{\Delta L}{L}\ J$$

The ratio of the gravitational energy that goes into the radiated energy is assumed 1/2. Equating then twice (9) to the (8) yields α straightforward. Finally taking α and the observed value for ΔL/L = 0.00088 one obtains from (5) ΔR/R and then ΔR = 8.9 km i.e. ΔR =12mas from 1 AU, that is the expected variation of the solar radius in half a solar cycle (in anti phase with solar activity). This calculation is self-consistent in the sense that the single observation of the variation in luminosity is sufficient to fix ΔR, ΔT and the thickness of the layer which expands. However two parameter are free: the shape of expansion and the amount of gravitational energy which goes into radiated energy. Although the assumptions made for these two parameters in this study seem reasonable, a variation of them can give significantly different results.

**1.4.3. Numerical Simulation.** Another work, based on three-dimensional numerical simulations, was made by Li et al. (2003). Their models include variable magnetic fields and turbulence. The magnetic effects are: (1) magnetic pressure, (2) magnetic energy, and (3) magnetic modulation to turbulence. The effects of turbulence are: (1) turbulent pressure, (2) turbulent kinetic energy, and (3) turbulent inhibition of the radiative energy loss of a convective eddy, and (4) turbulent generation of magnetic fields. Using these ingredients they construct many types of solar variability models. Here the results are presented:

| Group (1) | Model (2) | $f$ (3) | $f_1$ (4) | $M_{Dc}$ (5) | $\sigma$ (6) | $B_0$ (G) (7) | $\log P$ (8) | Depth (Mm) (9) | $\Delta \ln L$ (%) (10) | $\Delta \ln R$ ($10^{-6}$) (11) | $\Delta T$ (K) (12) | $\Delta R_{CZ}$ ($R_\odot$) (13) | $\Delta B$ (kG) (14) |
|---|---|---|---|---|---|---|---|---|---|---|---|---|---|
| I | A | 1 | ... | −1.45 | 0.35 | 5000 | 14.00 | 232.57 | 0.056 | 320 | −0.22 | −0.106 | 1200 |
|   | B | 1 | ... | −2 | 0.2 | 2800 | 13.22 | 138.69 | 0.061 | 350 | −0.13 | 0.067 | 690 |
|   | C | 1 | ... | −4.25 | 0.2 | 200 | 10.74 | 20.07 | 0.074 | 47 | 0.94 | 0 | 49 |
|   | D | 1 | ... | −6.5 | 0.2 | 17 | 8.46 | 4.45 | 0.078 | 10 | 1.09 | 0 | 4.2 |
|   | E | 1 | ... | −7.5 | 0.2 | 7 | 7.46 | 2.43 | 0.080 | 9.1 | 1.13 | 0 | 1.7 |
|   | F | 1 | ... | −9.5 | 0.2 | 0.6 | 5.53 | 0.38 | −0.076 | 1.4 | −1.11 | 0 | 0.15 |
| II | G | 3 | ... | −4.25 | 0.2 | 120 | 10.74 | 20.07 | 0.074 | 16 | 1.02 | 0 | 30 |
|   | H | 3 | ... | −6.5 | 0.2 | 10 | 8.46 | 4.45 | 0.080 | 3.2 | 1.15 | 0 | 2.5 |
|   | I | 3 | ... | −7.5 | 0.2 | 3.8 | 7.46 | 2.43 | 0.074 | 2.0 | 1.06 | 0 | 1.4 |
|   | J | 3 | ... | −9.5 | 0.2 | 0.55 | 5.53 | 0.38 | −0.083 | 0.45 | −1.20 | 0 | 0.15 |
| III | K | 1 | 0 | −9.5 | 0.2 | 0.7 | 5.53 | 0.38 | −0.109 | 2.2 | −1.59 | 0 | 0.17 |
|   | L | 3 | 0 | −9.5 | 0.2 | 0.6 | 5.53 | 0.38 | −0.109 | −0.04 | −1.46 | 0 | 0.15 |

Table 1. Models for which the magnetic field has a Gaussian profile. (Li et al. 2003)

| Group (1) | Model (2) | $f$ (3) | $f_1$ (4) | $f_2$ (5) | $f_3$ (6) | $\Delta \ln L$ (%) (7) | $\Delta \ln R$ ($10^{-6}$) (8) | $\Delta T$ (K) (9) | $\Delta R_{CZ}$ ($R_\odot$) (10) | $\Delta B$ (kG) (11) |
|---|---|---|---|---|---|---|---|---|---|---|
| I | M | 1 | ... | 0.04 | ... | −0.105 | 2.5 | −1.62 | 0 | 0.27 |
| II | N | 3 | ... | 0.022 | ... | −0.105 | 0.7 | −1.52 | 0 | 0.20 |
| III | O | 1 | 0 | 0.036 | ... | −0.099 | 0.2 | −1.42 | 0 | 0.26 |
|   | P | 3 | 0 | 0.11 | ... | −0.099 | 0.6 | −1.33 | 0 | 0.45 |
| IV | Q | 3 | 0.155 | 0.08 | 6.4 | −0.066 | −5.0 | −0.94 | 0 | 0.38 |
|   | R | 1 | 0.155 | 0.08 | 6.4 | 0.106 | −3.7 | 1.54 | 0 | 0.38 |

Table 2. Models for which the magnetic energy density is assumed to be proportional to the turbulent kinetic energy density. (Li et al. 2003)

The further subdivision into groups indicates the growing importance of turbulence. In both Tables 1 and 2, column (1) marks the group, column (2) names the model, column (3) shows the parameter f to indicate if turbulence affects the radiation loss of a convective eddy (1, no; 3, some), and column (4) indicates whether the turbulent kinetic energy (and turbulent pressure) is included or not (ellipses denote that they are not included).

In Table 1, column (5) lists MDc that is the depth of the peak of the applied magnetic field, column (6) lists the width of the Gaussian profile, and column (7) lists B0 proportional to the amplitude of the field. Columns (8) and (9) express MDc in terms of the pressure variable (used in turbulence simulations) and physical depth, respectively. Columns (10)–(12), respectively, list the variations of global solar parameters L, Teff , and R in half a solar cycle (+ for a variation in phase and - in antiphase with the solar cycle). The last two columns give the corresponding variations for the convective depth and the maximal values of the magnetic fields.

In Table 2, column (5) lists f2, which specifies the ratio of the magnetic energy over the turbulent kinetic energy. Column (6) lists f3, which indicates the depth dependence of the magnetic modulation of turbulence. Columns (7)–(11) are the same as columns (10)–(14) of Table 1. In this work the authors select the only model that fits all the observational constraints: the R-model, from which $\Delta R$ = 2.6 km i.e. $\Delta R$ = 3.5 mas in antiphase with the solar cycle, from 1 AU. However the constraints they used are uncertain and the simulation needs for improvement as stated by the authors.

## 1.5. Measures in the 17th century

It is interesting to know the global parameters of the Sun during the period of low solar activity in the 17th century (the Maunder minimum). Ribes and Nesme-Ribes (1993) reported the observations recorded at the Observatoire de Paris from 1660-1719 taken by French astronomers, including J. Picard, and the Italian director G. D. Cassini.

In addition to the collection of quantitative sunspot observations, they made some interesting observations about the solar diameter, even though the goal of the astronomer did not focus on the determination of the solar diameter. In fact the rotation rate of sunspots is derived from the angular distance between successive positions of the spot with respect to the solar diameter. Then it was necessary to assess the accuracy in the determination of the solar diameter first. Auzout (1729) wrote in the summer of 1666: *"we are able to measure the diameter with an accuracy of one arcsecond ...At its apogee, the horizontal solar diameter is 31'37 arcsec or 31'38 arcsec, never 31'35 arcsec. At the perigee, it can reach 32'45 arcsec or perhaps 32'44 arcsec"*. We know that ground observations are affected by the Earth's atmosphere, which increases the apparent size. This is known as "the irradiation effect". A correction of 3.1 arcseconds is usually assigned to ground observations to obtain the correct solar diameter (Auwers 1891). Applying this correction there is still a difference of about 7 arcseconds in the diameter for these 17th century data with respect to the current data. According to theoretical predictions, this value seems to be surprisingly large. For example, in analytical work seen in section 1.4.2, the authors argue for the Maunder Minimum $\Delta R$ = 88 mas assuming $\Delta L$ about 4 time larger than an ordinary solar cycle. The observed value is then about forty times higher than expected. This difference leads us to suspect and to investigate the correctness of the measures of the astronomers. Ribes and Nesme-Ribes describe in detail the measurement methods used at the time: *"Using a 9 1/2-foot quadrant fitted with telescopic sights, observations were carried out at the meridian, when the Sun was moving roughly parallel to the horizon. Filars were displaced by a screw. When the filars of the eyepiece were placed parallel to the diurnal motions, the observer could check that the Sun was moving parallel to the filars. The distance between the* *filars tangential to the edges of the solar image determined the vertical diameter. To measure horizontal diameters, the filars were placed perpendicular to the horizon. Both horizontal and vertical apparent diameters could be measured. The vertical diameters, for their part, were affected by atmospheric refractions. The horizontal measurements required great skillfulness because of the diurnal motion of the Earth... Although the corresponding diffraction disk was of the order of two arcseconds, the positions of the Sun's edges were defined with an accuracy of one arcsecond, for each measurement. This is the theoretical limit obtained with a micrometer having 40000 divisions per foot for such a telescope."*

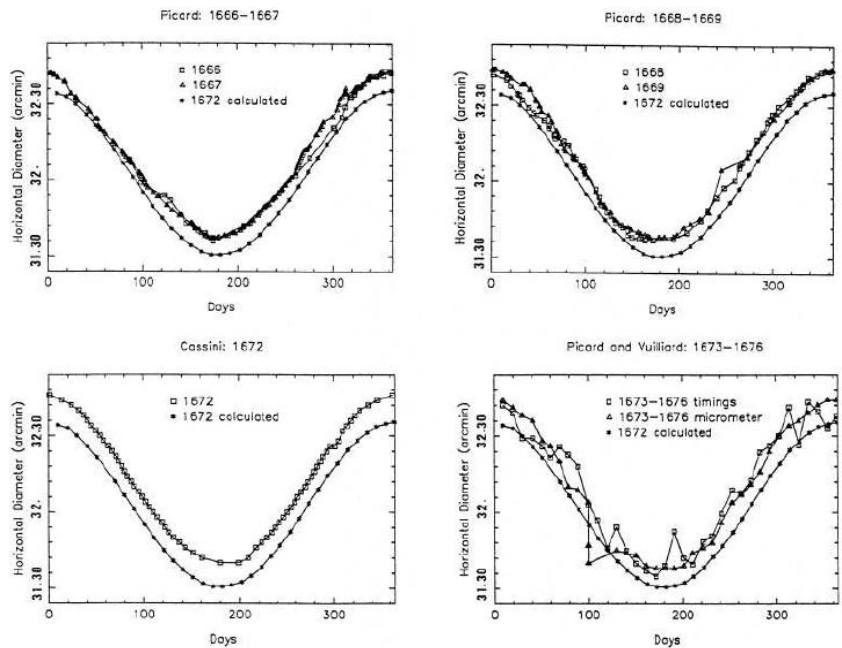

Figure 1.5.1. Apparent horizontal diameter observed throughout the year. A) by Picard, with the micrometer method, for the years 1666 and 1667. B) by Picard, with the micrometer method, for the years 1668 and 1669. C) by Cassini, with the micrometer method, for the year 1672. D) by Picard and his assistant Vuilliard, with both micrometer and noon transits methods over the time interval 1673 to 1676. For comparison, the solid line with stars represents the apparent solar diameter as it would appear to an Earth's observer throughout the year 1672, assuming for solar diameter a value of 961.18 arcseconds (close to recent measurements) corrected from the irradiation as estimated by Auwers. (Ribes and Nesme-Ribes 1993)

There are several ways for showing that this accuracy was in fact achieved.
• The difference of apparent diameter between perigee and apogee was about 65 arcseconds, whatever the year. This value is extremely close to the calculated variation of the diameter for the year 1672, i.e. between 64 and 65 arcseconds (independent of the absolute value).
• Solar diameter was measured with another method, the noon transit (see section 3.2). Although less accurate, it accords better with the stated diameter than calculated one
with the today's value (solid line in figure 1.5.1 – D).
• With the same micrometer method the observers were able to determine the planets' diameters with an accuracy of 1 arcsecond.
The intriguing question is whether these 7 arcseconds more of the solar diameter correspond to a real expansion of the Sun's surface in the 17th century, or if some larger irradiation correction should be applied to these observations. This leads us to the problem of the definition of the solar edge (called *solar limb*), discussed in the next chapter.

# CHAPTER 2
# Observing the Limb Darkening Function

## 2.1. Observation from the ground

**2.1.1. True limb and observed limb.** Solar images in the visible wavelength range show that the disk centre is brighter than the limb region. This phenomenon is known as the 'limb darkening function' (LDF) or as 'centre-to-limb variation (CLV)'. The lack of linearity of the photographic plate response used for the first measurements of the LDF (Canavaggia and Chalonge 1946) was subsequently overcome by photoelectric detectors, and more recently by CCDs. Moreover a variety of sophisticated methods were employed to derive the exact profile of the solar limb.

In spite of the improvements in the measurements, there is always the uncertainty resulting from the effect of Earth's atmosphere (seeing) and this effect is always considered to be the main source of the discrepancies among the diameter determinations. The effect, as showed in figure 2.1.1, is the smoothing of the steepness of the profile on the edge, introducing a fictitious inflection point.

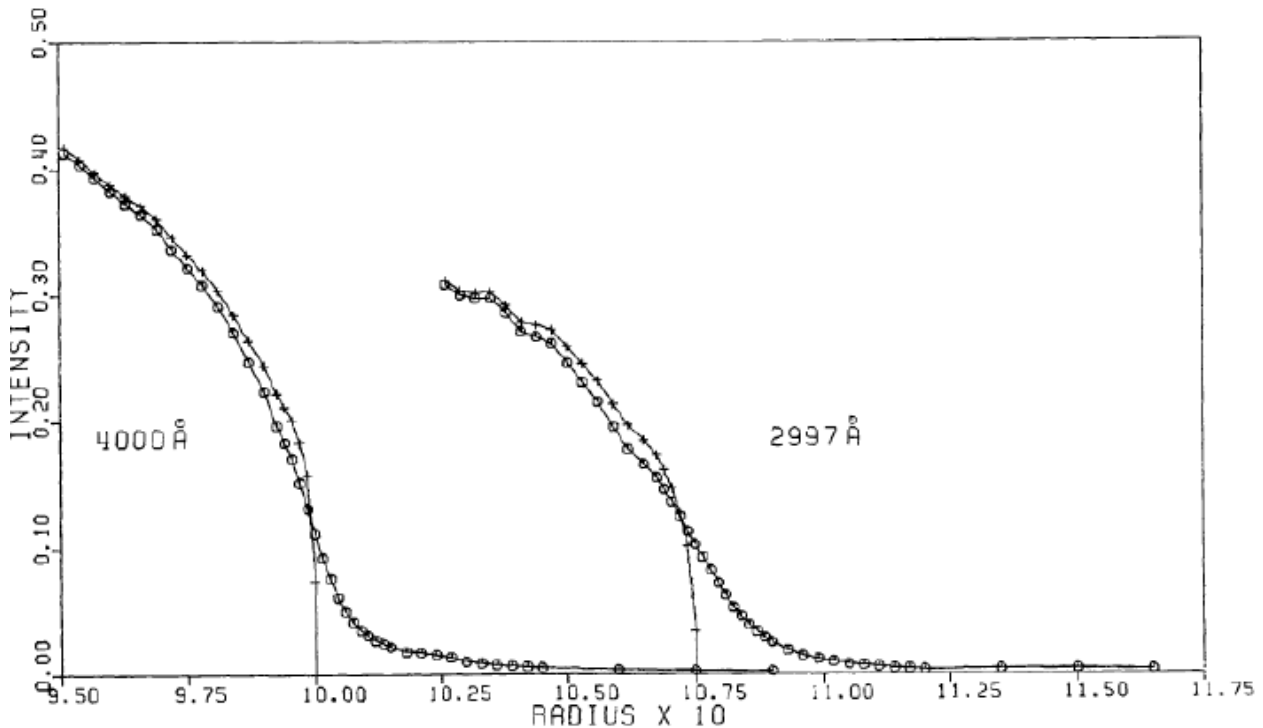

Figure 2.1.1. Sample limb intensities: Observed (O) and Corrected (+) for two wavelength: 400 nm and 299,7 nm. (Mitchell 1981)

The observed profile $O(x)$ is the convolution of the true profile $T(x)$ and the Point Spread Function (PSF) of the atmosphere (and the apparatus function) $A(x)$, where x is the axis parallel to the ray of the solar disk.

$$O(x) = \int_{-\infty}^{+\infty} A(x')T(x-x')dx'$$

To obtain the true profile sometimes it is used the convolution theorem: under suitable conditions the Fourier transform of a convolution is the pointwise product of Fourier transforms.
*F {f * g} = F {f} · F {g}*

The true profile obtained by the more detailed works presents an inflection point, which must be distinguished from the observed inflection point that is a result of the atmospheric seeing. The uncertainty produced by the atmosphere makes impossible the direct measure of the Inflection Point Position (IPP) from the ground, and sometimes even the methods for obtaining the true profiles are unable to calculate its location, such as it can seen in the profile of Figure 2.1.1. The solar models confirm the presence of this true inflection point and there is now a general consensus in identifying **the IPP as the definition of the solar edge**.

The limb shapes provided by the measurements have been modeled as a function of $\mu = \cos \Delta$, where $\Delta$ is the angle between the Sun's radius vector and the line of sight. Most of the analyses are based on a fitting of fifth-order polynomial functions:

$$P_5(\mu) = \sum a_k \mu^k \; ; \; \sum a_k = 1$$

Each of the empirical models has different values in the coefficients $a_k$. According to Thuillier et al. (2011) the most reliable are those of Allen (1973), Pierce and Slaughter (1977), Mitchell (1981), Neckel and Labs (1994) and Hestroffer and Magnan (1998). The last, hereafter named HM98, combined the measurements of Pierce and Slaughter (1977) and Neckel and Labs (1994), reconstructing the limb shape by a complex mathematical model to smooth the differences between the two data sets.

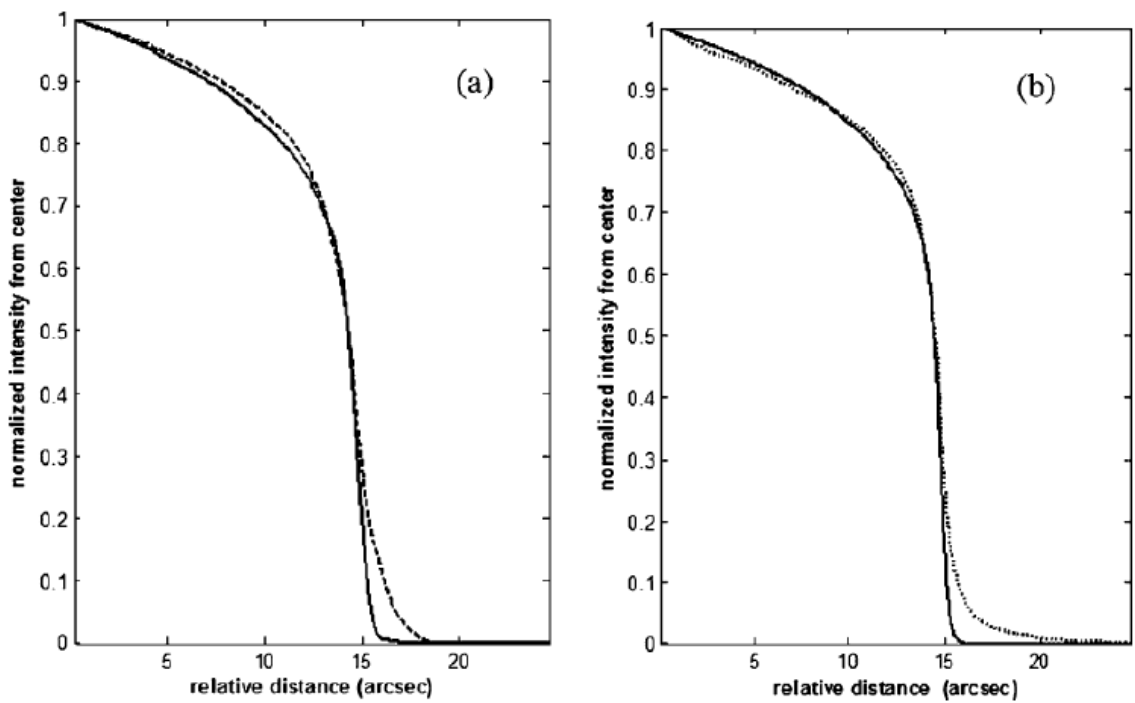

Figure 2.1.2. Measured solar limb (dashed line) of (a) SDS and (b) Pic du Midi compared with the predicted one (solid line) from the HM98 model. Intensities are arbitrarily normalized to unity (here at 0 arcsec). (Thuillier et al. 2011)

The difficulty of adapting the measurements obtained with empirical models is shown in Figure 2.1.2. The model HM98 is compared with the measures of the SDS (Egidi et al. 2006) and Pic du Midi Observatory.

A quantitative analysis for comparing the solar limb model and the different measures is made by comparing the FWHM of the first derivative of the solar limb, that is a measure of the steepness of the limb shape. The differences observed in Figure 2.1.2 are quantified in table 1.

| | λ (nm) | Resol. | FWHM Obs. | ΔFWHM Obs. – HM |
|---|---|---|---|---|
| SDS | 620.0 | 0.128 | 1.422 | 0.060 |
| Pic | 782.0 | 0.1 | 0.843 | 0.046 |

Table 1. Comparison between the solar limb FHWM predicted by the HM98 model with the measurements (fourth column). Differences between observation and model prediction FWHM are displayed in fifth columns. Unit is arcsec. (Thuillier et al. 2011)

These differences can be attributed to several aspects including residual scattered light, possible residual effects remaining from the correction for the Fraunhofer lines within the filter bandpass, and further atmospheric effects by faint aerosols and turbulence, which is corrected, but only to a certain extent, by modeling using the Fried parameter. These aspects are discussed in the next sections.

**2.1.2. Effect of the seeing on the edge.** The atmospheric turbulence produces the well known effects of blurring and image motion, animated by different timescales. With a single parameter $r_0$ (Fried parameter) it is possible to describe this phenomenon. A simple formula for the angular resolution $\rho$ is:

$$\sin \rho = 1.22 \frac{\lambda}{r_0}$$

in which is explicit the dependence on the wavelength too.
The atmospheric effect is better represented by the Kolmogorov model (Lakhal et al. 1999). To simulate this effect on the shape of LDF Djafer, Thuillier and Sofia (2008) used the HM98 solar model and the characteristics of the Définition et Observation du Rayon Solaire (DORaySol) instrument of Calern (see section 3.2). Figure 2.1.3 shows the PSFs of the atmosphere and their effect on the displacement of the position of the inflection point for several values of Fried's parameter r0. Figure 2.1.3 b shows that the position of the inflection point of the measured limb is subject to a displacement that increases with turbulence. This displacement is on the order of 0.123 arcsec for $r_0$ = 5 cm, and 1.21 arcsec for $r_0$ = 1 cm.

Attempts were made to define the solar limb bypassing the problem to measure the true IPP. The most common defines the edge to be that radius where the second derivative of the LDF is zero i.e. the inflection point of the observed LDF (different from the true LDF).
A different type of edge definition, called 'integral definition' was employed by Dicke and Goldenberg (1974) for relative diameter measurement in their solar oblateness observation. They integrated the intensity through each of two slots positioned in front of diametrically opposed edges of a solar image. These apertures extended from inside the limb outward beyond it. The slots were rotated, and the integrated intensity was interpreted as the protrusion of the edge beyond the inner radius of the slot.
A very used edge definition, called Fast Fourier Trasform Definition (FFTD), was developed by Hill, Stebbins and Oleson (1975). The FFTD seeks a point on the extreme limb of the Sun by centering an interval so that one term of the finite Fourier transform of the limb darkening function over this interval is zero.
These three definition are compared in Figure 2.1.4 in which is shown their dependence on PSF.

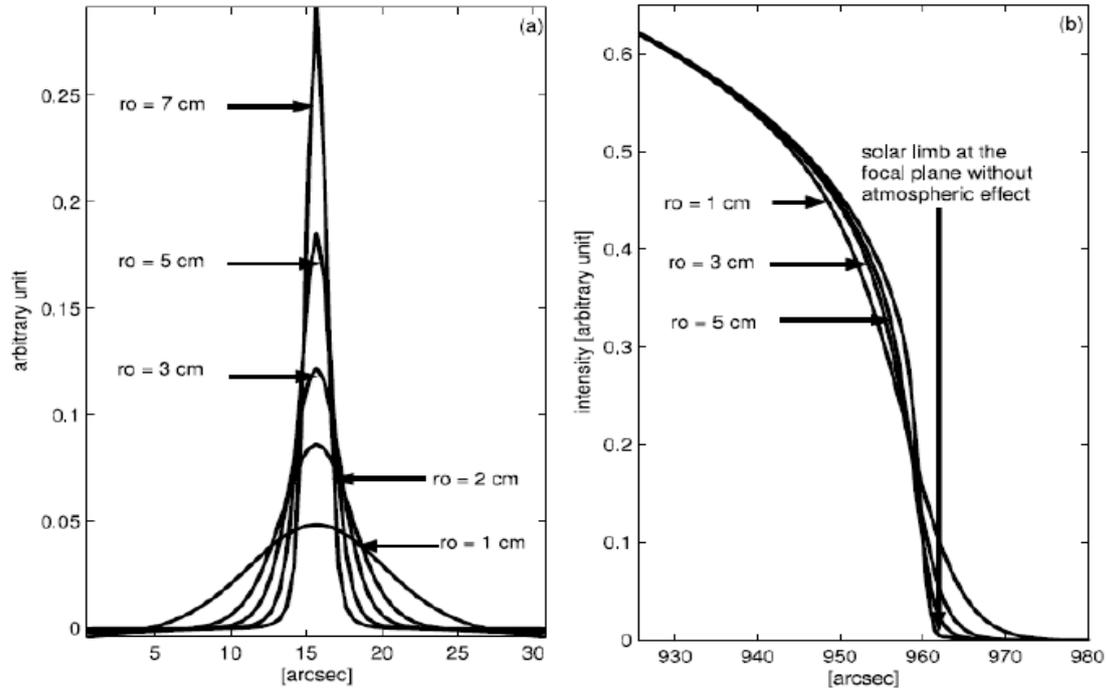

Figure 2.1.3. (a) PSF for several values of Fried's parameter r0 according to the Kolmogorov turbulence model, through a telescope having the instrumental characteristics of the DORaySol instrument; (b) effect of atmospheric turbulence on the solar limb. (Djafer, Thuillier and Sofia 2008)

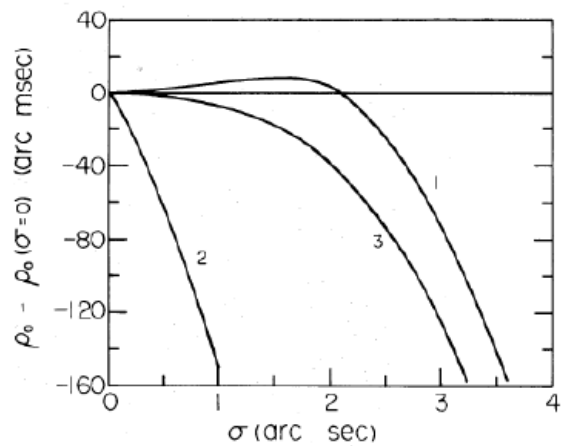

Figure 2.1.4. Instrumental and atmospheric effects on the edge location. The relative displacement of the edge is shown for a Gaussian transfer function and three edge definitions: (1) the FFTD, (2) the second derivative technique, and (3) the integral definition. (Hill, Stebbins and Oleson 1975)

### 2.2. Dependence on instrumental effects

Instrumental effects depends on the characteristics of each telescope. The FWHM of the PSF is proportional to $\lambda/D$, where $\lambda$ is the wavelength of observation and D is the pupil diameter. This is similar to the FWHM of the atmospheric PSF, that is proportional to $\lambda/r_0$. Then for a given value of the wavelength, if D increases, the FWHM of the PSF decreases, and consequently the calculated diameter increases (see figure 2.1.3). Thus, two instruments with different D will measure different solar diameters if this instrumental effect is not taken into account. However, for a given instrument (D), if $\lambda$ increases, the FWHM of the PSF increases, and consequently the calculated diameter decreases.

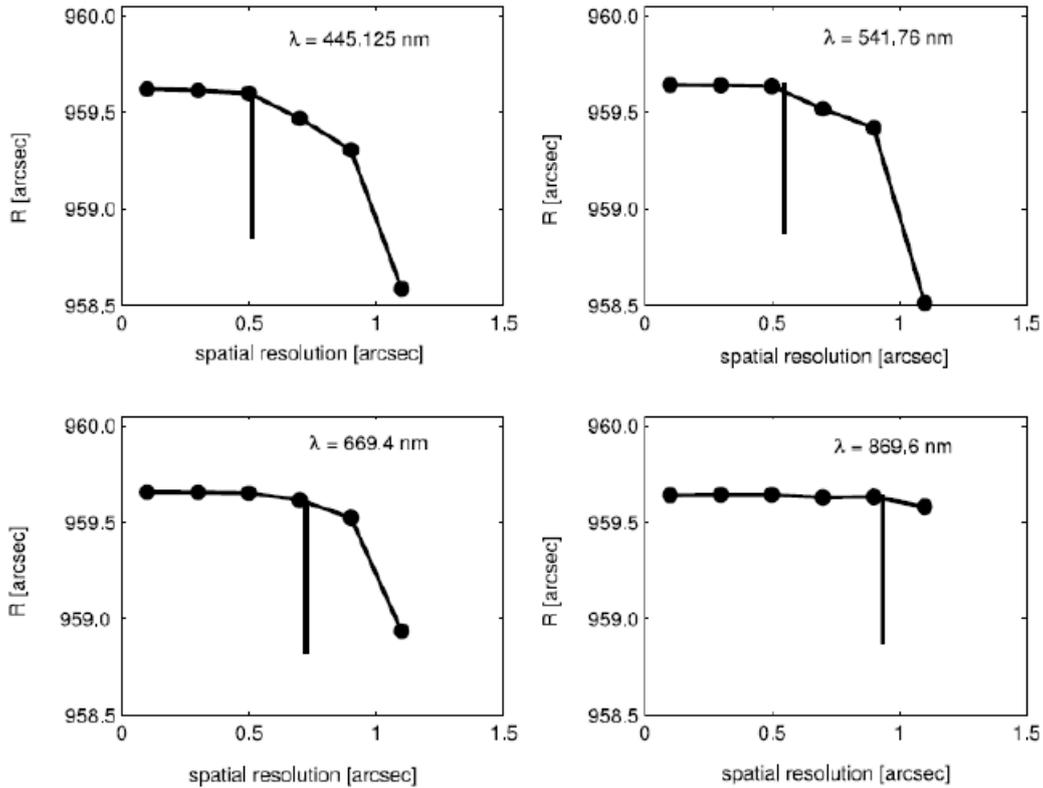

Figure 2.2.1. Effect of spatial resolution on solar diameter measurement. The vertical lines indicate the diffraction limit. The reference radius used in the simulation is 960 arcsec (Djafer et al. 2008)

Spatial resolution is another characteristics to consider. For any instrument, the Shannon (1949) condition must be satisfied. This requires that the PSF must cover at least two sensing elements, that is to say, 2 pixels when using a CCD. Furthermore, the spatial resolution must be less than or equal to F($\lambda$/2D), where F is the instrument's focal length.

Djafer, Thuillier and Sofia (2008) illustrated a case in which the Shannon condition is not satisfied. They simulated a solar image according to the HM98 model, for four wavelengths (445.125, 541.76, 669.4, and 869.6 nm), seen through an optical system having a pupil diameter D equal to 10 cm. These values correspond respectively to diffraction-limit values of 1.072 arcsec, 1.366 arcsec, 1.687 arcsec, and 2.192 arcsec. For each wavelength, they vary the spatial resolution between 0.1 arcsec and 1.1 arcsec and they determine the mean solar radius of the corresponding image (Figure 2.2.1). Whenever the Shannon condition is not satisfied we lose resolution, and the measured diameter decreases and loses precision. However, this effect decreases with wavelength.

## 2.3. Dependence on wavelength

The position of the solar limb depends on wavelength not only through the PSF (instrumental or atmospheric) but also through the solar physics that forms different shapes of true LDF for different wavelengths.

**2.3.1. Observations.** The few measurements of the LDF at different wavelengths do not yet allow a clear description of the dependence of the solar diameter on the wavelength. This situation is complicated by the fact that the various measures do not have the same bandwidth, in addition to the use of different instruments (see Table 2).

| Site/Instrument | λ (nm) | Δλ (nm) | Period | D (cm) | r (arcsec pixel$^{-1}$) | R (arcsec) | Reference |
|---|---|---|---|---|---|---|---|
| Mount Wilson | 525.02 | 0.014 | 1970–2003 | 30.48 | 9.6 × 13.1, 12.9 × 20.1 | 959.486 ± 0.005 | Ulrich & Bertello 1995, Lefebvre et al. 2006 |
| Calern: | | | | | | | |
| 11 prisms | 540 | 200 | 1989–1995 | 10 | 0.60 | 959.590 ± 0.010 | Laclare et al. 1999 |
| 11 prisms | 538 | 200 | 1996 | 10 | 0.74 | 959.360 ± 0.030 | Sinceac et al. 1998 |
| Prism | 850 | 160 | 1996 | 10 | 0.74 | 959.385 ± 0.035 | Sinceac et al. 1998 |
| 11 prisms | 538 | 200 | 1996–1997 | 10 | 0.74 | 959.630 ± 0.080 | Chollet & Sinceac 1999 |
| DORaySol | 548 | 60 | 2001 | 10 | 0.50 | 959.509 ± 0.014 | Andrei et al. 2004 |
| Rio de Janeiro: | | | | | | | |
| Prism | 563.5 | 168 | 1997–1998 | 10 | 0.50 | 959.200 ± 0.020 | Jilinski et al. 1999 |
| | | | 2001 | 10 | 0.50 | 959.190 ± 0.013 | Andrei et al. 2004 |
| Antalya: | | | | | | | |
| 2 prisms | 550 | 180 | 1999–2000 | 10 | 0.78 | 959.030 ± 0.070 | Gölbaşı et al. 2001 |
| | | | 2001–2003 | 10 | 0.78 | 959.290 ± 0.010 | Kılıç et al. 2005 |
| Boulder (SDM) | 800 | 10 | 1981–1986 | 10 | 1 | 959.680 ± 0.018 | Brown & Christensen-Dalsgaard 1998 |
| Locarno/Izaña | 475.8 | 2.1 | 1997 | 45 | 0.179 | 959.73 ± 0.050 | Wittmann 1997 |
| | 486 | 2.1 | 1997 | 45 | 0.179 | 959.81 ± 0.030 | Wittmann & Bianda 2000 |
| | 583 | 2.1 | 2000 | 45 | 0.179 | 960.09 ± 0.040 | Wittmann & Bianda 2000 |
| Kitt Peak | 329.8–660 | 2 | 1981 | 152 | 0.925 | 959.620 ± 0.030 | Neckel 1995 |
| SDS | 620 | 80 | 1992–1996 | 13 | 0.128 | 959.561 ± 0.111 | Egidi et al. 2006 |
| | | | | | | 959.658 ± 0.091 | Djafer et al. 2008 |
| MDI | 676.78 | 0.0094 | 1996–2006 | 15 | 2 | 959.283 ± 0.150 | Kuhn et al. 2004 |

Table 2. Listed are wavelength l, bandpass Dl, telescope diameter D, spatial resolution r and the observed Sun's radius R. (Djafer, Thuillier and Sofia 2008)

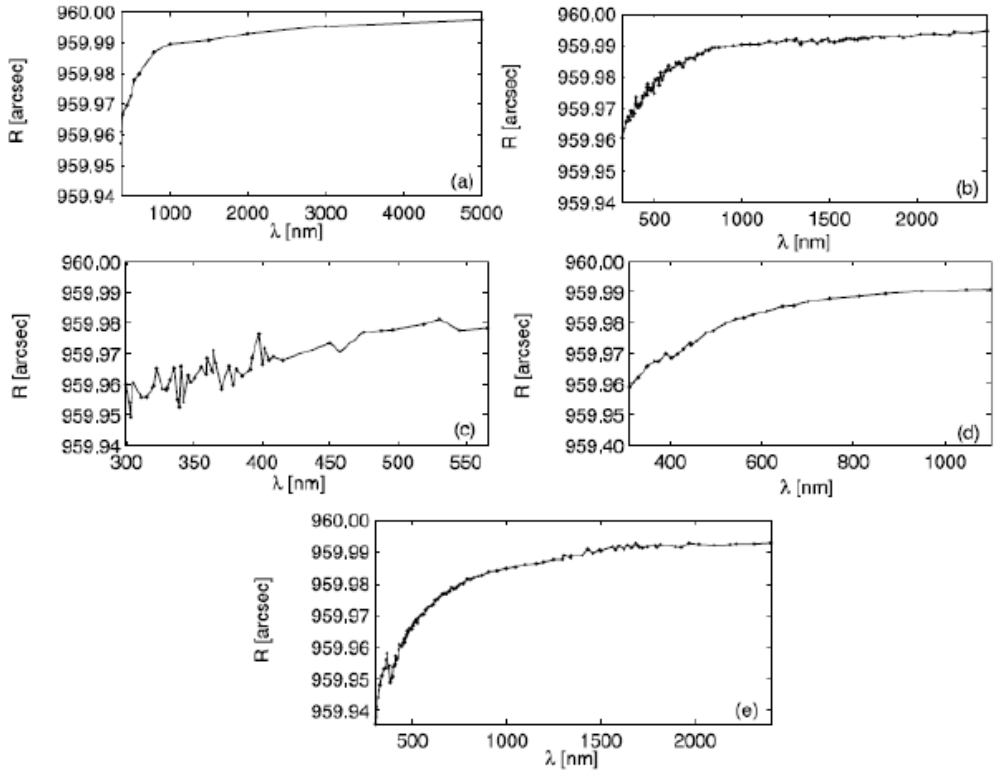

Figure 2.3.1. Variation of the calculated solar radius (R) as a function of wavelength suggested by five solar models: (a) Allen (1973), (b) Pierce & Slaughter (1977) and Pierce et al. (1977), (c) Mitchell (1981), (d) Neckel & Labs (1994), and (e) HM98. (Djafer, Thuillier and Sofia 2008)

**2.3.2. Continuum.** To simulate the dependence of the LDF on wavelength several solar limb models were derived, such as those of Allen (1973), Pierce & Slaughter (1977) and Pierce et al. (1977), Mitchell (1981), Neckel & Labs (1994), and HM98. For comparing their prediction, Djafer, Thuillier and Sofia (2008) simulated a solar image seen through an optical system with constant PSF (independent of wavelength) and with a CCD detector that permits a spatial resolution of 0.01 arcsec. The results are presented in Figure 2.3.1 and show that for all models, the calculated solar radius increases with wavelength.

The difference on the solar limb for these models is discussed by Thuillier et al. (2011) (Figure 2.3.2). They noted the HM98 model differs by as much as 1 arc second from the measurements at the limb region.

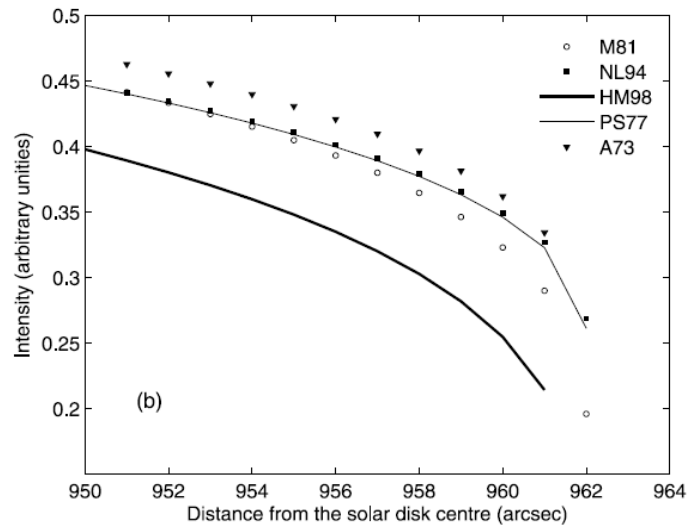

Figure 2.3.2. LDF for l= 559.95 nm, A73: Allen (1973), PS77: Pierce & Slaughter (1977) and Pierce et al. (1977), M81: Mitchell (1981), NL94: Neckel & Labs (1994), and HM98. (Thuillier et al. 2011)

The dependence of the solar limb on the wavelength has a sign opposite to that seen for the PSF. The combined effect is shown by Djafer, Thuillier and Sofia (2008). They simulated a solar image according to the HM98 solar model for four wavelengths (445.125, 541.76, 669.4, and 869.6 nm), observed through an optical system having a pupil diameter of 10 cm and a sampling spatial frequency of 0.1 arcsec (Figure 2.3.3).

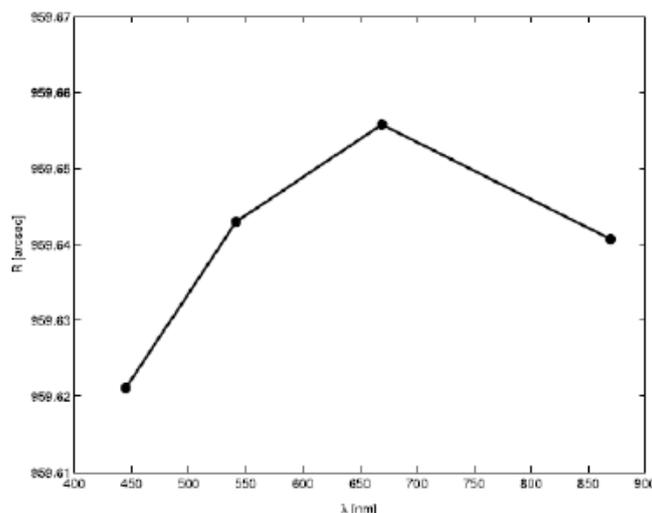

Figure 2.3.3. Instrumental and physical effects of wavelength on solar diameter measurement. The value of the reference radius used in the simulation is 960 arcsec. (Djafer, Thuillier and Sofia 2008)

**2.3.3. Fraunhofer lines.** The absence of consensus on the choice of the spectral domain of observation shown in Table 2 is probably the main source of differences in solar radius obtained. Certain instruments observe in the continuum at different wavelengths, while others observe in the center of a Fraunhofer line, as does the Mount Wilson instrument (Ulrich & Bertello 1995; Lefebvre et al. 2006).

In addition, there are instruments that use a narrow bandpass, such as MDI (0.0094 nm) and MountWilson (0.014 nm), whereas others use a wide spectral domain on the order of hundreds of nanometers, as do, for example, the CCD astrolabes. The effect of the presence of Fraunhofer lines is illustrated by Thuillier, Sofia, and Haberreiter (2005). The authors reconstructed the limb profile for different wavelengths, including the continuum and the centre of spectral lines, with the Code for Solar Irradiance (COSI). The results on the prediction of inflection point position is shown in figure 2.3.4.

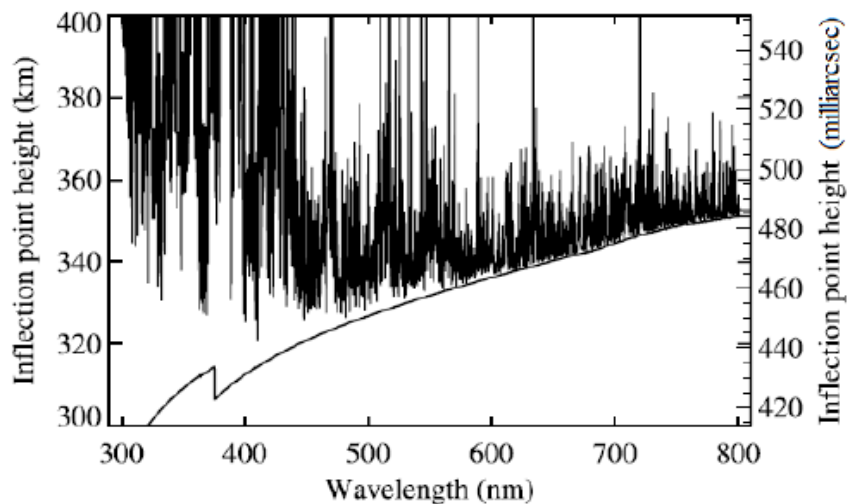

Figure 2.3.4. COSI predictions of the inflection point position as a function of wavelength for the continuum (lower solid line) and for the case when Fraunhofer lines are taken into account, running the model at high resolution. (Thuillier, Sofia, and Haberreiter 2005)

Neckel (1996) concluded after calculating the solar radius in 1981 and between 1986 and 1990 that the solar radius differences will exceed 0.1 arcsec if the observations are undertaken in spectral bands that contain Fraunhofer lines.

Moreover the behavior of the intensity of Fraunhofer lines over the solar cycles is not the same for all lines (Livingston 1992; Livingston & Holweger 1982; White & Livingston 1978). Therefore, in the case of wide-bandpass observations, where Fraunhofer lines are present the chromospheric emission associated with these lines may modify the value of the solar diameter determination and thus affect the study of its variability as a function of solar activity.

## 2.4. Dependence on solar features

**2.4.1. Asphericity.** The asphericity is the observed variability of the radius with the latitude. The asphericity of the Sun, and in particular the oblateness: $f = (R_{eq} - R_{pol})/R_{eq}$, have a great implications on the motion of the bodies around the Sun. It can also provide important informations on solar physics as the other global parameters. Some authors have suggested that oblateness has a variation in phase with the solar cycle (see Rozelot). Therefore the changes seen in the first chapter of the TSI could be explained also with a different shape of the Sun rather than a variation of the solar radius. For all these reasons the studying of the asphericity is a topic of great importance.

Measuring the diameter of the Sun for several solar latitudes may provide an indirect measure also on asphericity. Conversely, for monitoring the variation of the solar diameter, one must take into account that different solar latitudes may correspond to a different solar radius due to the asphericity.

The measurement of Rozelot et al. (2003) is reported. They used the scanning heliometer installed at the Pic du Midi Observatory (described by Rosch et al. 1996). Using the FFTD (see section 2.1.2) they achieved the precision of few milliarcsecond in the solar radius observing in good conditions of seeing (mean r0 = 18 cm).

The best fit adjusting the data can be computed through a polynomial expansion of the radius contour of the form:

$$R(\psi)|_{\rho=constant} = R_0 + \sum_n d_n(R_0) P_n(\psi)$$

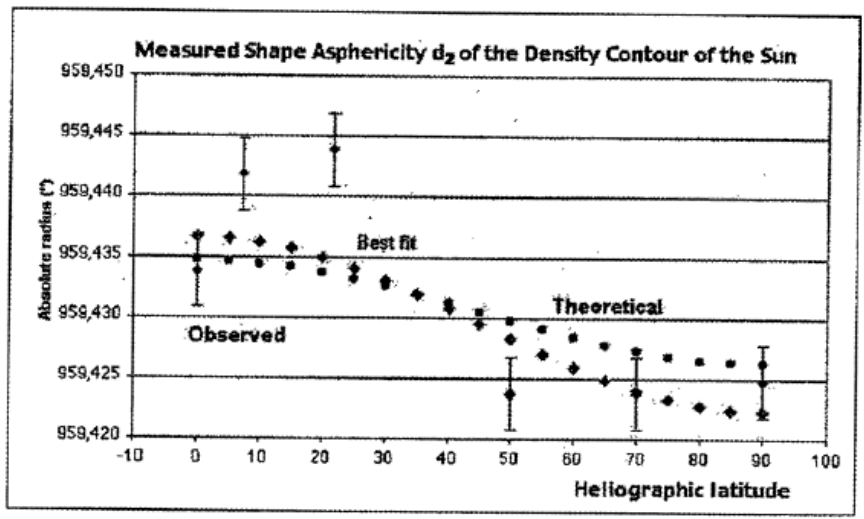

Figure 2.4.1. The solar limb shape variations as observed at the Pic du Midi Observatory from September 3rd to 6th, 2001. Diamonds: observed data (in mas) and their best fit. Circles: theoretical curve as deduced from Armstrong and Kuhn (1999). The amplitude of the variation is 20 mas, and 24 mas between the highest and the lowest error bar. (Rozelot et al. 2003)

Where $\psi$ is the colatitude, $P_n$ the Legendre polynomial of degree $n$, $R_0$ the mean contour radius and $d_n$ the shape coefficients.

Results put in evidence a distorted solar shape, the departures to a pure spherical body not exceeding 20 mas (24 mas taking account the highest and the lowest part of the error bars). The Sun shows an equatorial band up to about 27° of latitude followed by a depressed zone from around 45° to 70°. The best data fit gives $d_2 = -1.00 \cdot 10^{-5}$, a value that can be compared to the solar oblateness (different from $d_2$) that is $f = 9.41 \cdot 10^{-6}$, that account for a difference between the equatorial radius and the polar one of 9.0 ±1.8 mas. Moreover the authors (Rozelot et al. 2003) proposed a model to interpret such results which consists of a nearly uniform rotating core combined with a prolate solar tachocline and an oblate surface.

The recent space mission RHESSI (Fivian et al. 2008) provide a new measurement for oblateness. The full data of the solar aspect sensors (SAS) on board gives an oblateness of 10.74 ±0.44 mas. By restricting the data base to avoid faculae, including a component outside the active regions, it obtains a lower value for the oblateness of 7.98 ±0.14 mas.

**2.4.2. Solar surface magnetic structure.** The effect of different solar surface magnetic structures on the IPP is discussed in detail by Thuillier et al. (2011). Because of the lack of observations on this topic they used three models that provide the solar atmosphere structure for different active regions (see also section 7.2.2): VAL-C (Vernazza, Avrett and Loeser 1981), FCH09 (Fontenla et al. 2009) and COSI (Haberreiter et al. 2008, Shapiro et al. 2010). The results are summarized in Table 3.

The VAL81 model predicts the sunspot as the dominant effect in the IR region and the faculae in the visible region. The FCH09 model predicts a quasi wavelength independent displacement, while for the sunspot case this model predicts an inward displacement with increasing wavelength. Furthermore, we observe that the effect of faculae in the FCH09 is significantly smaller than the effect of sunspots. For COSI, the effects of faculae and sunspots have nearly no dependence on wavelength. However, the displacement of the inflection point is very small for faculae (of the order of a few mas) and much larger for sunspots (of the order of 300 mas). Despite all the differences it's clear a common trend: for all the models the presence of faculae displaces the inflection point outward with respect to its location predicted by the quiet Sun models (C and QS), while the sunspot displaces it inward.

| Model types | $\Delta$IPP (400 nm) | $\Delta$IPP (600 nm) | $\Delta$IPP (800 nm) |
|---|---|---|---|
| P-QS for FCH09 | 35.86 mas | 55.82 mas | 46.95 mas |
| S-QS for FCH09 | $-50.40$ | $-183.32$ | $-197.60$ |
| A-C for VAL81 | $-11.74$ | $-6.96$ | $-1.63$ |
| F-C for VAL81 | 6.70 | 13.49 | 15.39 |
| P-QS for COSI | $-4.10$ | 4.10 | 1.40 |
| S-QS for COSI | $-285.10$ | $-290.80$ | $-285.30$ |

Table 3. Difference of the inflection point position (DIPP) for different atmospheric structures as a function of wavelength. The atmospheric structures used are: the VAL81 sunspot models (A), quiet Sun (C), and very bright network (F); FCH09 and COSI quiet Sun (QS), the sunspot model (S), and faculae model (P). The differences between the positions of the inflection points are given in mas. (Thuillier et al. 2011)

The observation of the IPP in conjunction of these phenomena can help to discern between the models, providing useful information on the solar atmosphere. Conversely, the measure of the solar diameter observed in the active regions can significantly distort the measurements. Therefore this detail have to be discussed in every measurement.

# CHAPTER 3
# Measurement methods

## 3.1. Direct measure

**3.1.1. Heliometers.** Direct measurements of the Sun with a single pinhole were already made by Tycho Brahe in 1591 and Johannes Kepler in 1600-1602. They calculate:

$$\theta_\odot = (d_m - d_p)/f$$

Where $\theta_\odot$ is the apparent solar diameter, $d_m$ is the diameter of the image on the screen, $d_p$ is the diameter of the pinhole, and $f$ is the focal length.

They recognized the limitations of this type of measure: geometric effects (the shape of the image is influenced by the shape of the pinhole) and the motion of drawn solar disk on the screen due to the rotation of the Earth, which makes this measurement a challenging task. For f large the geometric effects become less important, but the effects of refraction became important.

All these constraints led to the development of the heliometers that exploit the idea of bringing into contact two opposite limbs of the Sun. The principle of this method is summarized in the two-pinhole geometry (figure 3.1.1): the two pinholes are separated by a fixed distance *d*. A flat mirror projects onto it the light of the Sun. The pinholes produce two images of the Sun on a screen that is parallel to the mask. The centers of these images are also separated by *d*. When the two limbs are in contact the solar diameter is given by:

$$\theta_\odot = (d - d_p)/f_c$$

The uncertainty in the resolution of the image is now transferred to the determination of $f_c$. In this way it is overcome the uncertainty arising from measuring the position of two moving limbs of the solar image, and one need to measure only the focal length $f_c$, being *d* and $d_p$ measurable with the accuracy possible in the laboratory (Sigismondi 2002).

The determination of the contact between the two images can be very accurate even with the naked eye, and this experiment gives results better than the classical Rayleigh limit calculated for a single pinhole ($\rho = 1.22 \lambda/d_p$).

The development of heliometers, has led to new techniques to duplicate the solar image, such as the use of prisms, or solution of bisecting refracting objectives, but one have to take into consideration chromatic effects.

A new approach is developed by Victor d' Avila et al. (2009) which consists of bisecting a primary mirror of a telescope.

For this method, the lack of accuracy in the single measurement can be overcome by the high number of measures that decrease the statistical error.

**3.1.2. Space observations.** Because of the simplicity of this method, it is adopted by the few space missions for the astrometric measurement of the Sun. Space missions have the great advantage to overcome the uncertainty introduced by the atmosphere.

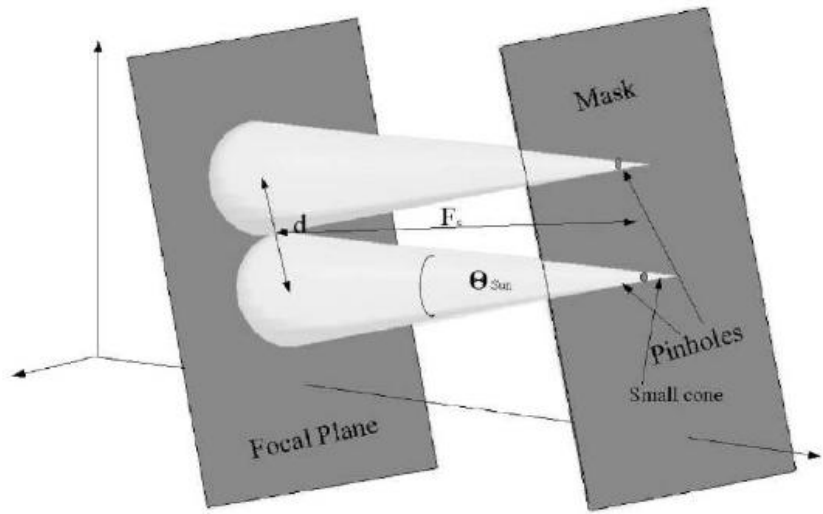

Figure 3.1.1. Two pinhole geometry. (Sigismondi 2002)

The Solar Disk Sextant (SDS), a balloon-born telescope, can be considered a modern heliometer. The basic principle of the SDS instrument is the use of a prism in front of the objective to form a double image of the Sun separated by slightly more than the angular diameter of the Sun.

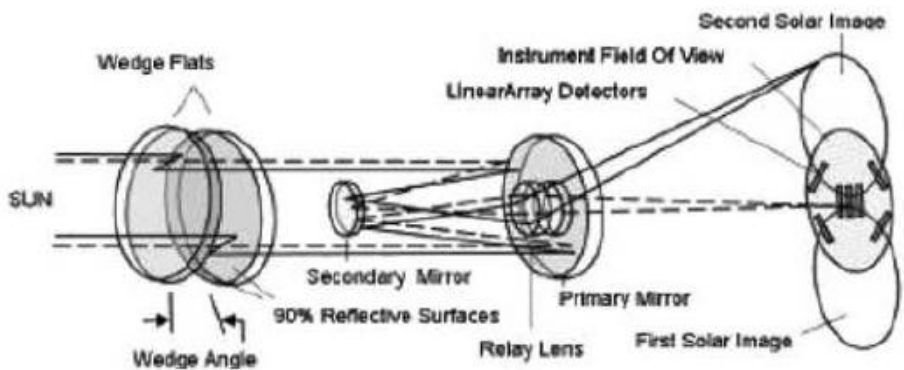

Figure 3.1.2. Optical layout of the SDS. (Egidi et al. 2006)

As shown in Figure 3.1.2, two opposite solar edge arcs are imaged near the center of the focal plane; this position minimizes, in the reconstruction of the two solar images, the errors due to cylindrically symmetric optical distortions. The distance between the centers of the two reconstructed images provides the scale calibration of the images in the focal plane against variations of the focal length. The accuracy of the SDS derives from the system design, which uses a single optical train to transfer the split solar images to the detectors.
The solar limbs are determined by means of the Fast Fourier Transform Definition (FFTD). An accuracy of few mas was expected (Egidi et al. 2006).

PICARD is the last space mission dedicated to the study of the Earth's climate and Sun variability relationship. It was launched in June 2010. It consists in three instruments: SOVAP for the study of Total Solar Irradiance, PREMOS to perform helioseismologic observations, and SODISM for the measure of solar diameter and limb shape in three wavelength: 535, 607, 782 nm in spectral domains without Fraunhofer lines. The expected accuracy of one milliarcsecond is based on very good dimensional stability, which is ensured by use of stable materials. Four prisms are used to generate four auxiliary images placed in each corner of the CCD. The distance between a point of the limb of the central solar image and the corresponding point of the auxiliary image only depends on the angle of the prism and of the temperature which will be measured with the appropriate accuracy. These measurements enable to check the relationship between the angular distance of two points on the Sun and the distance of their images on the CCD (Figure 3.1.3). The solar diameter will be referenced to the angular distances of doublets of stars so that the measurements which will be achieved in the next decades referenced to the same doublets, enable to evaluate the long term evolution of the Sun (Assus et al. 2008).

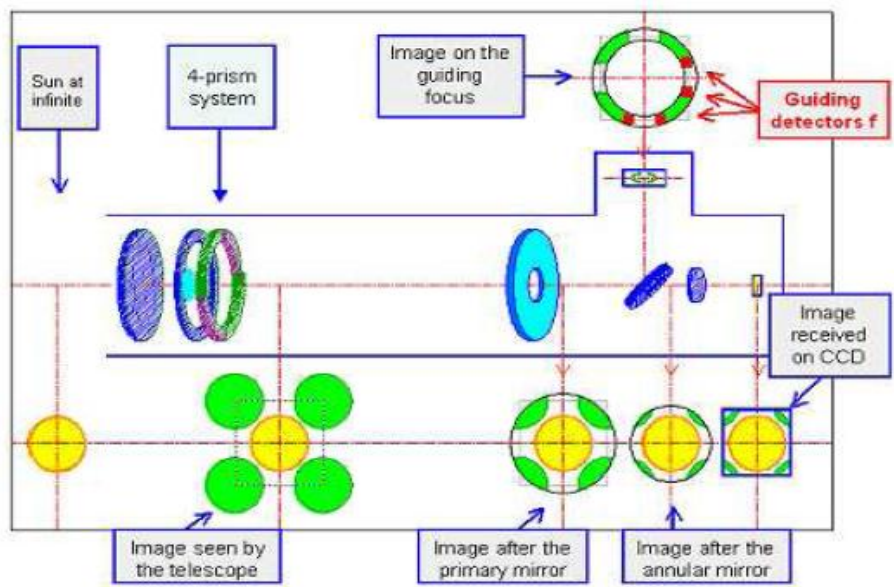

Figure 3.1.3. Optical diagram of SODISM on board of PICARD. (http://smsc.cnes.fr/PICARD/)

The development of astrometric space missions to measure the Sun does not make obsolete the methods of measurement from the ground for at least three reasons:
• balloons and satellites can observe only in a limited number of wavelengths
• space missions are limited in time
• for certainty about the measures it is better to compare different approaches, especially in this topic where the results seem inconsistent.

## 3.2. Drift-scan methods

Although the method of direct measurement of the solar diameter can offer a good accuracy, the construction of optical systems used for this purpose is often difficult. In fact, the optical instrument have to minimize optical aberrations and also must have excellent thermal and mechanical stability. If these phenomena are not taken into required account, the final result can be misleading.

The drift-scan methods succeed in bypass a good part of these phenomena. These methods consist in the use of a fixed telescope and the observation of the drift of the solar image through a meridian or a given almucantarat (circle of the same atitude above the horizon). Knowing with accuracy the speed of transit of the Sun and measuring the transit time, it can give an accurate measure of the diameter. Should also be noted that the measured time is not affected by atmospheric refraction. The advantage of drift scan method is that it is not affected by optical defects and aberrations, because both edges are observerd aiming in the same direction, moreover with parallel transits one can gather N observations, at rather homogeneous seeing conditions, during the time of a single one of past measurements (Wittmann, Alge and Bianda 1993; 2000). The measurements of the solar diameter by meridian transit were monitored in Paris by Jean Picard from 1666 to 1682 (see section 1.5), and on a daily basis since 1851 at Greenwich Observatory and at the Campidoglio (Capitol) Observatory in Rome since 1877 to 1937. Figure 3.2.2 show the inconsistency of the measures taken maybe due to atmospheric phenomena.

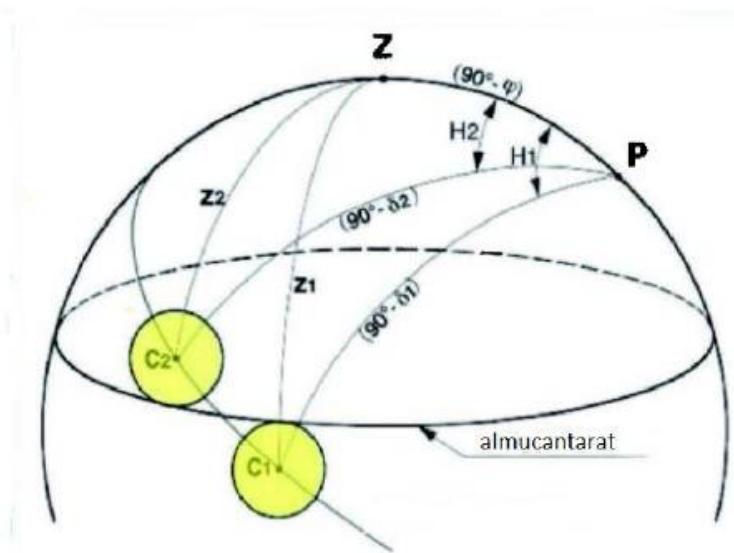

Figure 3.2.1. Transit of the Sun through a given almucantarat. (Fodil et al. 2010)

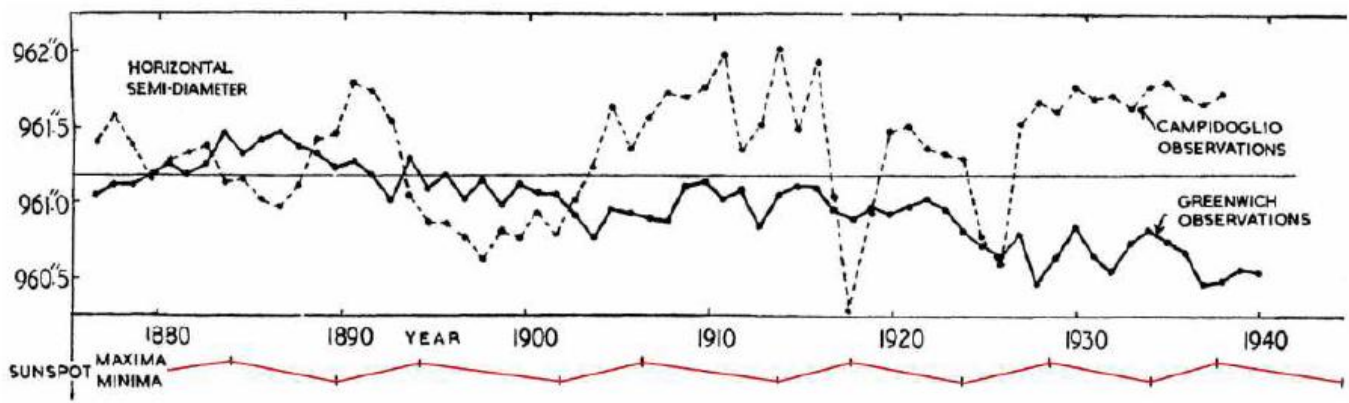

Figure 3.2.2. Horizontal semi-diameters of the Sun measured at Greenwich with Airy's meridian circle. Campidoglio observations are superimposed (Gething 1955). Straight lines correspond to a radius of 961.2 arcsec. (Sigismondi 2011)

Some projects using this method are currently underway.
DORAYSOL (Definition et Observation du RAYon Solaire) is an alt-azimuthal instrument working as DanJon astrolabe, where all refractors have been replaced by reflectors. The instrument is composed of:
• the filter: a silice entrance window keeping the solar magnitude close to that of the Moon;
• the mercury mirror: making the horizontal reference, and forming the reflected image

of the solar edge;
• the reflector variable prism: allowing measurements on different zenithal distances;
• the cassegrain telescope: a primary mirror of 120mm, a secondary of 20mm, coupled to a CCD of 640x480 pixels mounted on the focus to get an image of the Sun edge. The crossing point of the two edges will define the exact zenithal distance of the solar border;
• a set of filters will enable observations in different wavelengths; and
• the rotating shutter will switch between the direct image and the reflected image on the CCD (Fodil et al. 2010).

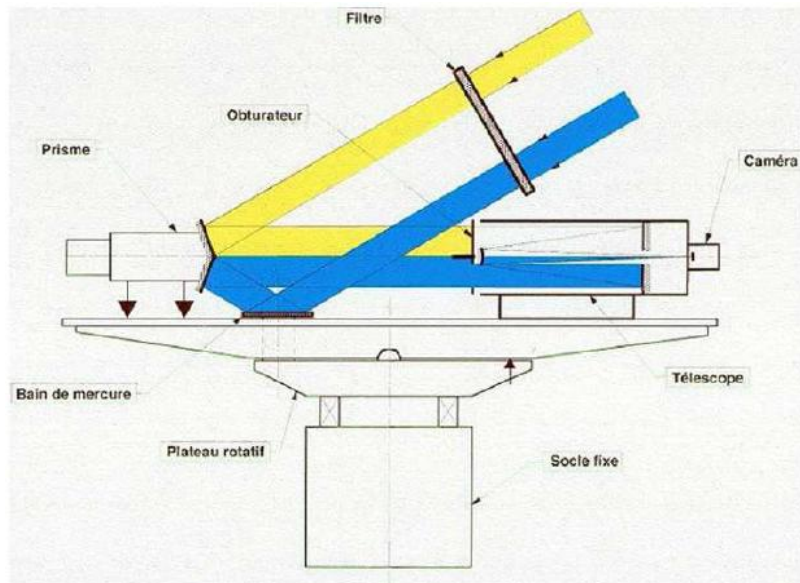

Figure 3.2.3. DORAYSOL: opto-mechanical architecture. (Morand et al. 2010)

DORAYSOL is part of the PICARD-SOL mission: by comparing the limb shape and diameter measured in orbit with the ground based measurements it will enable after the end of PICARD space mission to continue the measurements from the ground with the possibility to interpret them, in principle without ambiguity.

CLAVIUS is another ground measurement project that take advantage of the PICARD space mission observing in the same wavelength of PICARD satellite. A CMOS camera used for this purpose give the possibility to understand in a very deep way how the photons are converted in a digital number and at which time. This allows to precisely study the limb affected by seeing effects (image motion and blurring) and by instrumental effects induced by the telescope. The instrument consists in an optical device dividing a solar portion (about 100 arcsec x 200 arcsec) in two images filtered by different interference filters. The two images are projected on the CMOS sensor of the camera, and are digitalized with a high cadence (frequencies higher than the typical seeing frequencies of few hundreds Hertz) (Sigismondi et al. 2008).

### 3.3. Planetary transits
The observation of planetary transits across the solar disk allows a measurement of the solar dimension largely independent from the effects of atmospheric seeing. Through an accurate knowledge of the apparent motion of the planets (Mercury and Venus) on the Sun, and by measuring the transit time, it can determine the path traversed by the planet and so infer the solar dimension. Since the geometry of the planet-Sun is defined outside the Earth's atmosphere, the light curve in the contact area is not affected by seeing effects.

During the transit of Venus in 1761 dozens of international expeditions attempted to observe this transit all over the world because of the suggestion by Edmond Halley (1716) that transits of Venus can be used to determine the distance to Venus from its parallax, and thus the Sun's parallax, and thus the Astronomical Unit. During the instants of contact between the planet and the Sun edge, astronomers reported that a ligature joined the silhouette of Venus to the dark background exterior to the Sun. This dark "black drop" meant that observers were unable to determine the time of contact to better than 30 s or even 1 min.

To date, the black drop is a problem in determining the exact moment of contact between the edge of the planet and the solar limb. To eliminate the effects it was necessary to know the causes. Pasachoff et al. (2004) show that although many works attributed the black drop mainly to the atmosphere of Venus, TRACE satellite (Transition Region and Coronal Explorer) observed a Black Drop also for the transit of Mercury in 1999. But TRACE was above the Earth's atmosphere and Mercury has essentially no atmosphere, so any Mercurian black drop could not arise from atmospheric factors.

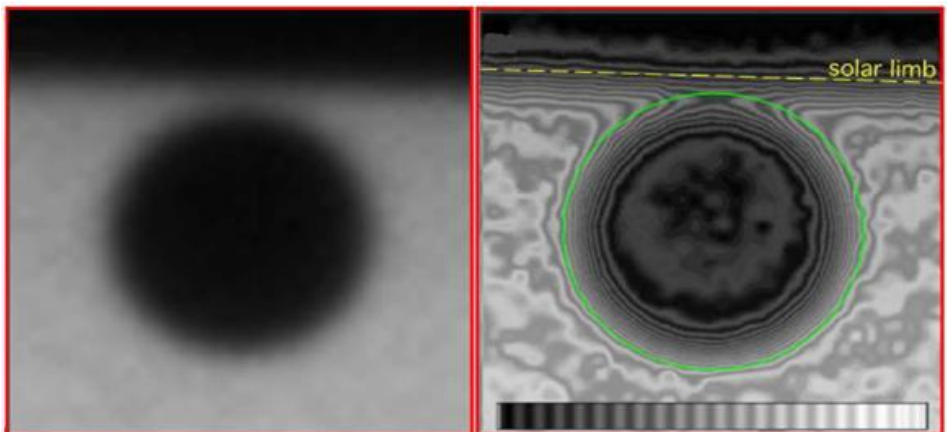

Figure 3.3.1. The black-drop effect on a TRACE image from the 1999 transit of Mercury, observed from NASA's Transition Region and Explorer (TRACE) spacecraft. It is shown both as an image and as isophotes. (Schneider, Pasachoff, and Golub/LMSAL and SAO/NASA)

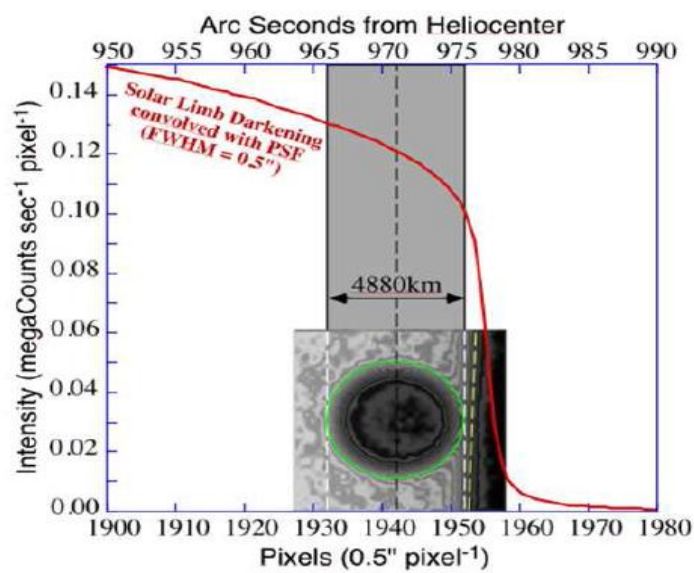

Figure 3.3.2. Modeling the contributions of the telescope's point-spread function and of solar limb darkening on TRACE images from the 1999 transit of Mercury, observed from NASA's Transition Region and Explorer (TRACE) spacecraft. (Schneider, Pasachoff, and Golub/LMSAL and SAO/NASA)

Extensive modeling by Pasachoff et al. revealed that the point-spread function was not sufficient to explain the observed form of the black drop, and that the effect of the solar limb darkening had to be included (Fig. 3.3.2).

Removing the two contributions left a circularly symmetric silhouette for Mercury, indicating that all causes of a measurable black-drop effect were accounted for (Figure 3.3.3)

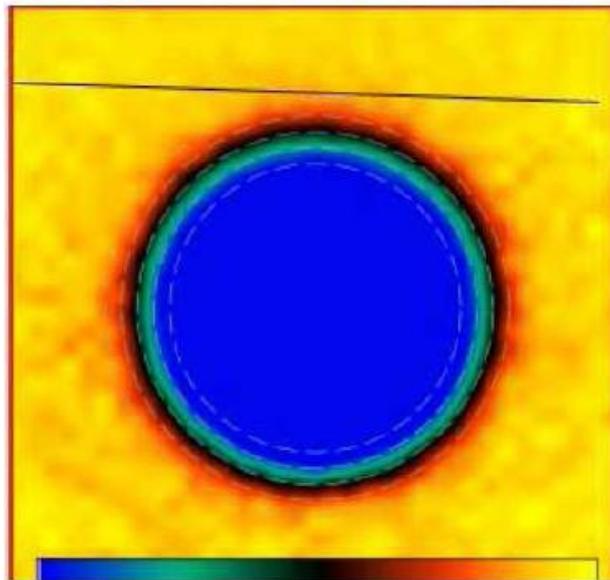

Figure 3.3.3. A TRACE image from the 1999 transit of Mercury, observed from NASA's Transition Region and Explorer (TRACE) spacecraft. The effect of the telescope's point-spread function and the solar limb darkening have been removed, revealing a symmetric Mercury silhouette. The position of the solar limb is marked. (Schneider, Pasachoff, and Golub/LMSAL and SAO/NASA)

Among missions that performed a measurement of the solar diameter with this method, there was also a measure of the SOHO satellite, during the transit of Mercury in May 2003. The result for the semi-diameter was 959.28 ±0.15 arcsec to 1 AU. This result should be compared with measures of SDS (see section 1.3.3 and 3.1.2) that between 1992 and 1996 observed a semi-diameter ranging between 959.5 and 959.7 arcsec. This gap seems to be far from a real variation of the solar diameter. Since these measurements were made with modern technology and through space missions, it emphasize that the problem is still far from a clear solution.
Mercury and Venus are not the only bodies that can transit between the Earth and the Sun. From the ground it can see in fact also transits of the Moon, better known as eclipses.
The eclipses were observed since ancient times for their charm, and physical aspects related to the Sun and the Moon were inferred from them since the birth of modern science. Different methods to infer a measure of the solar diameter were also developed by observations of eclipses. This topic is referred to the next chapter, where it is also proposed a new method to exploit the observation of eclipse with interesting results for this study.

# CHAPTER 4
# The method of eclipses

Total or annular eclipses can be treated as planetary transits regarding the measure of the solar diameter. But differently to the others planetary transits the occultation of solar light makes more simple measuring the instants of contact between the lunar and solar limb, even from the ground. Moreover, this phenomenon occurs more frequently than other planetary transits, being visible about every year somewhere on Earth.

## 4.1. Historical eclipses

The high visibility of the eclipse has produced a series of observations, even in times prior to the birth of the telescope. Our interest on the centenary variations of the solar diameter therefore make extremely important the analysis of these observations, even if made with the naked eyes.

**4.1.1. C. Clavius, 1567, Rome.** Stephenson, Jones and Morrison (1997) taken into account the observation of an annular eclipse made by the Jesuit astronomer Christopher Clavius in April 9, 1567 from Rome in order to derive limits to the Earth's rotational clock error. They attributed the appearance of ring to the "inner corona" of the Sun. But Pasachoff (2005) gives a description of the inner corona far from a circular symmetry. If the ring of the annular eclipse was instead the last layer of the photosphere, the average angular radius of the Sun would have been some arcsec larger than its standard value[*] of 959.63 arcsec (this correction is called as $\Delta R$ hereafter). Figure 4.1.1 shows that for the solar limb being higher than the mountains of the Moon, should be $\Delta R > +4.5$ arcsec. This result is even more surprising when one considers that measures of J. Picard a century later confirmed this greater measure (see section 1.5).

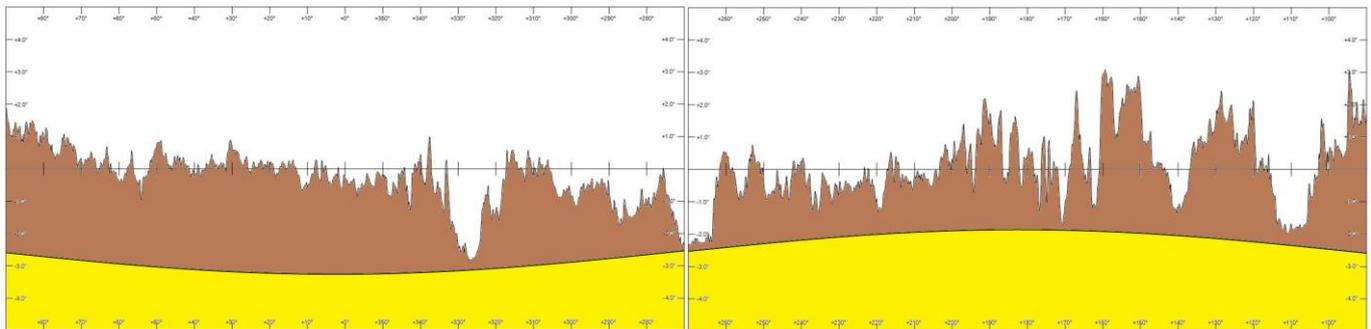

Figure 4.1.1. Eclipse of April 9, 1567 simulated with Occult 4 4.0 software. View from Collegio Romano (lat = 41.90 deg N, long = 12.48 deg east) where probably he would have made his observation. The lunar limb's mountains are plotted in function of the Axis Angles (the angle around the limb of the Moon, measured eastward from the Moon's north pole). The height scale is exaggerated. The solid line is the standard solar limb. The figures are the northern (left) and the southern (right) semicircle. To have a complete ring as Clavius reported, the radius of the Sun should be increased by further 4.5 arcsec above the lunar mean limb with respect to the standard value of 959.63 arcsec.

---

*defined by Auwers, 1891.

The observation of Clavius was the subject of several studies and publications: Kepler asked Clavius to confirm it was a solar ring, rather than diffuse appearance, that he would attribute to the lunar atmosphere, but always Clavius confirmed what he already wrote.

Because this observation was made with naked eyes, a more careful study on this eclipse has to take in account the angular resolution of an eye pupil. According to the formula for the angular resolution $\rho = 1.22 \lambda/D$, and taking into account a pupil diameter $D \sim 2$ mm (day vision), one gets $\rho \sim 50$ arcsec. Details wider than this limit are not visible on the Moon profile. This means the ring of the annular eclipse could have been not complete but divided by mountains no more than 50 arcsec, that is $\sim 3°$ in Axis Angle. For explain the observation of a complete ring with naked eye, for this eclipse, is thus sufficient $\Delta R \sim +2.5$ arcsec, that remains a surprising value. According to this proof of a larger solar diameter Eddy et al. (1980) assumed a secular shrinking of the Sun from the Maunder Minimum to the present.

The interpretation of this eclipse is still debated.

**4.1.2. Halley, 1715, England.** Edmund Halley attempted to measure the size of the umbra shadow by observing the total eclipse of 1715 in England.

Halley collected the numerous reports of this observation. His idea was to associate the data time of duration of the eclipse with the position for each observer, in order to assess the size of the shadow of the Moon on Earth. But from these observations we can obtain also interesting informations about the solar diameter. Morrison, Parkinson and Stephenson (1988) showed in their work the exact points of view in order to correctly infer the ephemeris, and then to give a measure of the solar diameter at the time.

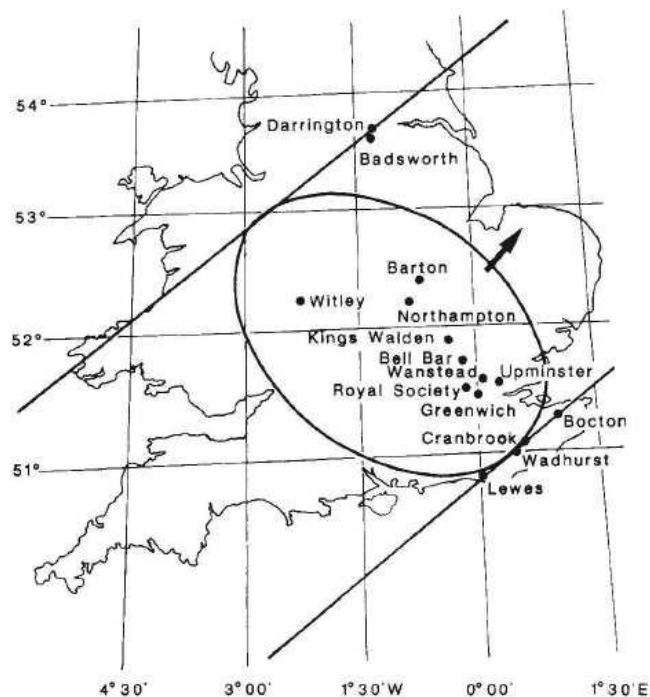

Figure 4.1.2. Path of umbra shadow of 1715 eclipse, showing the position of the oval shadow at one instant. Observations were made from the places marked on the map. (Morrison, Parkinson and Stephenson 1988)

In the present work, thank to the Occult 4 software and the new data on lunar profile (see section 4.2) we are able to reanalyze the 1715 eclipse data. In particular we consider the observations on the southern and northern edges of the shadow, where the eclipse is nearly grazing (the North or the South pole of the Moon moves nearly tangent to the photosphere). The advantage of these observations is that we gain a useful information only from the eyewitness about a total or partial eclipse observed. Instead the only way to get a measure of the solar diameter by an observation near the centerline[*] is measuring the duration time of the totality, which can be affected by errors if made with naked eyes. In fact the resulting solar diameter for two observer near the center line (Rev. Pound and Halley) differs for about 0.8 arcsec. It was not easy to identify the positions of the observers, and some uncertainties remain. Table 1 shows the eyewitness and the correspondent ephemeris simulated by Occult 4.

| Location of observation | Position coordinates | How appeared the Sun in the instant of maximum occultation | Standard Sun position by Occult 4 |
|---|---|---|---|
| **Darrington** | 52°, 40', 33.5" N 358°, 43', 41.1" E | "Point like Mars" | The standard solar limb is visible for about 0.1 arcsec above the bottom of a lunar valley |
| **Bocton Kent** | 51°, 17', 16.8 N 0°, 56', 16.8" E | "Point like Star" | The standard solar limb is about 0.85 arcsec under the lunar surface |

Table 1. Eclipse in 1715, England. Observation in the northern (Darrington) and southern (Bocton Kent) limit of the umbra shadow (position coordinates courtesy of D. Dunham).

Both observer we consider never observed a total eclipse. The first saw a "point of light like Mars" and the latter a "point of light like a star" in the instant of maximum occultation. According to Occult 4 and considering those points like Sun's photosphere we have a $\Delta R$ observed not lower than -0.1 arcsec for the first, and not lower than +0.85 arcsec for the latter. Adjusting for the possible errors on ephemeris (see section 6.3) we obtain $\Delta R > +0.38 \pm 0.1$ arcsec (the error is due to the uncertainties on the observation's coordinates).
This result is compatible with the first study developed by Dunham et al. (1980) on this eclipse: $\Delta R = 0.34 \pm 0.2$ arcsec.
An alternative explanation can be given considering that the regions immediately above the solar photosphere can have a brightness that become important when observed during an eclipse. The question is whether the brightness of this marginal lines-emission region (see figure 7.5) could explain also the Clavius observation. In section 5.3 we infer a consideration about this point.
It is thus emphasized the importance of measuring the limb darkening function well outside of the inflection point for evaluating the brightness in the visible.

## 4.2. Modern observations

**4.2.1. Baily's Beads.** The idea of Halley of using the Moon as a rule for evaluating the solar diameter was taken by Ernest Brown at Yale University for the observing campaign of the eclipse of January 25, 1925, total in New York.
Kubo (1993) subsequently found values for the solar diameter during the eclipse of 1970, 1973, 1980 and 1991, using for the first time theWatts lunar profile, and exploiting the same geometry of the Moon after a Soros cycle. But the error bars evaluated by Kubo did not consider larger systematic effects, due to the inaccuracy of Watts' profile of lunar limb for each libration phase (see section 4.2.2). Moreover Kubo used a photometer, thus the data were not spatially resolved. A decisive breakthrough was made thanks to D. Dunham that proposed to observe the Baily's beads in connection with lunar profile data. The Baily's beads, by F. Baily who first identified them during the annular eclipse of 1835 (Baily 1836), are beads of light that appear or disappear from the bottom of a lunar valley when the solar limb is almost tangent to the lunar limb. In grazing eclipses their number N can be high, providing N determinations of photosphere's circle. It is not their positions to be directly measured, but the timing of appearing or disappearing. In fact the times when the photosphere disappears or emerges behind the valleys of the lunar limb, are determined solely by topocentric ephemeris of the Sun and the Moon, their angular size and lunar profile at the instant, bypassing in this way the atmospheric seeing.

---

[*]Centerline is the path of the center of the umbra shadow. The umbra is the cone in which the Moon completely covers the Sun.

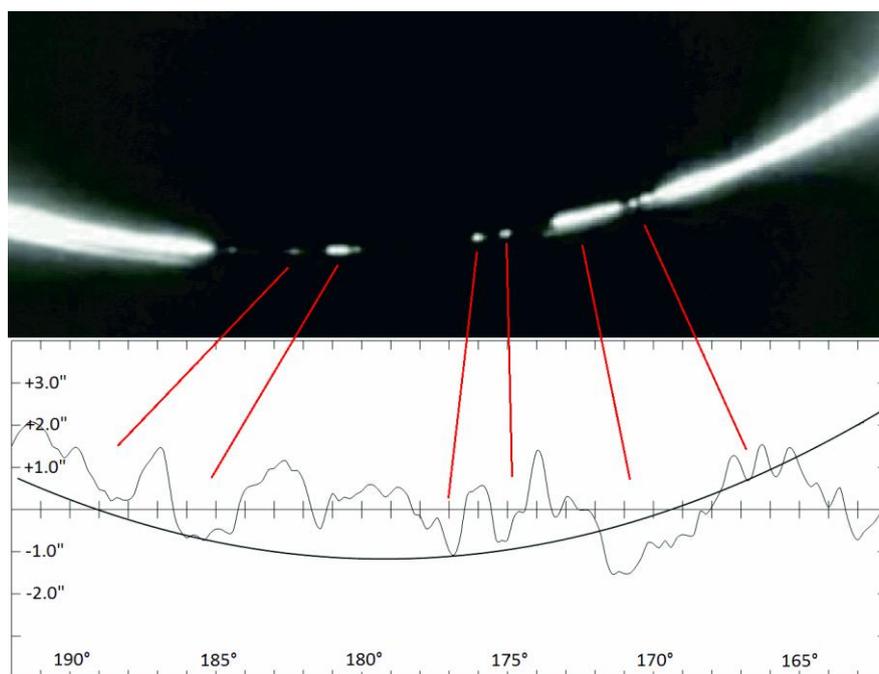

Figure 4.2.1. Eclipses of 15/01/2010, 5:25:45 UTC, Uganda (2°41' 19.8" N, 32° 19' 2.9"E) filmed by R. Nugent (up), lunar profile by Kaguya (low res.) by Occult 4 software (down).

**4.2.2. Lunar profile.** Until November 2009, the atlas of C. B. Watts (1963) was the most detailed map on the mountain profiles of the Moon at all libration phases. The spatial resolution of Watts' profile is 0.2° in axis angle (defined in figure 4.1.1) corresponding to ~ 3.2 arcsec at mean lunar distance from Earth. Since 1 arcsec at the same distance corresponds to 2 km, details on lunar limb are sampled each 6.4 km, with an accuracy of ±0.2 km. The intrinsic uncertainty on Watts limb's features is ±0.1 arcsec from mean lunar limb. Morrison and Appleby (1981) have extensively studied lunar occultations of stars and they found systematic corrections to be applyied to Watts' profiles. Their maximum amplitude is 0.2 arcsec. After that systematic correction random errors up to 0.1 arcsec are still possible. As an example the presence of the Kiselevka valley was discovered during the total eclipse of 2008 in a place where the Watts atlas of the lunar profiles did not show any alley (Sigismondi et al. 2009).
In November 2009 was publihed the new lunar profile performed by the laser altimeter (LALT) onboard Japanese lunar explorer KAGUYA (SELENE) (Araki et al., 2008), launched on 14 September, 2007. With Kaguya a new era for Baily's beads analysis is opened.

It was designed to perform the topography of the Moon from the altitude of 100 km. The number of geolocated points over the entire lunar surface is about $1.1 \times 10^7$. The radial topographic error is estimated to be ±4.1 m (1σ), where the range shift (<12m) caused by the distortion of the bounced laser signals due to the sloped and/or rough target terrain is dominant and incorporated to be 4 m (1σ). The horizontal topographic error is estimated to be ±77 m (1σ), where the orbit and pointing errors are the principal error sources and incorporated in the root sum square sense. The cross track spacing is less than 0.5°. The mean number density is 1/(220 or 240) m$^2$ with the largest gap about 2x2 km$^2$ [*]. An error of 4.1 m on lunar surface correspond to 2.1-2.3 mas at the distance of the Earth, a great improvement with respect to the error of ~100 mas of the Watts profile.

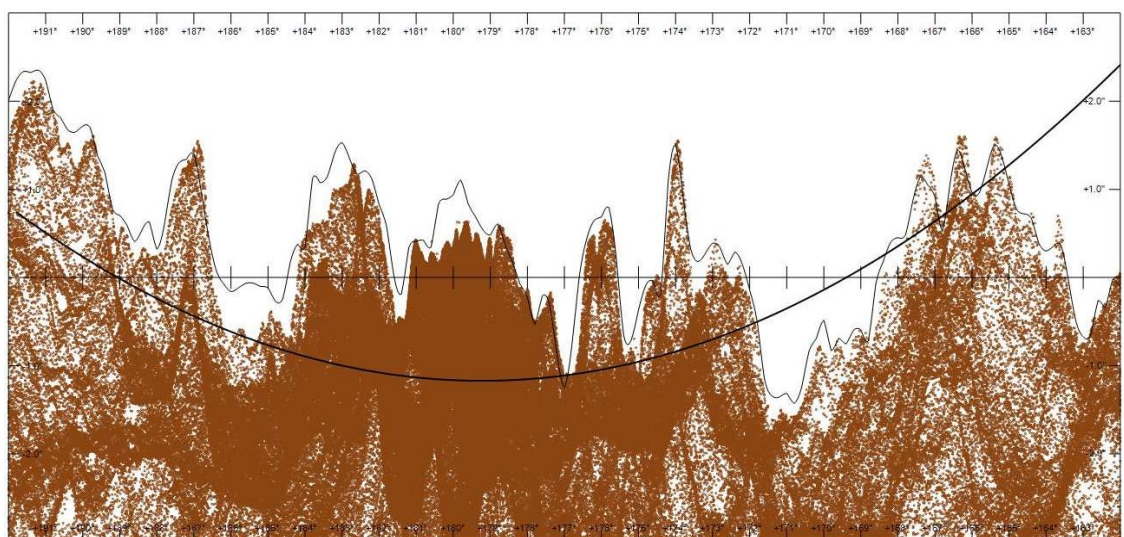

Figure 4.2.2. A comparison between Watts profile (solid line) and Kaguya profile (points) for the same configuration of Figure 4.2.1

**4.2.3. Recent analysis and new questions.** The International Occultation Timing Association (IOTA) is currently engaged to observe the eclipses with the aim of measuring the solar diameter. This is facilitated by the development of the software Occult 4[+] by David Herald.

The technique was to look at the time of appearance of the beads and calculating the relative position of the ephemeris with software Occult 4. In the instant of appearance or disappearance of the bead the Sun is tangential to the bottom of the valley of the Moon. The simulated Sun by Occult 4 has the standard radius (959.63 arcsec). The difference between the simulated edge of the Sun and the bottom of the lunar valley at the bead's instant is a measure of the radius correction with respect to the standard radius (ΔR).

Many improvements have been made during last years on the technique of observation. In particular, the choice to select the beads generated mainly in areas near the lunar poles. The reasons are as follows:
• the difference in latitude's libration among two eclipses is rather small and the same valleys produce the same beads at each eclipse;
• the polar beads can be observed for a time even longer than the duration of totality;
• Kaguya's profile in the polar regions has a higher sampling;

---
[*] JAXA; http://wms. selene.jaxa.jp/selene viewer/index e.html
[+] The version used in this study permits to plot both the Watts and Kaguya lunar profile. This version is no more available.

- the beads generated in the polar regions also correspond to the polar regions of the Sun: the maximum offset of the two axes of rotation is about 8.8°. Thus there is no risk to observe active regions that could affect the measurements (see section 2.4.2);
- a possible systematic error in the measurement of time is minimized in the polar regions (see section 6.3).

The method of calculating solar diameter through the observation of the eclipse seems to be more like a space method: the atmospheric seeing is bypassed . However, if one go to look at the results it seems that they are inconsistent, like for other methods of measurement. But recently the consideration of the mountain profile of the Moon allowed to revise the error bars (see last points in Figure 4.2.3); and now thanks to the Kaguya profile we have the possibility to improve the accuracy of the measures.

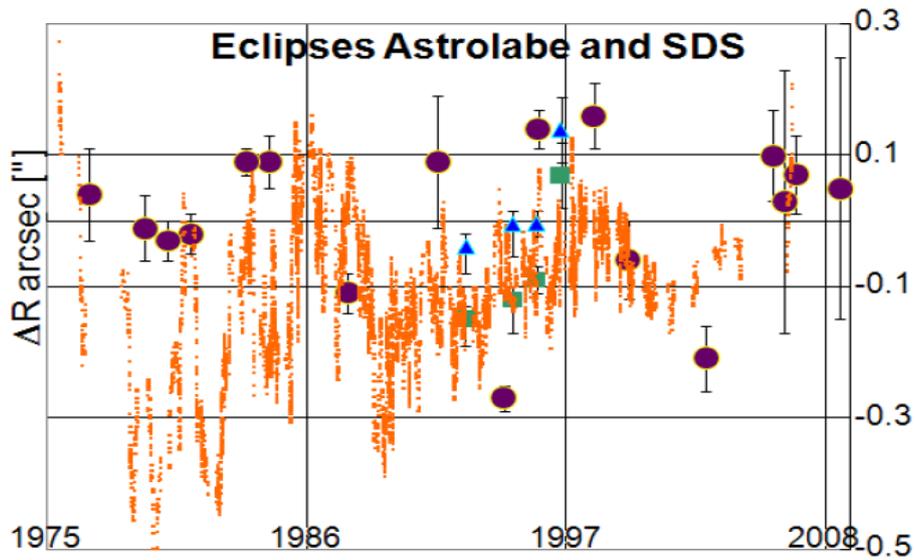

Figure 4.2.3. Corrections to mean solar standard radius of 959.63 arcsec: comparison between eclipse measurements of solar radius, astrolabe and SDS in the last three decades. The eclipse data are taken from Dunham et al. 2005, the last eclipses from Sigismondi 2006 and 2008. Black circles represent data of eclipses. Dots are astrolabe visual data courtesy of Francis Laclare. SDS data are represented with triangles (analysis, from Djafer et al. 2008) and with squares (analysis from Egidi et al. 2006).

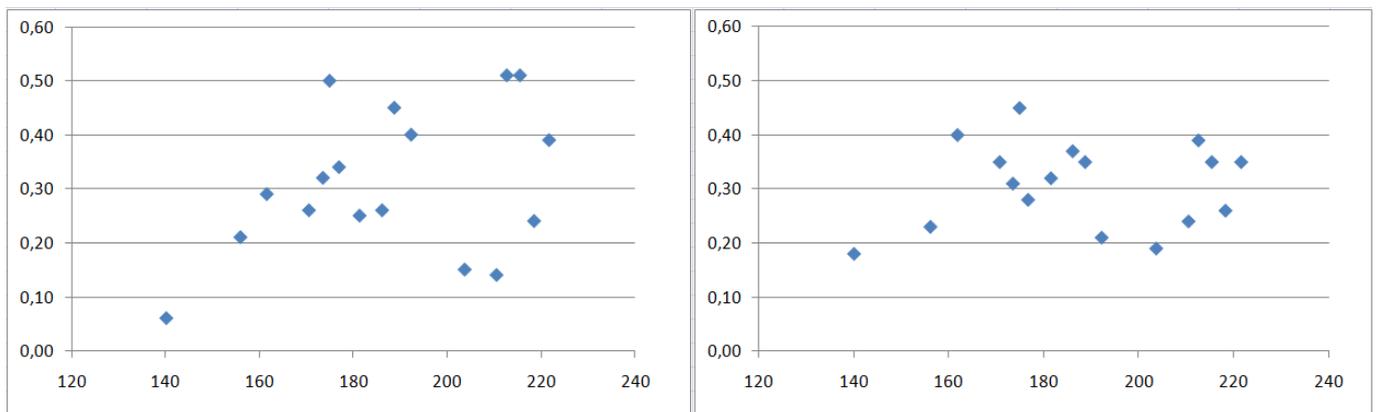

Figure 4.2.4. Baily's beads reported during the eclipse of 3/10/2005 in Spain by O. Canales. Only polar beads are selected (180°±40°). ΔR points in dependence of the axis angle. The left panel is the analysis with the Watts profile, the right panel with the Kaguya profile.

We recently analyzed the beads recorded during the 2005 eclipse reported in *Baily's beads atlas in 2005-2008 Eclipses* (Sigismondi et al. 2009). ata were first analyzed with the Watts' profile and later with the Kaguya profile.

The beads analyzed with the Watts profile always showed a greater scatter compared to the error attributed to the profile (see section 4.2.2). Some other source of error should be assessed. The situation was confirmed by analysis using the Kaguya profile.

An example is given in figure 4.2.4. The beads reported by O. Canales during eclipse of 2005 in Spain are analyzed. With Watts profile we obtain the result $\Delta R = 0.31 \pm 0.13$ (the error is the standard deviation), with Kaguya profile $\Delta R = 0.31 \pm 0.08$. With the Kaguya profile one expected a very lower scatter. By analyzing the beads with the Kaguya profile the contribution due to the errors on the mountain profile (that affected the Watts profile) was eliminated, but a large amount of scatter remains.

The source of error have to be found in the way of understanding the signal of the beads. The above analysis conceived the bead as a ON-OFF signal. But it should be treated rather like **a light curve of the bead event** (hereafter called light curve). Many factors can affect the amount of light received. Receiving more light on equal ephemeris situation can mean a higher calculated $\Delta R$, and vice versa. The scatter of the DR for one observer is then mainly due to:
• the arbitrariness of the observer in determining the instant ON;
• the shape of the lunar valley that produce the bead;

In addition there are other significant factors that could systematically vary the Signal/Noise ratio and thus the DR for different observers for the same eclipse or different eclipses:
• the transmittance of the telescope with filters;
• the characteristics of the detector;
• the transmittance of the atmosphere (could vary significantly for different eclipses for which the Sun has a different Zenith distance);
• the importance of the background noise, higher during the annular eclipse than the total one.

The total eclipse of 2008 in China was an occasion on which these differences became clear. Chuck Herold observed with a 13 cm Celestron telescope, and Richard Nugent using a 9 cm Questar was 1.6 km further out toward the shadow limit. It came out that Herold recorded the marginal line emission region (see figure 7.1.5) connecting all active beads, and it remained visible also during the totality, while Nugent did not. Since Herold was closer to the eclipse central path than Nugent, totality was longer as expected, but the begining of totality occurred without the usual zero luminosity signal (Sigismondi et al. 2009).

All these issues bring into question the earlier definition of the solar limb for observations of the eclipse. Moreover, this definition was not even comparable to other methods that instead use the profile of the LDF. The understanding of the bead as a light curve, instead, makes possible the use of the LDF profile through the observation of the beads.

In the next section we propose a new method for eclipse observations which defines the solar limb as the IPP (Inflection Point Position) of the LDF like other methods already seen. In this way the method becomes more reliable and the results are comparable.

### 4.3. A new method is proposed

The method proposed here is based on the assumption that the observed light curve of the bead is a linear transformation of the real light curve. This assumption seems reasonable for the following reasons:
• the transmittance of the instrument of observation remains constant (within the limits

of linearity of the detector);
• the transmittance of the atmosphere, during the short path of the Sun and the Moon in the sky, can be assumed constant (the zenith distance never changes more than 1°);
• even if seeing conditions may change rapidly the light curve is not affected.
The shape of the light curve is determined by the shape of the LDF (not affected by seeing) and the shape of the lunar valley that generates the bead.
Calling w(x) the width of the lunar valley (i.e. the length of the solar edge visible from the valley in function of the distance from the bottom of the valley), one could see the light curve L(y) as a convolution of LDF(x) and w(x), being |y| the distance between the botton of the lunar valley and the standard edge (where x = 0)*.
$L(y) = \int LDF(x)\, w(y - x)\, dx$
L(y) is given by its correspondence with L(t), the light curve in function of time (obtained from the observations). This correspondence is provided by Occult 4.

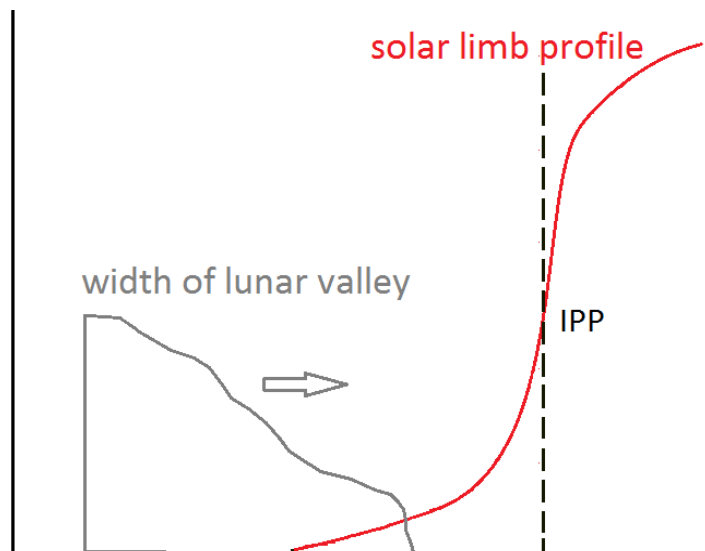

Figure 4.3.1. The light curve as a convolution of the solar limb profile and the width of lunar valley.

If the observed light curve is a linear transformation of the real light curve, as assumed, one could see the solar limb profile convolved as a linear transformation of the real solar limb profile:
$a \cdot L(y) = \int a \cdot LDF(x)\, w(y - x)\, dx$
where the position of the inflection point (IPP) in LDF(x) is conserved in a · LDF(x). Our goal is to deconvolve this relation in order to obtain a · LDF(x), i.e. the solar limb profile in arbitrary unit. A simple way to do it, is the transformation of the convolution into a discrete convolution:
$a \cdot L(m) = \sum a \cdot LDF(n)\, w(m - n)$
This is useful also to keep under control the errors of the light curve and of the lunar profile (see next chapter).
Thus we discretize the profile of the LDF in order to obtain solar layers of equal height and concentric to the center of the Sun. In the short space of a lunar valley these layers are roughly parallel and straight.

* we don't set equal to 0 the Inflection Point Position because its position is our goal

We also divide the lunar valley in layers of equal angular height. During a bead event every lunar layer is filled by one solar layer every given interval of time. Step by step during an emerging bead event, a deeper layer of photosphere enters into the profile drawn by the lunar valleys (see Figure 4.3.2), and each layer casts light through the same geometrical area of the previous one.

Being: $A_1, A_2.. A_n$ the area of lunar layers, from the bottom of the valley going outward;
$B_1, B_2.. B_n$ the surface brightness of the solar layer (our goal) from the outer going inward;

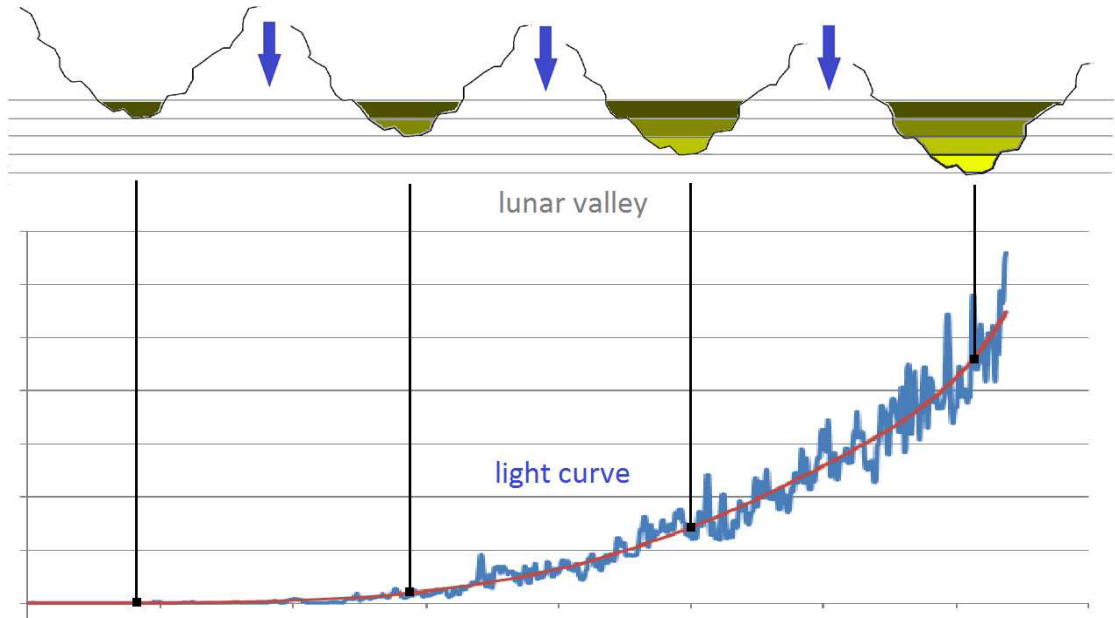

Figure 4.3.2. Every step in the geometry of the solar-lunar layers (up) correspond to a given instant in the light curve (down). The value of the light curve is the contribute of all the layers.

$L_1, L_2.. L_n$ the value of the observed light curve from the first signal to the saturation of the detector or to the replenishment of the lunar valley. We have:

$B_1 = L_1/A_1$
$B_2 = [L_2 − (B_1 \cdot A_2)] /A_1$
$B_3 = [L_3 − (B_1 \cdot A_3 + B_2 \cdot A_2)] /A_1$
$B_4 = [L_4 − (B_1 \cdot A_4 + B_2 \cdot A_3 + B_3 \cdot A_2)] /A_1$

and so on..

The situation described above relates to an emerging bead. The same process can run for a disappearing bead, simply plotting the light curve in lookback time (as explained in the next chapter).

The LDF obtained in this way is a discrete LDF. The smaller is the height of each layer, the more is the resolution of the LDF.

The angular height of the layers, i.e. the sampling, depend on the error of the lunar profile (as explained in the next chapter).

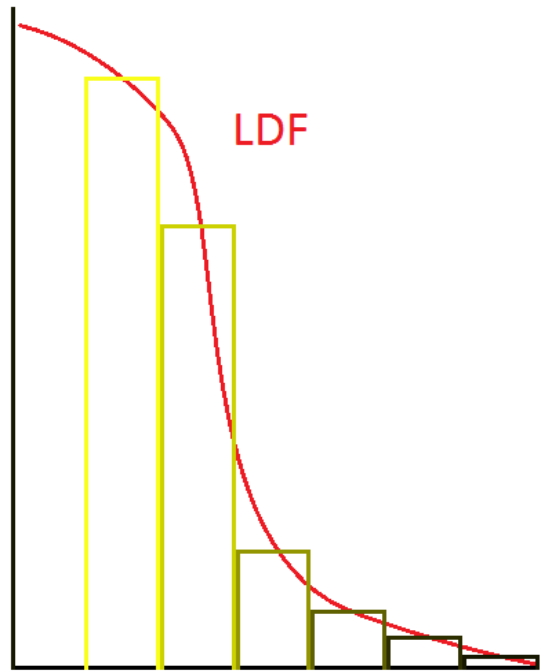

Figure 4.3.3. The discrete LDF. On y axis the brightness of each layer (per solid angle), on x axis the radial distance from the Sun's center.

The discrete LDF obtained in this way keeps the same shape of the real LDF, being its linear transformation. The position of the inflection point (IPP) is then conserved. The error on the IPP depend thus on the sampling and on the eventual error on the ephemeris (see chapter 6). This procedure require the assumption that the LDF profile maintain itself sufficiently constant during the time of the light curve. In section 7.3 we discuss about this assumption.
In the next chapter this method is applied for an observation in order to verify the feasibility.

# CHAPTER 5
# Applications of the method proposed

We studied the videos of the annular eclipse on 15 January 2010 obtained by Richard Nugent in Uganda and Andreas Tegtmeier in India. The equipment of Nugent was: a CCD camera (Watec 902H Ultimate); aMatsukov telescope (90 mm aperture, 1300 mm focal length); and a panchromatic ND5 filter (Thousand Oaks).
The equipment of Tegtmeier was: a CCD camera (Watec 120N); a Matsukov telescope (100 mm aperture, 1000 mm focal length); and an IOTA/ES green glass-based neutral ND4 filter.

**5.1.1. Light curve analysis.** Two beads located at AA = 171° and 177° are analyzed
for the two videos. They were disappearing for Nugent's video and appearing for Tegtmeier's video. For the analysis of the light curve a software specially developed for this purpose is used: the Limovie free software[*] (see Figure 5.1.1)
The standard deviation $\sigma$ of the background noise is calculated. Then a constant value of $5\sigma$ is subtracted from the light curve. The light curves of the disappearing beads are plotted back in time. The first positive value is considered the first signal. The negative values are set equal to 0. Then some polynomial fits are performed in order to reduce the electronic noise (see Figure 5.1.2).

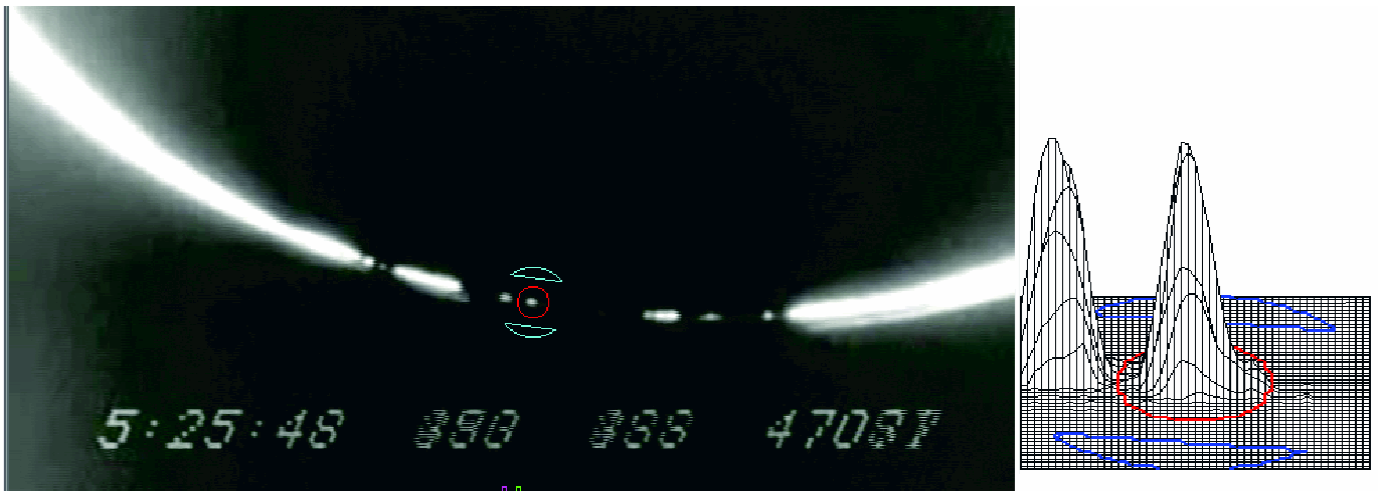

Figure 5.1.1. Limovie software. An instant of the light curve analysis is shown. The red circle selects the area where the light intensity is recorded as a function of time. The number of pixels involved in this area are $\pi r^2 = 380$, where r is a settable radius of the red circle. The two blue sections of the circle select the area where the background noise is recorded. On the right panel a 3D graph shows the intensity of the bead. This is useful in the choice of the radius of the red circle and in the evaluation of the saturation of the CCD.

---

[*] http://www005.upp.so-net.ne.jp/k_miyash/occ02/limovie_en.html.

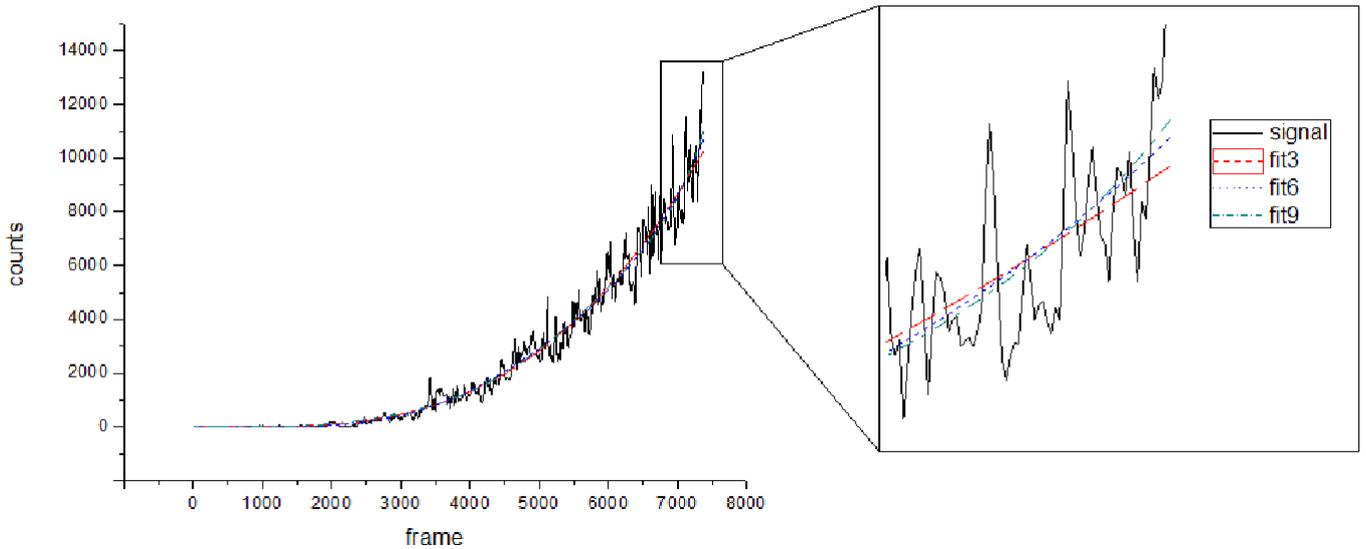

Figure 5.1.2. Light curve of the disappearing bead at AA = 177°. Increasing frame numbers correspond to decreasing time. The velocity of acquisition is 30 frames per second. The counts on the y-axis are the sum of the intensities of 380 pixels involved in the bead. Each pixel has 256 levels of intensity (eight bits). A detail of the light curve is zoomed in to on the right-hand-side panel. The polynomial fits (fit3–fit9, for the third- to ninth-order polynomials) are shown.

**5.1.2. The thickness of the layer.** The lunar valley analysis is performed with the new lunar profile obtained by the laser altimeter (LALT) onboard the Japanese lunar explorer Kaguya (see section 4.2.2).

The lunar valley has to be divided into layers as explained in Section 4.3 (see Figure 5.1.3). The layer at the bottom of the valley ($A_1$) is special, because i) it defines the thickness of all the other layers as well, ii) its area, being most heavily in contact with the lunar profile, is the most uncertain, and iii) according to the algorithm in Section 4.3, $A_1$ affects determinations of all $B_n$'s. The choice of the thickness of the layer has to be optimal; large enough to reduce uncertainties in $B_n$'s, but small enough to have a good resolution of the LDF. We chose h = 30 mas for the lunar valley at AA = 177° and h = 73 mas for the lunar valley at AA = 171°.

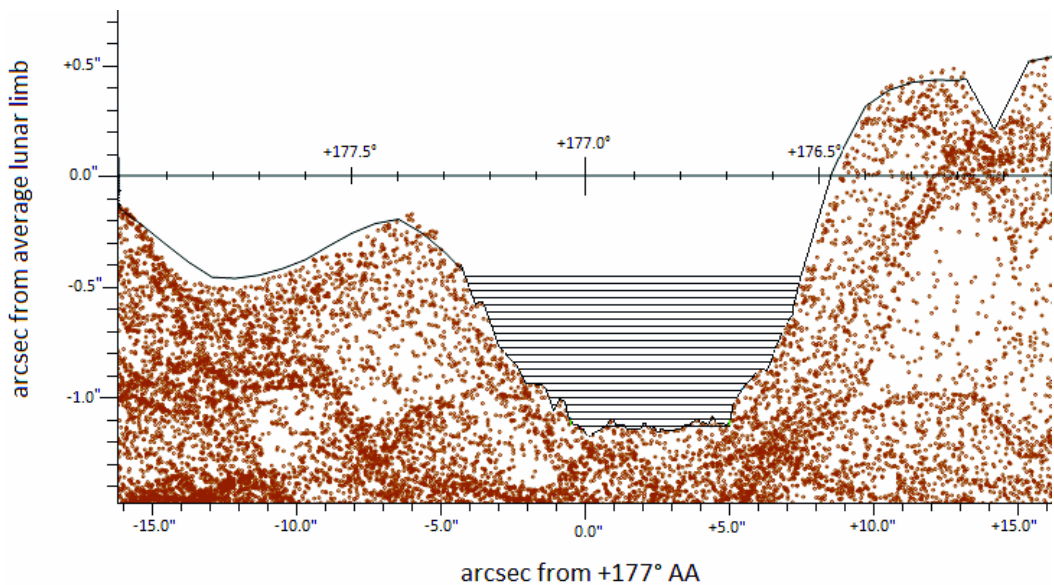

Figure 5.1.3. The lunar valley at AA = 177° is divided into 21 layers. The thickness of each layer is 30 mas.

**5.1.3. The motion of the solar limb.** The correspondence between time and y (defined in section 4.3) is not exactly defined, because the standard solar edge is not exactly parallel to the lunar mean limb during its motion (see figure 5.1.1).
We then calculate the mean position for 4 points: AA = 176.7°, 176.8°, 176.9°, 177.0°(see figure 5.2.2). The mean position in function of the time is a linear relation as expected (figure 5.2.3). The slope of the linear relation is the relative velocity with which the Sun fills the lunar valley. As an example we obtain for Nugent's video $w_y$ = 27.4 mas/s. Being the layer height h = 30.5 mas, every layer is filled in *t = h / $w_y$ = 1.11 s*. Given the time of the first signal: 5:26:8.47, and the time of the saturation of the CCD ~ 5:25:45, the total light curve lasts about T = 23.5 s. The number of the layers have to be *int(T / t) = 21* (see figure 5.2.2).

The motion of the Sun in the lunar valley have the two component: perpendicular to the mean limb $w_y$ (calculated above) and parallel $w_x$.
It could be interesting to know the total relative velocity w. It is obtained as follows: with the Occult 4 software we find the position of the standard solar limb with respect to the lunar limb in the initial and in the final instant of the light curve for many points along the 360°. The difference of these two positions is divided by the total time of the light curve (T). We plot the result (wy) against AA. The best sinusoidal function that interpolates these points give the direction (from the phase) and w (from the amplitude).
We obtain: direction AA = 268.9°, w= 0.354 arcsec/s.
For the lunar valley in AA=177° we obtain $w_x$ = 0.353 arcsec/s, being $w_y$ = 0.027 arcsec/s. (Figure 5.2.4).
Therefore during the light curve the surface of the Sun flows for T · $w_x$ = 8.2 arcsec, that corresponds to ~ 6000 km on the surface of the Sun (~0.5°in axis angle, that is about the width of the valley). There is thus another implicit assumption we made in the deconvolution procedure: the LDF has to remain sufficiently constant along the width of the valley. This assumption is discussed in section 7.3.

**5.1.4. Results.** A program in Fortran90 is performed to calculate the points along the LDF ($B_m$'s). The program takes into account the polynomial fits of the light curve and three values for each lunar layer area (A, A + ΔA, A − ΔA), giving a distribution of values for each point $B_n$. Figure 5.1.4 shows the results.
The resulting points show that the inflection point is clearly between the two profiles obtained for each of the two beads. The saturation of the CCD pixels precluded measuring inward the luminosity function for Nugent's video, while a low sensitivity made it impossible to measure outward the luminosity function for Tegtmeier's video. Therefore, it is impossible to infer an exact location of the inflection point, but it is possible to deduce upper and lower limits to the location of the infection point: -0.19 arcsec < ΔR < +0.05 arcsec.
The same eclipse was recorded by Adassuriya, Gunasekera, and Samarasinha (2012). The video analysis they performed with the classical approach (explained in Section 4.2) led to a value of ΔR= +0.26 ± 0.18 arcsec. This result does not seem compatible with the possible range for the position of the inflection point we found. This shows that the solar radius defined by the classical method is different from that defined by the inflection point of the LDF.
The LDF outside the inflection point seems to be very important in terms of brightness during an eclipse. Perhaps historical eclipses which gave a high value of the solar radius (examined in section 4.1) could be explained in this way. Anyway the eclipse observed by Clavius remains enigmatic.

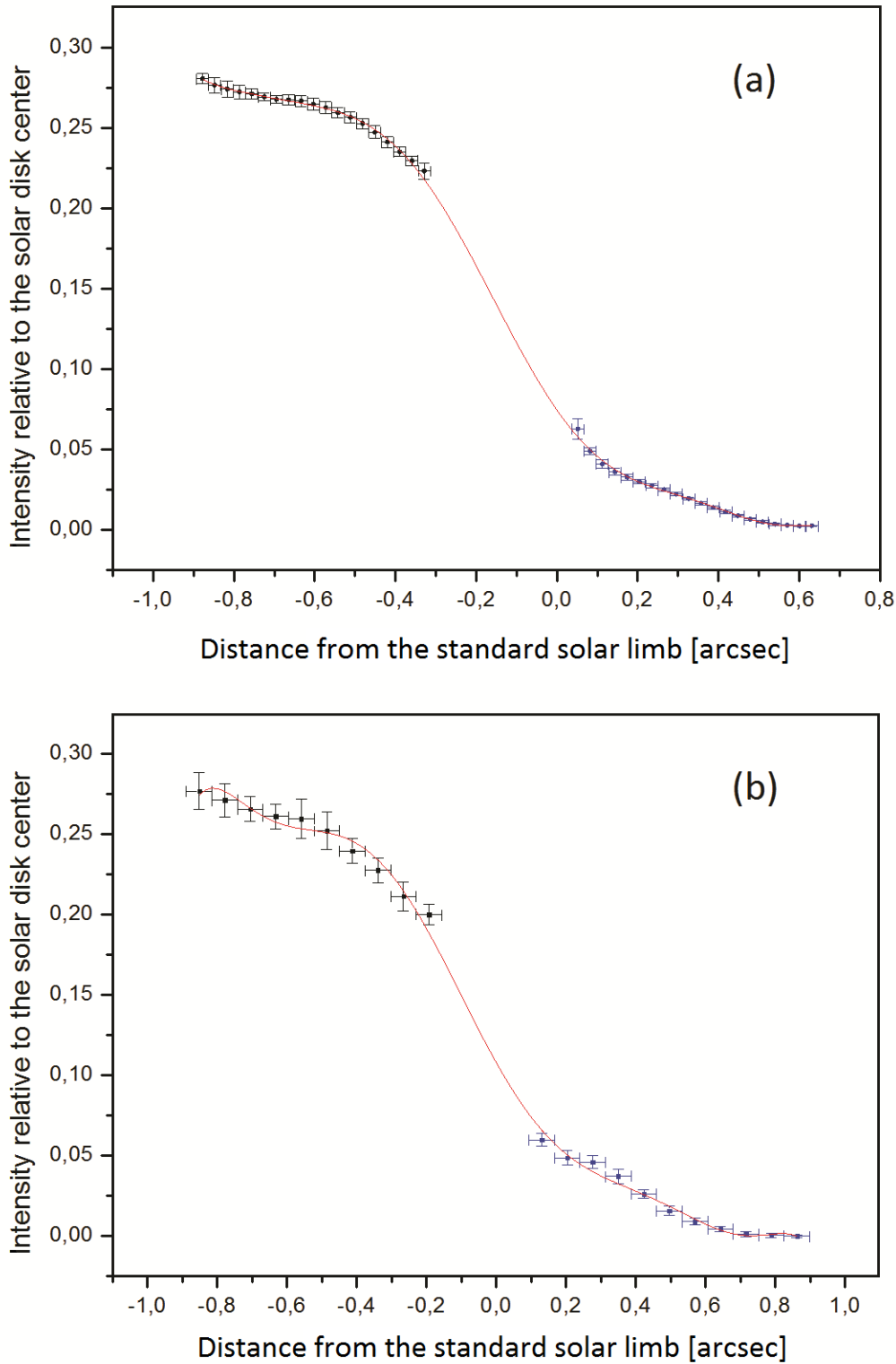

Figure 5.1.4. The luminosity profiles obtained are plotted and put together. The inner and brighter parts are obtained from Tegtmeier's video; the outer and weaker parts are obtained from Nugent's video. Panels (a) and (b) show, respectively, the profiles of the beads at AA = 177° and AA = 171°. The luminosity profiles are normalized to the center of the solar disk according to Rogerson (1959) for the inner parts, and in an arbitrary way for the outer parts. The zero of the abscissa is the position of the standard solar limb with a radius of 959.63 arcsec at 1 AU. The error bars on the y-axis are the 90% confidence level. The error bars on the x-axis are the thicknesses (h) of the lunar layers. The solid line is an interpolation between the profiles and gives a possible scenario for the position of the inflection point.

## 5.2. Simplifying the method

The procedure, explained so far, is involved and require a careful analysis of the light curve and the shape of the valley, in order to proceed with the deconvolution and to obtain an accurate evaluation of the errors. However this method could be greatly simplified by assuming for the lunar valley a linear shape, namely the width of the lunar valley increases linearly from the bottom outwards. This assumption does not introduce a significant bias on the results. In fact it was verified that the light curve is much more important for the final form of the LDF, compared to the shape of the valley lunar. Considering the formulas in section 4.3, this means one can obtain the shape of the **LDF by the first derivative of the light curve of the bead.** This operation is straightforward with any software for calculation. Obviously with this simplified procedure what one gets is just a shape of the LDF, losing any information about photometry. But it is a sufficient information if the target is the position of the inflection point to obtain the solar radius.

**Step by step.** What follows is a list of steps for the simplified method to detect the LDF. It differs with respect to the one exposed in section 5.1 because the analysis of the lunar valley is bypassed, and the analysis of the errors is also simplified:

1) Recording the beads, e.g. with Limovie software. The amount of light has not to depend by instrumental effect, like variable shutter or automatic gain. The Limovie data to be taken into account concern the signal recorded by the red circle (see figure 5.1.1) without any further elaboration, namely the column "Detect", and not the column "Result".

2) Subtracting the background noise. In some video of eclipses the background noise is significant and the instant of the first signal of an appearing bead could not be clear. The data concerning the background noise are obtainable toward the subtraction: Detect-Result columns. The resulting column represents the counts taken by the two background areas (blue), normalized for the detection area (red, see figure 5.1.1).

3) Excluding the saturated light curve. In some video the light could saturate the CCD. The saturation of the CCD has to be evaluated thanks to the visualization of the light curve of the bead in 3D (option of Limovie software, see figure 5.1.1). When the 3D bell shape of the bead reaches the top, the CCD is saturated.

4) Eliminating the electronic noise. To eliminate the electronic noise is sufficient to perform a polynomial fit of the light curve of the bead. The shape of the fit could be considered the shape of the light curve. Different grades of the polynomial fit could give different shapes. The difference between the polynomial fits could give an estimation on the error induced by the electronic noise (see figure 5.1.2).

5) Obtaining the LDF. The first derivatives of the polynomial fits have to be performed. They give the shape of the LDF. They could be different each other because they come from different functions. Those too much oscillating are not good and have to be throwed out. A plot of the profiles and an estimate by eye is sufficient to do the evaluation. In general the higher is the grade of the polynomial fit, the more is the oscillating behavior of its first derivative. Now one can perform the average and the standard deviation of the profiles for each point. These are the values and the errors of the piece of LDF concerning the path covered by the lunar valley during the light curve, usually less than one arcsecond for acquisitions with 8 bit of dynamic range.
6) Drawing the LDF along the solar radius. What we got in the previous point is a piece of the LDF in function of the frames of the video. However this independent variable does not make any sense. Any frame should be linked with time, thanks to the Limovie software, and time should be linked with the position of the solar limb (bottom of the lunar valley) with respect to the standard solar limb, thanks to Occult 4 software. Finding the correspondence between frame and position, one can plot the LDF in function of the solar limb position, being the standard solar limb in the zero point. This is necessary to locate the piece of LDF in a certain position along the solar radius.

## 5.3. Recommendations for future observations

**5.3.1. Dynamic range.** An immediate improvement of the method could be reached increasing the dynamic range of the detector from 8 to 12 bits, that is from 256 to 4096 levels of intensity. In this way the light curve could be observed over a longer path of the LDF before being saturated, revealing a longer piece of LDF with respect to those showed in figure 5.1.4. As stated in section 5.1, it was necessary to add different pieces of LDF coming from different observers, with different equipments, to obtain a composite profile longer enough to constrain the inflection point. However, in theory, with a 12 bit CCD it is possible to bypass this step obtaining a single profile useful for our purpose.

**5.3.2. Filters.** The LDF, and its inflection point position, change in function of the wavelength because both the continuum and lines behavior of the solar spectrum (see section 2.3). This fact makes evident that comparisons between different profiles taken by different filters have to made with care. The standardization of the filters between different observers is thus required. In section 7.2.2 it is highlighted how a measure in different bands can constrain models of solar atmosphere.

**5.3.3. Calibration.** As stated, this method permits to obtain the shape of the LDF, losing any information about photometry. These further information are not useful for the purpose of this paper, but they could be really useful for the study of the solar atmosphere. The potential of this method is evident on this direction. An observation in a certain bandpass and with calibrated equipment permits to obtain important spectrophotometric information.

# CHAPTER 6
# Ephemeris errors

## 6.1. Parameters used by Occult 4

The accuracy of the exact IPP of the LDF, depend on the sampling and on the brightness errors (see previous chapter), but also on the accuracy of the ephemeris. An error on the ephemeris would give a shift on the x axis of the LDF profile obtained in figure 5.3.2 and 5.3.3. In previous chapter we implicitly assumed as negligible the errors on the ephemeris. In this chapter we will verify if this assumption is reasonable.

Occult 4 uses various updatable data packets, to control all the necessary parameters to define the exact topocentric coordinates of the Sun and the Moon: the geocentric ephemeris, the reference geoid for the GPS, the exact Coordinated Universal Time (UTC) and its daily corrections, the polar motion and the atmospheric model.

**Geocentric ephemeris.** Occult 4 uses the JPL DE4xx ('JPL DE' stands for Jet Propulsion Laboratory Development Ephemeris). The ephemerides of the DE4xx series are expressed in coordinates referred to the ICRF (International Celestial Reference Frame). In this study we used the release DE423. These ephemerides provide geocentric or heliocentric coordinates of relevant bodies. The data have been made publicly available, in the form of data files containing Chebyshev coefficients, along with basic data-processing to recover positions and velocities. Evaluation and interpolation of the Chebyshev polynomials can give planetary and lunar coordinates to high precision. Accuracy for the inner planets is about 1 mas and no worse than 1 m for the Moon (http://ssd.jpl.nasa.gov/).

**Geodetic datum.** The latitude and longitude of observation sites are necessarily referred to a reference system - the Geodetic datum. The World Geodetic System is a standard for use in cartography, geodesy, and navigation. It comprises a standard coordinate frame for the Earth, a standard ellipsoid reference surface (the datum or reference ellipsoid) for raw altitude data, and a gravitational equipotential surface (the geoid) that defines the nominal sea level. The latest revision is WGS 84 that is the reference coordinate system used by the Global Positioning System (GPS) and is used by Occult 4. Presently WGS 84 uses the 1996 Earth Gravitational Model (EGM96) geoid, revised in 2004. This geoid defines the nominal sea level surface by means of a spherical harmonics series of degree 360, which provides about 100 km horizontal resolution (www1.nga.mil, http://cddis.nasa.gov/).

The precision of a GPS is about 2 m, and the accuracy allowed for the Earth coordinate by Occult 4 is 0.1" that is about 3 m. The resulting maximum error on the position of lunar limb is about 2 mas.

**Coordinated Universal Time.** The Coordinated Universal Time (UTC, replacing GMT) is the reference time scale derived from the Temps Atomique International (TAI) calculated by the Bureau International des Poids et Mesures (BIPM) using a worldwide network of atomic clocks. UTC differs from TAI by an integer number of seconds; it is the basis of all activities in the world. UT1 is the time scale based on the observation of the Earth's rotation. It is now derived from Very Long Baseline Interferometry (VLBI). The various irregular fluctuations progressively detected in the rotation rate of the Earth lead in 1972 to the replacement of UT1 as the reference time scale . However, it was desired by the scientific community to maintain the difference UT1-UTC smaller than 0.9 second to ensure agreement between the physical and astronomical time scales.

Since the adoption of this system in 1972 it has been necessary to add 23 s to UTC, due to the general slowing down of the Earth's rotation. The decision to introduce a leap second in UTC is the responsibility of the Earth Orientation Center of the International Earth Rotation and reference System Service (IERS).

The difference UT1-UTC, called DUT1, is daily updatable by Occult 4. The accuracy is 0.001 s, which allows for a negligible error on the (ephemeris.http://hpiers.obspm.fr/).

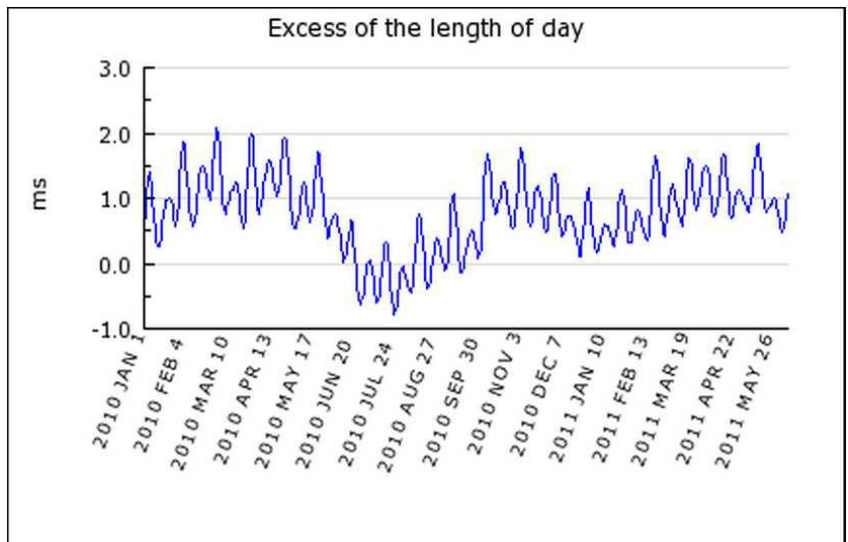

Figure 6.1.1. The not uniform slowing down of the Earth's rotation (Earth Orientation Center). (http://hpiers.obspm.fr/eop-pc/)

**Polar motion.** The physical mass of the Earth moves about the stable axis of rotation of the Earth. As well, the rotation of the Earth about its axis is non-uniform. The Earth Orientation Parameters (Occult 4 resource file daily updatable) define the location of the pole of rotation with an accuracy of 0.01" allowing for a negligible error on the ephemeris (http://hpiers.obspm.fr/).

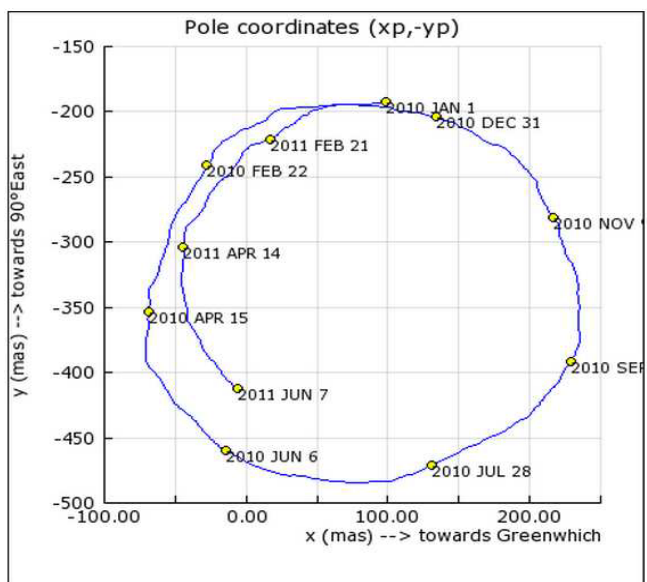

Figure 6.1.2. The moving terrestrial axis of rotation: polar motion (Earth Orientation Center). (http://hpiers.obspm.fr/eop-pc/)

## 6.2. Atmospheric model

Many programs for the calculation of ephemeris, like Occult 4, are based on the US1976 standard atmosphere. This model is useful for calculating the ray tracing.

US1976 is a static model, but different atmospheric parameters could mean different ray tracing in comparison to the calculation based on US1976 standard atmosphere. Regarding the measure of the solar diameter through the observation of the eclipse, the ray-tracing is important for knowing the position of the limit of the Moon's shadow on the surface of Earth at a certain time. The more are the variations of the atmospheric parameters compared to the standard atmosphere, the more could be the distance of the umbra shadow compared to the one provided by Occult 4. To resolve this problem we use a program developed by van der Werf (2003): Solemio. It is a flexible ray-tracing simulation for calculating astronomical refraction based on the US1976 standard atmosphere. This software allows for a free choice of the temperature, pressure and water-vapour at sea level, and allows for an adjustment of the vertical temperature profile.

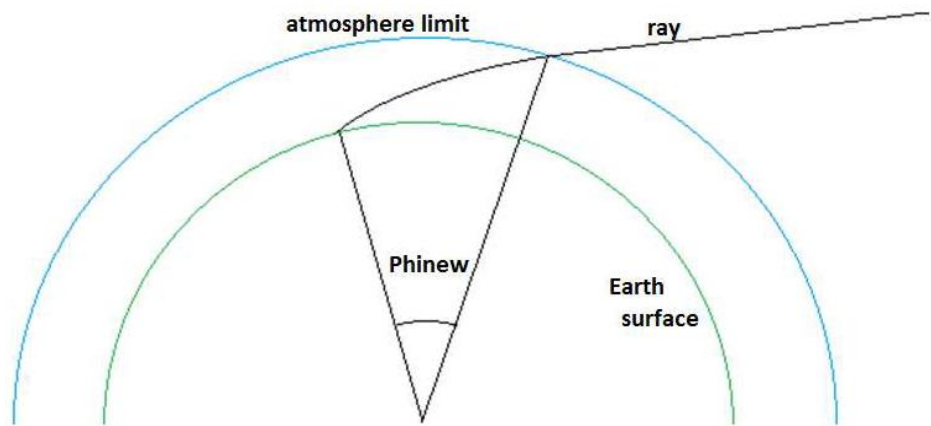

Figure 6.2.1. The ray-tracing described by the variable Phinew in the Solemio program by van der Werf (2003).

Among the variables useful in the text of the program, there is Phinew: the angle between the point of the ray at the conventional above limit of the atmosphere (85 km) and the ray at the height calculated in a certain step. The center of this angle is the center of the Earth. If one takes Phinew at the end of the trajectory of a ray (i.e. when the ray of light hits the Earth surface) and multiply it by the radius of the Earth, one will get the projection of the ray path inside the atmosphere, on the surface of the Earth.

The difference between the projections obtained with the true and the standard atmospheric parameters (true projection minus standard projection) gives the shift of the umbra shadow (antumbra for annular eclipse) with respect to the one provided by Occult 4. The direction of this shift is the Sun's azimuth, and the sense depends on the sign: toward the Sun's azimuth for negative values and in the opposite direction for positive one. Hereafter we call this shift as the vector **s**.

Solemio program allows to change the parameters on the ground, with respect to the parameters of the standard atmosphere: 288.15 K (15 °C), 1013.25 hPa, 0% RH (Relative Humidity). Many simulations have been carried out, and we found a simple formula linking $|\mathbf{s}|$ to the variation of atmospheric parameters.

$$|\mathbf{s}| = A \cdot (T\,[°C] - 15) + B \cdot (1013.25 - P\,[hPa]) + C \cdot RH\,[\%]/10$$

The coefficients A, B, C depend on the altitude angle of the Sun-Moon during the eclipse and are tabulated in Table 1.

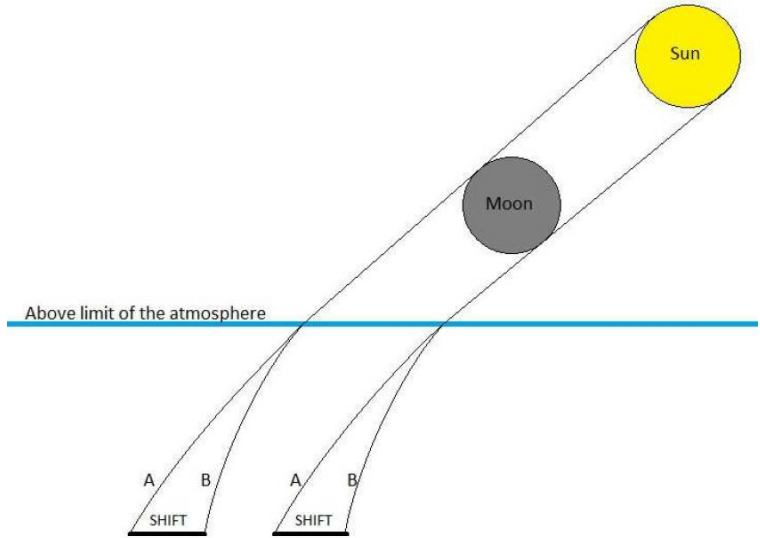

Figure 6.2.2. A and B correspond to 2 different configuration of the atmospheric parameters (say standard atmosphere and real atmosphere) and thus to different values of Phinew. This difference causes a shift of the eclipse shadow.

| altitude (°) | A | B | C | altitude (°) | A | B | C |
|---|---|---|---|---|---|---|---|
| 2 | 482,7 | 78,5 | 34,5 | 28 | 1,3 | 0,2 | 0,1 |
| 3 | 294,0 | 49,4 | 19,2 | 29 | 1 | 0,2 | 0,1 |
| 4 | 189,0 | 32,5 | 11,6 | 30 | 1 | 0,2 | 0,1 |
| 5 | 128,0 | 22,5 | 7,5 | 31 | 0,9 | 0,2 | 0,0 |
| 6 | 89,8 | 16,0 | 5,1 | 32 | 0,7 | 0,1 | 0,0 |
| 7 | 64,5 | 11,6 | 3,6 | 33 | 0,6 | 0,1 | 0,0 |
| 8 | 47,8 | 8,7 | 2,6 | 34 | 0,6 | 0,1 | 0,0 |
| 9 | 36,2 | 6,6 | 2,0 | 35 | 0,5 | 0,1 | 0,0 |
| 10 | 27,7 | 5,1 | 1,5 | 36 | 0,5 | 0,1 | 0,0 |
| 11 | 21,7 | 4,0 | 1,2 | 37 | 0,4 | 0,1 | 0,0 |
| 12 | 17,3 | 3,2 | 0,9 | 38 | 0,4 | 0,1 | 0,0 |
| 13 | 13,8 | 2,6 | 0,7 | 39 | 0,3 | 0,1 | 0,0 |
| 14 | 11,3 | 2,1 | 0,6 | 40 | 0,3 | 0,0 | 0,0 |
| 15 | 9,2 | 1,7 | 0,5 | 41 | 0,3 | 0,0 | 0,0 |
| 16 | 7,7 | 1,4 | 0,4 | 42 | 0,2 | 0,0 | 0,0 |
| 17 | 6,4 | 1,2 | 0,3 | 43 | 0,2 | 0,0 | 0,0 |
| 18 | 5,4 | 1,0 | 0,3 | 44 | 0,1 | 0,0 | 0,0 |
| 19 | 4,6 | 0,9 | 0,2 | 45 | 0,1 | 0,0 | 0,0 |
| 20 | 3,9 | 0,7 | 0,2 | 46 | 0,1 | 0,0 | 0,0 |
| 21 | 3,4 | 0,6 | 0,2 | 47 | 0,1 | 0,0 | 0,0 |
| 22 | 2,8 | 0,5 | 0,2 | 48 | 0,1 | 0,0 | 0,0 |
| 23 | 2,5 | 0,5 | 0,1 | 49 | 0,1 | 0,0 | 0,0 |
| 24 | 2,2 | 0,4 | 0,1 | 50 | 0 | 0 | 0 |
| 25 | 1,8 | 0,3 | 0,1 | 51 | 0 | 0 | 0 |
| 26 | 1,6 | 0,3 | 0,1 | .. | .. | .. | .. |
| 27 | 1,4 | 0,3 | 0,1 | 90 | 0 | 0 | 0 |

Table 1. Listed are the $|s|$ (m) for a variation respectively of +1 K (A), -1 hPa (B), +10%RH (C) on the ground, in dependence of the altitude angle of the Sun, with respect to the values of the standard atmosphere ($|s|$ is the shift of the umbra shadow with respect to that calculated with the US1976 standard atmosphere).

These coefficients give only a rough estimate of the true | **s** |, as also the shape of the vertical profile of atmospheric parameters can be different from the standard atmosphere. In general we obtain a negligible | **s** | if the eclipse is observed over 50°(altitude angle). Anyway the effect of the shadow shift on the ephemeris calculated by Occult 4 have to be discussed case by case. With Occult 4 we can achieve the correct ephemeris by applying the shift **s** to the observer position. During the eclipse recorded by Nugent (see previous chapter) the altitude of the Sun was ~ 19°, then | **s** | could be significant. Taking into account only the ground temperature, that was the only registered parameter (18 ± 2 °C) we obtain from the Table 1 | **s** | = 13.8 ± 9.2 m. In considering the uncertainty deriving from other parameters we take twice the maximum value as a conservative estimate on the total | **s** | = 40 m. Moving the observation location of 40 m in the opposite direction of the Sun during the eclipse (Sun azimuth angle was 113,5°) one obtain the position that give a correct value of the ephemeris (figure 6.2.3).

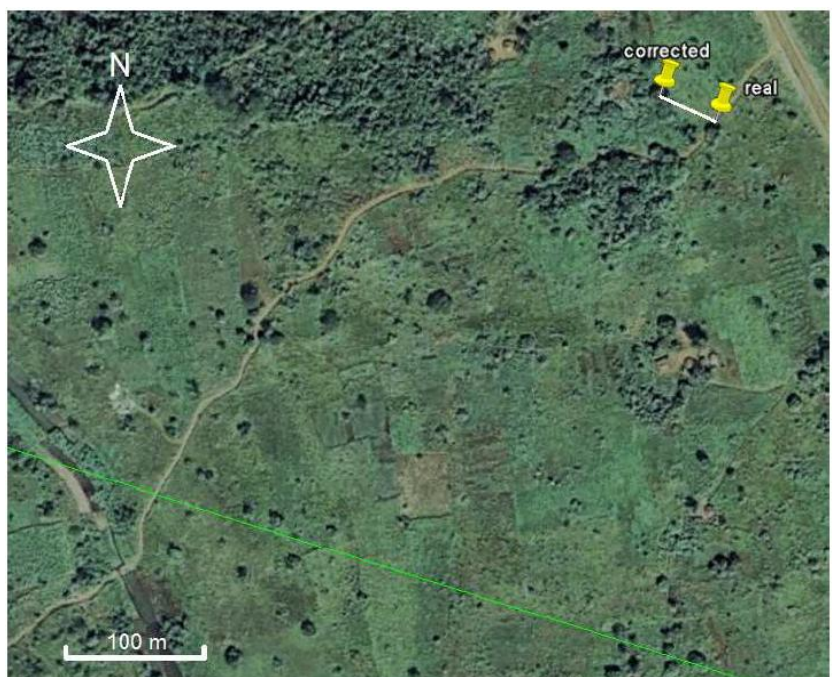

Figure 6.2.3. The real position of observation of R.Nugent during the eclipse examined in previous chapter (2°41' 19.8"N, 32°19' 2.9"E), and the corrected position for the effect of the shift of the eclipse shadow (241' 20.3"N, 32°19' 1.7"E). The length of the white solid line is a conservative estimate of | **s** | (40 m), its direction is the opposite of the Sun's azimuth during the eclipse. Google Earth provide the corrected coordinate position. The green solid line is the north limit of the antumbral path calculated by Occult 4.

Occult 4 gives the same results obtained before the correction: the effect of the **s** is negligible in this case. This is due to the fact that **s** direction is roughly parallel to the shadow path. In fact looking to a polar bead (as we made), a shift parallel to the shadow path have the same effect of a time shift ($\Delta t$ = | **s** | / shadow velocity). In this case $\Delta t = 10^{-2}$ s, negligible for a polar bead (see next section). Instead if **s** was perpendicular to the shadow path, would give a shift of about 20 mas on the solar profile obtained, and thus on the IPP, and thus on $\Delta R$. The magnitude of **s** is uncertain (we have only a rough estimate) but the direction depends only on the azimuth of the Sun during the eclipse. Since this direction is almost parallel to the path of the shadow in this eclipse we can assume that a significant bias is not introduced.

## 6.3. Assuming ephemeris bias

In the previous section we saw that a careful study must be done on ray tracing in the case of observations below 40°-50°altitude. But apart from that we have seen in general the ephemeris errors can be considered negligible. In this section we treat the problem without this assumption.

**Assuming a bias on time.** Given a bias on the time, for a polar bead the error is minimized. The figure 5.2.4 shows that the velocity wy of the solar limb in the valley is minimal for polar beads. In particular, for the bead analyzed in the previous chapter (wy = 27 mas/s) one get an error of 27 mas for a bias of 1 s. While for an equatorial bead, 1 s bias may give a bias of 350 mas on the IPP.

**Assuming a bias on longitude.** A bias on the longitude is also minimized by the polar beads. In fact the bias effect is:
$\Delta R_i = \Delta R + \sin(C + AA) \cdot B$
$\Delta R_i$ is the $\Delta R$ resulting from the $i^{th}$ bead.
AA is the axis angle of the bead.
B is the amplitude of the bias.
C is the position angle of the Moon's north pole measured from true north. It oscillate between -24° and 24° and during the light curve analyzed in previous chapter was -8.79° (automatically calculated by Occult 4).

**Assuming a bias on longitude and latitude.** Assuming an ephemeris error also on the latitude, the problem could be determined if a second observer catches the beads in the opposite pole. A good estimate on total $\Delta R$ is obtained averaging the $\Delta R$ obtained by the two observers.

**Assuming a bias on time, longitude and latitude.** In general, an unknown bias on longitude, latitude and time means that the Sun is shifted with respect to the standard Sun, being the magnitude and the direction of this shift unknown. To study this case we perform a simulation on a excel worksheet. The simulation has to take into account:
• the position of the standard Sun with respect to the center of the Moon, and its radius $R_{st}$;
• the radius of the real Sun $R_{re}$;
• the shift of the center of the real Sun with respect to the center of the standard Sun (hereafter called shift).
We assigned the values: simulated Sun radius $R_{re}$ = 960.0 arcsec (and thus $\Delta R$ = 0.37 arcsec being the standard radius 959.63 arcsec); shift magnitude = 0.335 arcsec; shift direction = 63.4° position angle. The center of the Moon is placed 60 arcsec south of the center of the standard Sun. The distance from the bottom of a lunar valley and the IPP (that we called y in section 4.3) is measured along the radius of the Moon. In three different longitude position of the Moon we simulated 3 polar beads with the respective $\Delta R_i$ measured.
The same operation is made for the opposite pole (with the center of the Moon placed 60 arcsec north of the center of the standard Sun).
The resulting $\Delta R_i$ are plotted against the position angle in figure 6.3.1. We noticed they are arranged on a sine wave. We obtain the parameters of this sinusoidal function:
• Amplitude = 0.34. It is an estimate of the magnitude shift (in arcsec) of the center of the Sun with respect to the standard Sun.

- Phase, that provides an estimate of the shift direction: 61.7° position angle.
- Average = 0.364. It is the estimate on ΔR (in arcsec). The error is -6 mas with respect to the given value.

In general, we have good estimates for ΔR (error within 10 mas) assuming a shift magnitude no more than 1 arcsec, and at least 2 beads for each pole.

A correct procedure of interpolation for real beads has to take into account the errors and the sampling relative to each $\Delta R_i$.

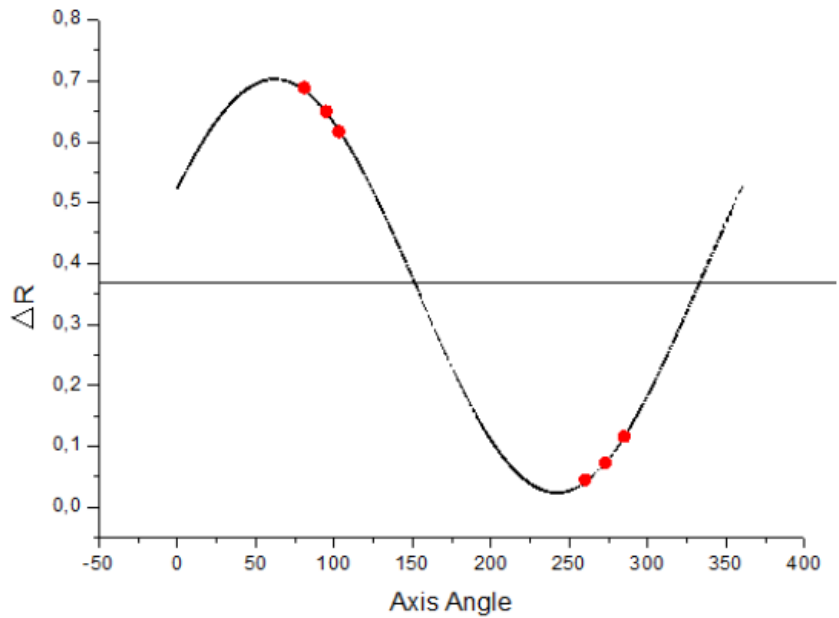

Figure 6.3.1. The simulated beads (red points) with the best sinusoidal fit.
The solid straight line is its average.

# CHAPTER 7
# Physical constraints by the LDF shape

## 7.1. Direct determinations by LDF observation

The limb shape contains information about the solar photosphere that cannot be derived from observations of the disk centre alone. Since the limb shape is determined by the density and temperature stratification, as well as the chemical composition of the solar photosphere, its measurement is important to validate models of the solar atmosphere.

This section shows how to get direct information from the observation of the LDF (we follow the Foukal 2004). However, for this purpose are necessary absolute measures, and not just a relative measure such as the one obtained in this study. To get the absolute calibration one can adjust the measure through an empirical or theoretical model. An example is the work of Faller and Weart (1969) who obtained a LDF by a method similar to that shown in this study. They compared their measure with the Bilderberg model (Gingerich and de Jager 1967,1968) to obtain the best fit with an absolute measure, and they found clues about a temperature minimum just above the photosphere (see section 7.1.2).

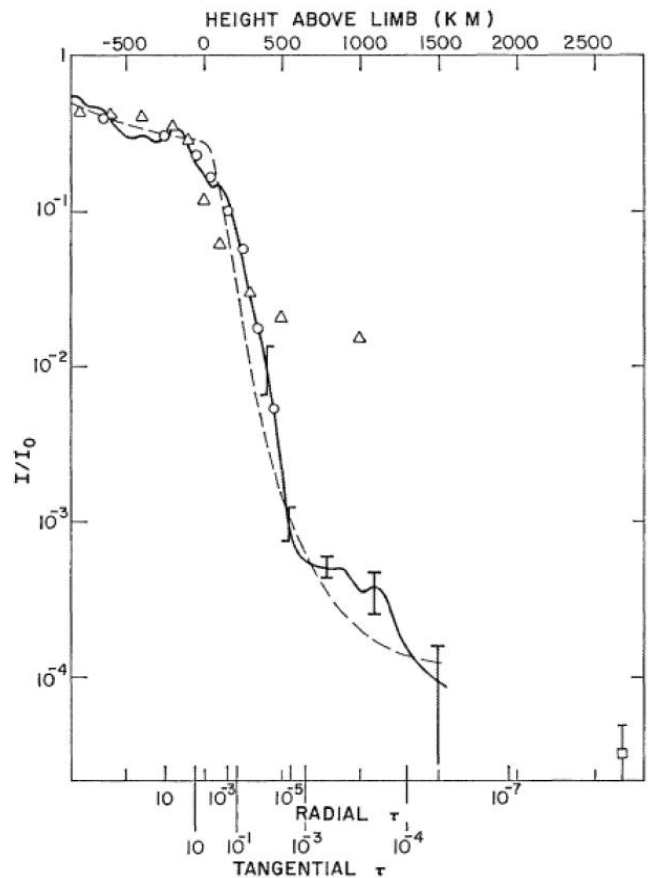

Figure 7.1.1. LDF in a narrow band around 527.8 nm, ratio to center-disk intensity.
Solid line and open square, observation. Dashed line, Bilderberg model prediction.
Open circle, Kristenson (1960). Open triangle, Dunn (1959). (Faller and Weart 1969).

It is difficult to calibrate the LDF obtained in Chapter 5 because we missed the point of reference that is the inflection point. But the detail obtained is much better than the previous measures like that of Faller and Weart, such that the method seems very promising for this task.

**7.1.1. The Radiative Transfer Equation.** To deal with radiation in a medium where both absorption and emission occur along a pencil of rays, we consider the transfer of a beam of intensity $I_\lambda(\theta, z)$ making an angle $\theta$ to the normal, through a plane-parallel element of atmosphere of thickness dz (Figure 7.2). The monochromatic extinction and emission coefficients per unit volume in the element are respectively $\kappa_\lambda(z)$ and $\varepsilon_\lambda(z)$. The change in beam intensity on passing through the element is:

1) $\mu \frac{dI_\lambda(\theta,z)}{dz} = \varepsilon_\lambda(z) - \kappa_\lambda(z) I_\lambda(\theta, z)$

being $\mu = \cos\theta$.

Introducing the optical depth, $d\tau = -\kappa_\lambda dz$, as a measure of the opacity of the element, we have from (1)

2) $\mu \frac{dI_\lambda}{d\tau_\lambda} = I_\lambda - S_\lambda$

The function $S_\lambda = \varepsilon_\lambda / \kappa_\lambda$ is called the source function.

The general solution of this linear first-order equation can be obtained by use of integrating factors when SΔ is given throughout a plane-parallel slab bounded by the surfaces $\tau_1 < \tau_2$.

3) $I_\lambda(\tau_1, \mu) = I_\lambda(\tau_2, \mu) \exp\left(\frac{-(\tau_2 - \tau_1)}{\mu}\right) + \frac{1}{\mu} \int_{\tau_1}^{\tau_2} S_\lambda \exp\left(\frac{-(\tau - \tau_1)}{\mu}\right) d\tau$

To obtain the emergent intensity SΔ emitted at the top of a semi-infinite atmosphere such as the photosphere, we have

4) $I_\lambda(0, \mu) = \frac{1}{\mu} \int_0^\infty S_\lambda(\tau_\lambda) \exp\left(\frac{-\tau_\lambda}{\mu}\right) d\tau_\lambda$

since the first term of equation (3) vanishes and $\tau_1 = 0$, $\tau_2 = \infty$.

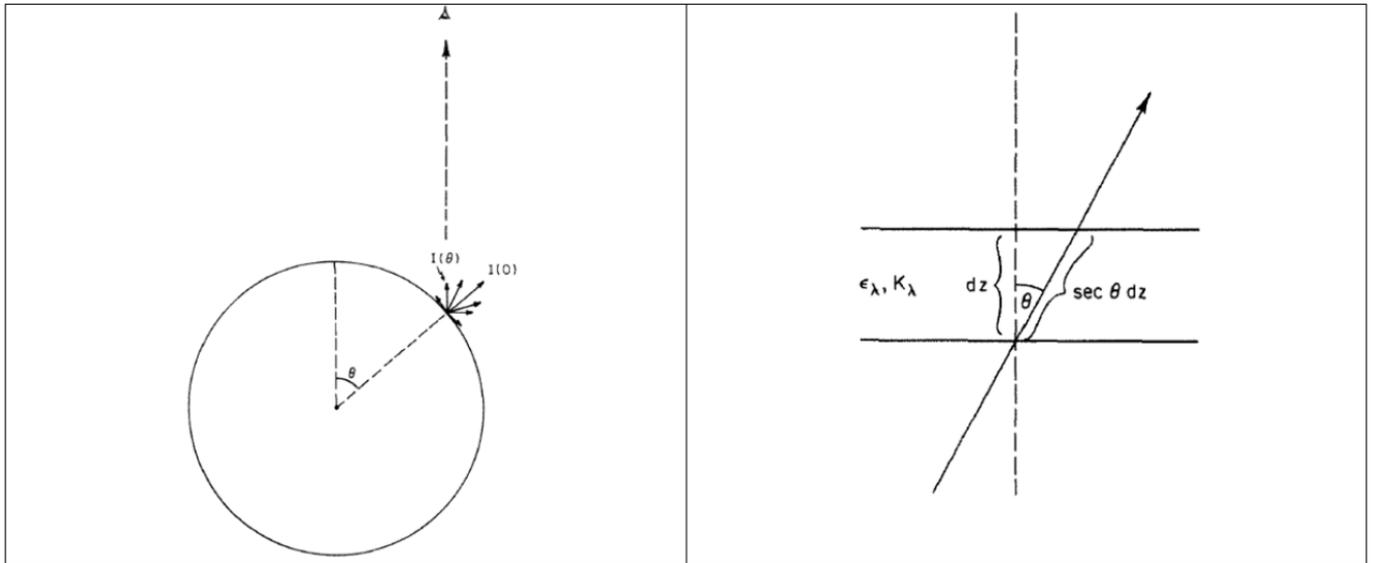

Figure 7.1.2. Left panel, illustration of how the angular distribution of atmospheric intensity is observed with measurements obtained at various heliocentric angles. Right panel, slab geometry used in deriving the radiative transfer equation. (Foukal 2004)

That is, the intensity emergent from the slab at an angle $\Delta = \arccos \mu$ is the sum along that slant path of all the contributions from the local source functions at each optical depth, each weighted exponentially by the optical depth between that point and the top of the slab.

**7.1.2. Temperature profile and opacity.** A direct determination of the temperature profile with optical depth can be obtained by inversion of the LDF.
From the equation (4), assuming LTE (Local Thermodynamic Equilibrium) and expressing the intensity relative to the disk-center intensity I(0, 1), we have for the LDF:

$$5)\ \phi_\lambda(\mu) = \frac{I_\lambda(0,\mu)}{I_\lambda(0,1)} = \frac{1}{\mu} \int_0^\infty \frac{B_\lambda(\tau_\lambda)}{I_\lambda(0,1)} \exp\left(\frac{-\tau_\lambda}{\mu}\right) d\tau_\lambda$$

The normalized source function can be expanded as a power series in $\tau_\lambda$

$$6)\ \frac{B_\lambda(\tau_\lambda)}{I_\lambda(0,1)} = a_\lambda + b_\lambda \tau + c_\lambda \tau^2 + \ldots$$

The coefficients of this expansion are evaluated by substituting equation (6) into equation (5) and integrating to yield the series

$$7)\ \phi_\lambda(\mu) = a_\lambda + b_\lambda \mu + c_\lambda \mu^2 + \ldots$$

whose coefficients can be evaluated by least-squares fitting to observations of the monochromatic limb darkening.

Once the coefficients are known, this expression gives the optical depth dependence of the normalized source function. We can then measure the intensity *I (0, 1)* at Sun center with an absolutely calibrated detector to obtain $B_\lambda(\tau)$ (the Plank Function) from equation (6). This yields the temperature as a function of optical depth, $T(\tau_\lambda)$, at that wavelength.

To determine the temperature profile with the geometrical depth needed to integrate the pressure in the hydrostatic equation, we require the opacity. A plot of the LDF $\phi_\lambda(\mu)$ for widely separated wavelengths, shows that the falloff of the intensity near the limb is much less pronounced in the red than in the violet (as in figure 7.1.3 and as already seen in section 2.3).

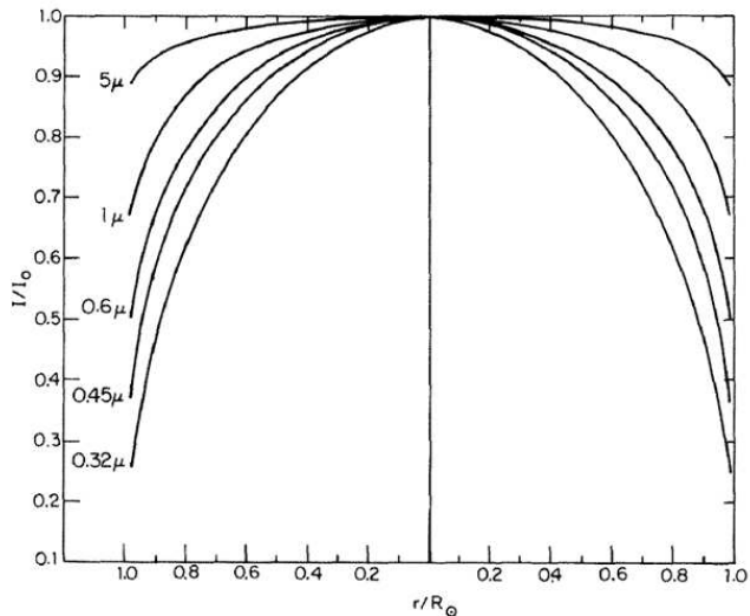

Figure 7.1.3. Photospheric limb-darkening curves measured at wavelengths between 0.32 mm in the ultraviolet and 5 mm in the infrared. (Foukal 2004)

The geometrical depth z, and thus the layer temperature $T_b$ (brightness temperature), visible at a given μ depends on wavelength.

To extract information on the absorption coefficient, one might plot the value of μ (or equivalently of $\tau_\lambda$) at which a fixed brightness temperature is observed, against the wavelength of the observation. Since a fixed $T_b$, defines a fixed geometrical depth, the variations of μ (or $\tau_\lambda$) in such a plot show the wavelength variation of the absorption coefficient.

It is most convenient to proceed by differentiating the expression for the optical depth to yield

8) $\frac{d\tau_\lambda}{dT} = \kappa_\lambda \rho(z) \frac{dz}{dT}$

where we have assumed that the wavelength dependence of $\kappa_\lambda$ is the same over the range of conditions found in solar layers. At a layer of given geometrical depth z, the temperature gradient *dT/dz* and the density *ρ(z)* will be fixed.

The left-hand side of expression (8) can be derived directly from observations ($d\tau_\lambda/dT = d\mu/dT_b$). It follows that a plot $d\tau_\lambda/dT$ against wavelength, obtained from LDF observations, should directly yield the wavelength dependence of $\kappa_\lambda$ and also a rough idea of its absolute value.

If the main opacity mechanism is known, the curve for *T(τ)* can be transformed to a curve of mass column density by starting at *τ = 0* and calculating the amount of material (in hydrostatic equilibrium) required to increase the monochromatic opacity by the amount $\tau_\lambda$. Comparison of the results obtained from this calculation independently at various wavelengths shows good agreement in establishing the curve of *T(z)*. Since the curve of *T($\tau_\lambda$)* changes shape with wavelength, the curve at λ = 0.5 μm is used as the reference profile and is usually denoted T($\tau_{0.5}$). The layer with $\tau_{0.5}$ = 1 is taken as the limit of the photosphere.

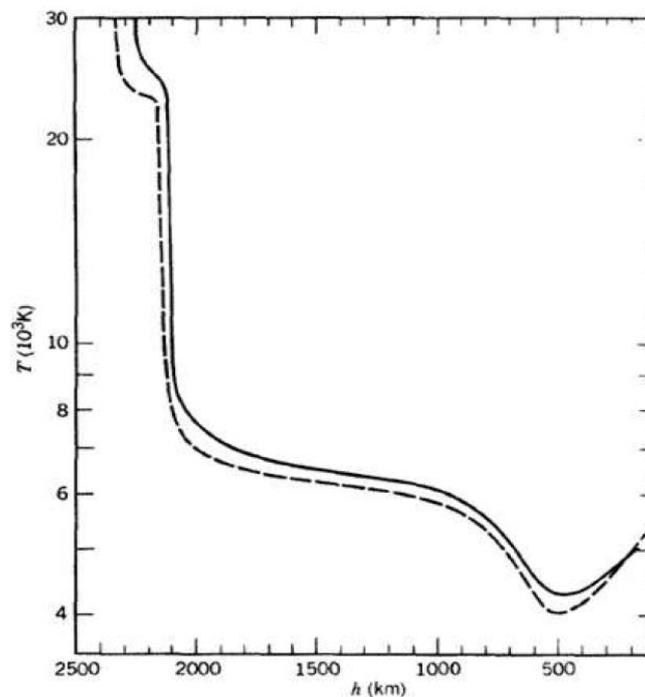

Figure 7.1.4. Plot of temperatures versus height for the cells (dashed) and network (solid). Photosphere limit (Δ0.5 = 1) is set to 0 km. (Vernazza, Avrett, and Loeser 1981).

Temperature profiles obtained (figure 7.1.4) show a minimum just above the photosphere. The layer between the top of the photosphere and the minimum is called mesosphere according to the Earth atmosphere definition. The chromosphere is usually defined to begin in the temperature rise just above the temperature minimum region. The model indicates a rapid temperature rise from minimum values around 4500 K located some 500 km above $\tau_{0.5} = 1$ to a broad temperature plateau of between $6 \cdot 10^3$ K and $7 \cdot 10^3$ K, about 500 km higher. This plateau extends out to about 2000 km and is followed by a very rapid rise toward coronal values within a few hundred kilometers.

It is worth noting that the surprising increase in temperature outwards poses the problem to find the mechanism responsible for heating these layers by non-thermal energy transport, such as waves or electric currents. Heating by radiation, convection, or conduction from the cooler photosphere is ruled out by the second law of thermodynamics.

**7.1.3. Flash Spectrum.** During an eclipse, the flash spectrum is the spectrum captured at the instants of beginning and end of totality by photographing over a moving plate the spectrum of the Sun.

Studies of limb spectra provide some of the most direct information about chemical composition, temperatures, and pressures of the solar atmosphere. Good "flash" spectrums, taken in the few seconds around eclipse totality, are the only unambiguous source of information on the height dependence of these properties.

Moreover our present recognition that the feeble lights of the chromosphere and corona are emitted by plasmas much hotter than those of the immensely brighter photospheric disk emerged in large part from the identification of the coronal forbidden lines by Edlen (1943).

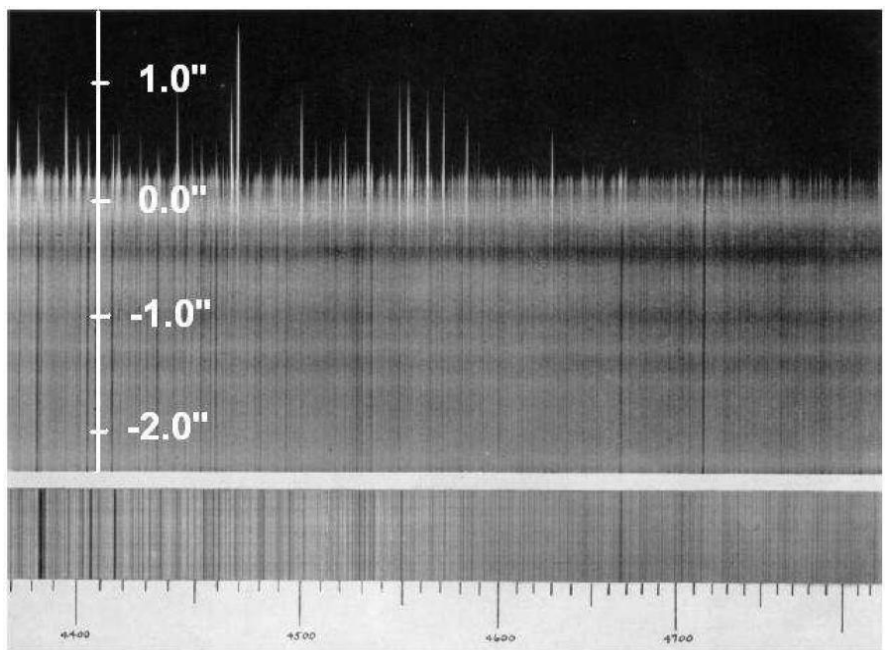

Figure 7.1.5. Flash spectrum of 1905 total eclipse by W.Campbell. The lower part of the figure is the spectrum before the totality, where the continuous of the photosphere and a lot of emitting lines. In the upper part there are only the lines of chromosphere; their blend is $10^{-4}$ times the continuum. After the beginnig of totally only the continuum of the corona can be detected. The blend of the small emitting lines is perceived as white light and it is $\Delta\ 10^{-3}$ times the intensity of the photosphere's continuum. (Sigismondi 2009).

The depths and profiles of strong Fraunhofer lines provide important constraints on T(z) in the upper photosphere and are usefully compared to the predictions of models that account for the non-LTE effects in these layers (see section 7.2.1).

Flash spectrum are taken since 1905 by W. Campbell. His work is a cornerstone on this topic.

## 7.2. Limb shape models

**7.2.1. The construction of a model.** The construction of an atmospheric model builds on three basic assumptions:
• the geometry of a plane-parallel, homogeneous layer, since the thickness of the photosphere is much less than the solar radius and the inhomogeneities of granulation can be neglected to a first approximation.
• The gas pressure distribution is assumed to be determined by the hydrostatic equation ($dP = -\rho g dr$) and the equation of state ($P = \rho RT/\mu$).
• The photosphere can be assumed to be in radiative equilibrium, although so-called radiative-convective models include a mixing-length representation of the convection that carries a substantial part of the heat flux through the deep photospheric layers.

Waves are neglected in most models because the interaction of large-amplitude waves with convection and radiation is a difficult subject that requires further theoretical effort.
The iterative procedure leading to a model begins with an estimate of the vertical temperature profile T(z), which is used to calculate a density distribution Δ(z) using the hydrostatic relation. Given a chemical composition, the opacities for this layer can be found from calculated tabulations as a function of ρ and T. In a LTE model, these opacities are based on calculations using the Saha and Boltzmann equilibrium equations to determine ionization and excitation of the important species. The LTE radiative transfer equation is solved to obtain a value of the emergent flux $F = \sigma T_{eff}^4$ corresponding to the temperature profile with optical depth. This total wavelength-integrated flux is a conserved quantity with height in a radiative equilibrium model.
The procedure used in deriving a non-LTE model is similar, but much more involved. The ionization and excitation needed to calculate the opacities must now be determined from the simultaneous solution of the coupled set of statistical equilibrium equations together with the non-LTE radiative transfer equation.
Purely theoretical models are iteratively adjusted toward the condition of radiative (or radiative-convective) equilibrium by changing the temperature profile. Each model is thus determined by a specific value of $T_{eff}$, the value of $g$, and by the assumed solar abundances.
Semi-empirical models are iterated instead toward agreement with certain observable quantities such as the continuum limb darkening and the disk-center emergent fluxes over a range of continuum wavelengths (Foukal 2004).

**7.2.2. A comparison between models.** Thuillier et al. (2011) evaluate in their work if the differences in predictions of the properties of the solar limb shape by some models can be discriminated by the measurements. They consider 4 models:
• VAL81 by Vernazza, Avrett and Loeser (1981). It is a semi-empirical one-dimensional model. The calculation of the limb shape for a given solar atmosphere structure requires a set of opacities and a radiative transfer model. In this work Thuillier et al. use a simplified radiative transfer model assuming spherical symmetry and local thermodynamic equilibrium (LTE).
• FCH09 is a semi-empirical atmospheric model developed by Fontenla, Balasubramaniam and Harder (2009). Calculation of the limb profile using the FCH09 atmospheric structures is performed with the Solar Modelling in 3D (SolMod3D) code. This code takes into account the spherical symmetry. The NLTE radiative transfer equilibrium is assumed.

• SH09 (PHOENIX) by Short and Hauschildt (2009) is a theoretical one-dimensional model for stellar atmosphere under the assumption of spherical symmetry, horizontally homogeneous layers and LTE.
• COSI (Code for Solar Irradiance, Haberreiter, Kosovichev, and Schmutz, 2008; Shapiro et al., 2010) is a semi-empirical model. Spherical symmetry and NLTE are assumed.

In general assuming LTE is not appropriate for the coronal and chromospheric layers. If the radiative transfer is solved in spherical symmetry the emerging intensities of the limb profile is more reliable.
The resulting limb shapes for three wavelengths are shown in figure 7.2.1.
Having the measured limb profile, one can compare it with the models to find the best fit and thus single out the best model. However, the reference point of the models is $\tau_{0.5} = 1$, not the center of the solar disk. There are two difficulties with this reference point: it is not accessible to the measures and its position with respect to the center of the solar disk could be different for different models.

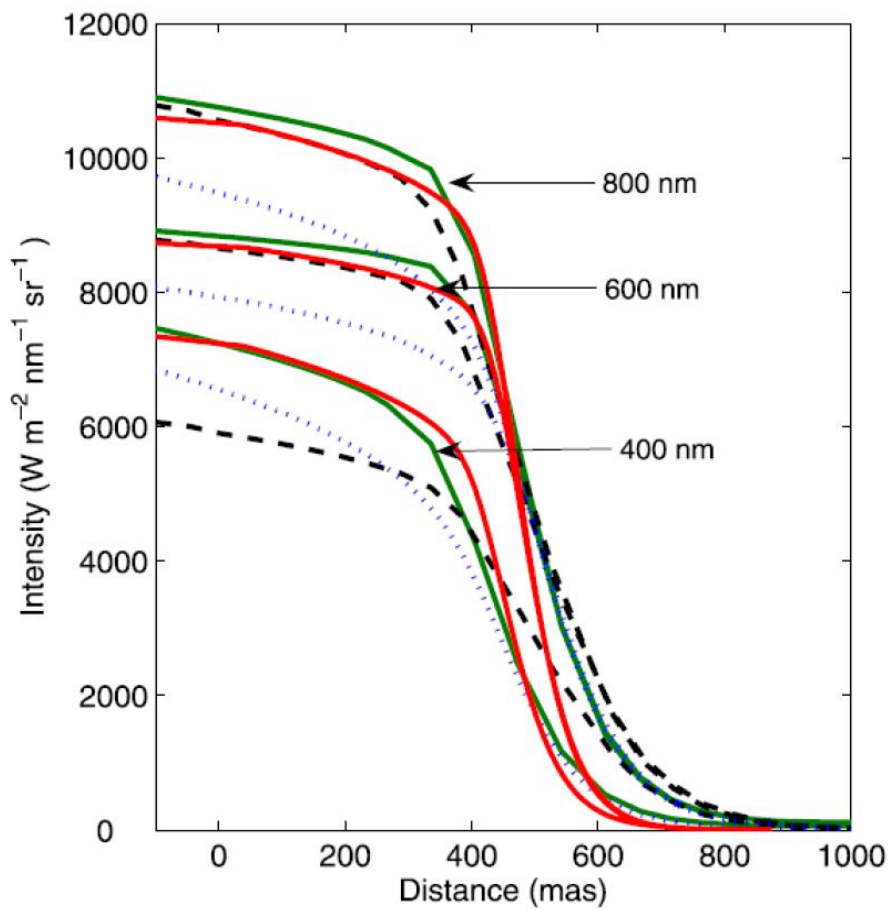

Figure 7.2.1. Limb shapes for three wavelengths predicted by SH09 (red), FCH09 (black), VAL-C (blue), and COSI (green) for 400, 600, and 800 nm. (Thuillier et al.2011)

To overcome this difficulties Thuillier et al. propose to discriminate the best model according to the variability of the IPP with wavelength.

| Models | IPP(400) | IPP(600) | IPP(800) | ΔIPP1 | ΔIPP2 |
|---|---|---|---|---|---|
| VAL-C81 with Straka06 opacities | 444.04 | 494.65 | 513.20 | 50.6 | 69.2 |
| COSI | 432.45 | 464.99 | 486.23 | 32.5 | 53.8 |
| FCH09 | 473.87 | 454.11 | 475.55 | − 19.8 | 1.7 |
| SH09 (ref: 958.7034 arcsec) | 959.34 | 959.36 | 959.37 | 26.0 | 37.0 |

Table 1. Comparison between the inflection point position (IPP) obtained with the FCH09, SH09, COSI, and VAL81 models. The second, third, and fourth columns provide IPP with respect to $\Delta 0.5 = 1$ at 400, 600, and 800 nm, respectively. The fifth and sixth columns give the variation of IPP using the position at 400 nm as reference (in units of mas). DIPP1 and DIPP2 show the difference of inflection point position at each wavelength (600 nm and 800 nm) with respect to its position at 400 nm. (Thuillier et al. 2011)

From Table 1 one can see that the variation of the position of the inflection point with wavelength can change considerably depending on the radiative transfer assumptions (LTE or NLTE) and opacities. Between the models, the prediction difference clearly increases with wavelength. Consequently, measurements in the near-IR facilitate the distinction between models.
A common feature of all the models is that the predicted slope of the limb is steeper than the observed one. This might be explained by the fact that the models represent the continuum and are based on one-dimensional static atmospheric structures. Indeed, the real solar atmosphere includes a variety of dynamical processes (such as granulation, gravity waves, oscillations, etc.) that inevitably broaden the observed limb.

### 7.3. Solar network pattern

As stated, the choice of the polar beads is right also for avoiding the observation of the limb in a solar active region. The measures of the solar limb in a quiet region is necessary if we are interested in comparable measures for monitoring the solar diameter. However, even the quiet region have a certain pattern due to some dynamic processes. Here we analyze the most relevant with regard to their influence on the measurement process introduced in this study.

**Granulation.** The characteristic scale of the bright granules is about 1000 km, ranging widely from structures as small as the ~ 300 km (resolution limit) up to the largest features extending to over 2000 km. The average center-to-center granule spacing is 1400 km with a wide range. The average photometric contrast measured near disk center in white light around 550 nm ranges between roughly 10 and 20%. This contrast increases with decreasing wavelength in the visible. The life pattern of an individual granule seems to consist of its formation from several smaller pre-existing components, its expansion to a maximum size of 3-5 arcsec, and its splitting up into several fragments which tend to fade away in situ (Figure 7.3.1). Figure 7.3.1 indicate that a typical granule lasts for about 5 minutes. The temperature variation pattern caused by granulation persists to at least 150 km above the $\tau_{0.5} = 1$ level (Foukal 2004).
For the purposes of the LDF measurement with the method shown in this study, we note that the granulation does not significantly change the measurement made at different points, as there could be many cells in the line of sight, and they maintains a configuration approximately constant in the duration of the light curve of one bead (less than a minute, see figure 7.3.1).

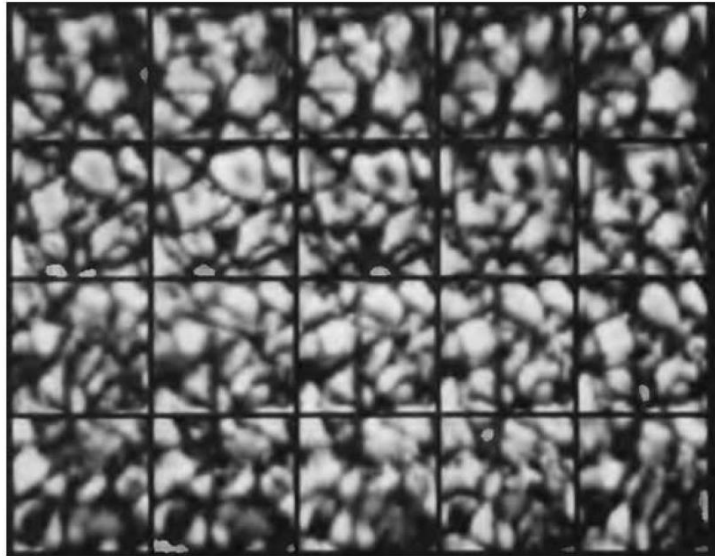

Figure 7.3.1. A time sequence showing granule evolution. The field of view in each of the 20 frames is 7 x 7 arcsec (about the width of the lunar valley we analyzed in chapter 5), and the time intervals are about 1 min. Photos taken at La Palma, Canary Islands, by C. Scharmerand P. Brandt and provided by T. Tarbell. (Foukal 2004)

**Supergranulation.** The Sun presents a much larger scale pattern than granulation, with velocity amplitude diminishing toward disk center. This velocity pattern, called supergranulation by Leighton (1962), consists of cells of characteristic scale around $3 \cdot 10^4$ km in which predominantly horizontal velocities of about 0.5 km s$^{-1}$ are observed. It is unclear whether they represent large scales of convection at some depth below the photosphere or, for instance, magnetic dynamo waves engendered by nonuniform solar rotation. The lifetime of the cells is of the order of a day. The photospheric network can be glimpsed in white light near the limb as a tracery of bright structures roughly outlining the supergranular cells. It becomes invisible in a photograph closer to disk center, on account of granulation noise, being the supergranulation photometric contrast between 0.1-1 % (Foukal 2004).

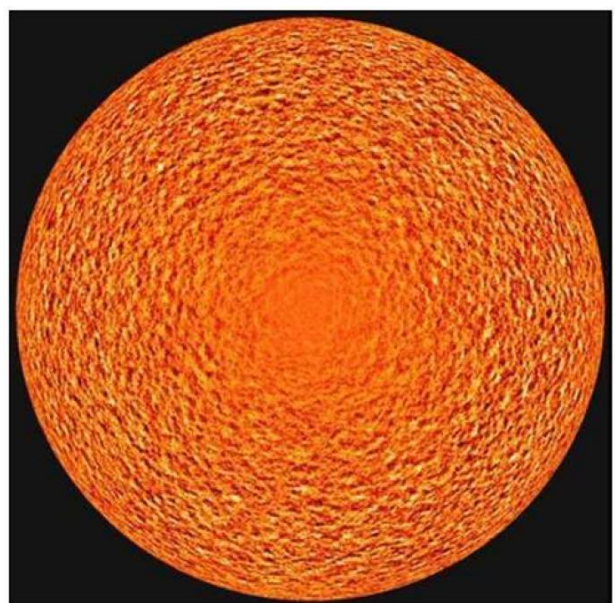

Figure 7.3.2. A SOHO/MDI Dopplergram showing supergranular speed pattern (the axisymmetric brightness gradient produced by solar rotation has been removed). (sci.esa.int)

The scale of the cells is larger than the scale of a lunar valley. It is difficult to estimate the influence of this pattern on the measures of the LDF in different points. The slight incompatibility of the two LDF profiles obtained in Chapter 5 doesn't seem to be due to this pattern, because the incompatibility occurred in regions outside the photosphere.

**Spicules.** Spicules dominate the chromosphere in non-active regions. They extend over roughly 10 arcsec, with thicknesses of between 1 and 2 arc sec. They seem to last typically between 5 and 10 min when followed in Hafilms. The largest and longest lived occur near the poles. They are a really common view on the solar surface and thus they can be considered an integral part of the solar surface regarding the LDF profile. The limb models seen in the section (7.2), place the inflection point approximately at the lower limit of the chromosphere. Thus according to this model the LDF measured in this study belongs entirely to the chromosphere. Regarding the different profiles shown in figure 5.3.3 we may have found a different pattern of spicules in the two profiles obtained.

The figure 7.3.3 summarizes the layers of the solar atmosphere, indicating the area where the spicules take place. We do not know exactly where the profile of the LDF obtained in this study is placed. The important brightness obtained would place it in the marginal line emission region, close to the photosphere, while the models shown above would put it entirely in the inner chromosphere.

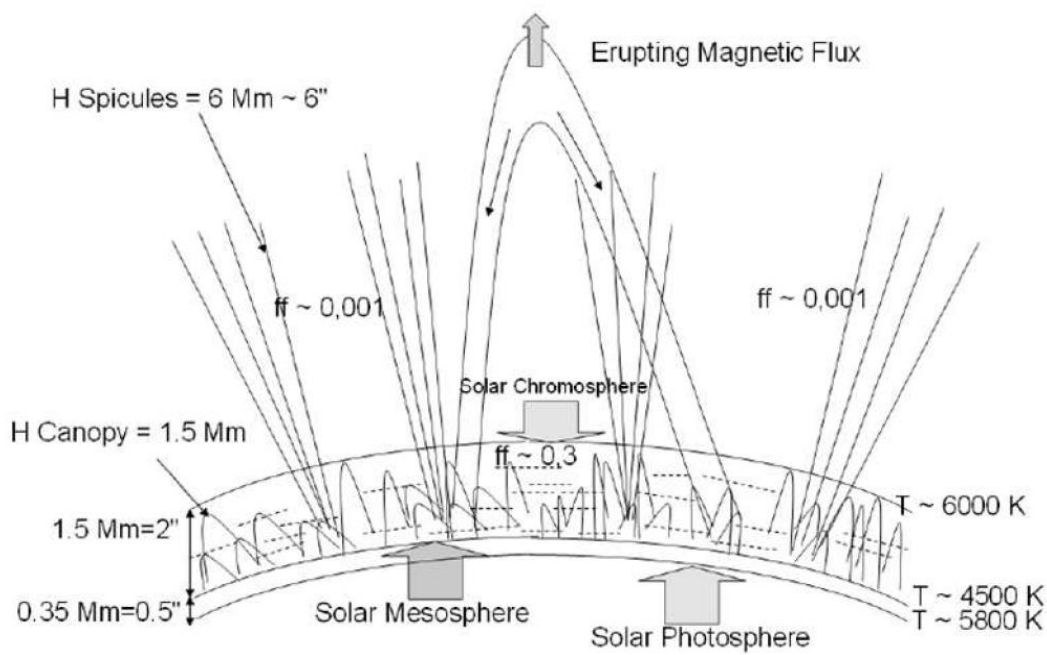

Figure 7.3.3. The Filling Factor (ff) or heterogeneity factor, is the ratio between the volume occupied by the plasma over the total volume, because the plasma is magnetically confined. The inversion of the temperature occurs around 350 Km, or 0.50 arcsec above the photosphere. (Sigismondi et al. 2011)

# Conclusion

In this study we take into account the potentiality of the observation of eclipses in defining the luminosity profile of the edge of the Sun.

From the first observations of Halley, huge developments in the method and means have been made, from the consideration of the Baily's beads, to the recent mapping of the lunar surface by the satellite Kaguya.

An improvement takes place in this study, considering the bead as a light curve forged by the LDF and the profile of the lunar valley. A first application has been described in this study. We obtained a detailed profile constraining the inflection point position. It shows the functionality of the method.

A further consideration is coming out: the inner region of the Sun's atmosphere is brighter than expected around the visible band, and can be easily confused with the photosphere. This could be a good track for the investigation of the enigmatic eyewitnesses of the historical eclipses.

Future observations should take into account these results to achieve the goal of obtaining the inflection point position, which is the conventional definition of the solar edge. In this study we recommended many possible improvements to obtain a detailed profile comparable to the space astrometry.

The current space mission Picard is a good opportunity to compare the results. It is therefore recommended observation of the next eclipses at the same bandpass at which the satellite Picard works.

The monitoring of the Sun is attracting increasing interest in recent years mainly because the maximum of solar activity in cycle 24 is lower than expected. It is possible that the Sun slips back into a new Maunder minimum. This has interesting implications on Earth's climate, but it is also a testing ground for theories that link the diameter of the Sun with its activity.